\documentclass{aa}
\usepackage{graphicx}
\usepackage{txfonts}
\usepackage{natbib}
\usepackage{verbatim}
\usepackage{longtable}
\usepackage{afterpage}
\usepackage{multirow}
\bibpunct{(}{)}{;}{a}{}{,}

\begin{document}

\title{A general model for the identification of \emph{specific} 
PAHs in the far\textendash IR}

\author{G. Mulas\inst{1,2}
        \and
        G. Malloci\inst{1,2}
        \and
        C. Joblin\inst{2}
        \and
        D. Toublanc\inst{2}
}

\institute{INAF~\textendash~Osservatorio Astronomico di Cagliari~\textendash~Astrochemistry 
Group, Strada n.54, Loc. Poggio dei Pini, I\textendash09012 Capoterra (CA), Italy 
\textemdash{} \email{[gmulas; gmalloci]@ca.astro.it}
\and
Centre d'Etude Spatiale des Rayonnements, CNRS et Universit\'e Paul 
Sabatier\textendash Toulouse~3, Observatoire Midi-Pyr\'en\'ees, 9 Avenue du colonel Roche,
31028 Toulouse cedex 04, France 
\textemdash{} \email{[giuliano.malloci; christine.joblin; dominique.toublanc]@cesr.fr}
}

\date{Received ?; accepted ?}

\abstract
{\emph{Context}. 
In the framework of the interstellar PAH hypothesis, far\textendash IR 
skeletal bands are expected to be a fingerprint of single species in 
this class.\\
\emph{Aims}. 
A detailed model of the photo\textendash physics of interstellar PAHs is 
required for such single\textendash molecule identification of their far-IR features in 
the presently available Infrared Space Observatory data and in those of the
forthcoming Herschel Space Observatory mission.\\
\emph{Methods}.
We modelled the detailed photophysics of a vast sample of species in
different radiation fields, using a compendium of Monte\textendash Carlo techniques
and quantum\textendash chemical calculations. This enabled us to validate the use
of purely theoretical data and assess the expected accuracy and
reliability of the resulting synthetic far\textendash IR emission spectra.\\
\emph{Results}.
We produce positions and intensities of the expected far\textendash IR features
which ought to be emitted by each species in the sample in the considered 
radiation fields. A composite emission spectrum for our sample is computed 
for one of the most favourable sources for detection, namely the Red 
Rectangle nebula. The resulting spectrum is compared with the estimated 
dust emission in the same source, to assess the dependence of detectability 
on key molecular parameters.\\
\emph{Conclusions}. 
Identifying specific PAHs from their far\textendash IR features is going to be a 
difficult feat in general, still it may well be possible under 
favourable conditions.\\

}

\authorrunning{Mulas, Malloci, Joblin \& Toublanc}
\titlerunning{A model for the identification of PAHs in the far\textendash IR}

\maketitle

\section{Introduction}\label{introduction}
The hypothesis of the ubiquitous presence of free gas\textendash phase polycyclic 
aromatic hydrocarbons (PAHs) in the interstellar medium (ISM) originated 
about 20 years ago \citep{leg84,all85}. Due to their spectral properties 
and their high photostability, these molecules were suggested as the most 
natural interpretation for the so\textendash called ``Aromatic Infrared Bands'' 
(AIBs), a set of emission bands observed near 3.3, 6.2, 7.7, 8.6,  
11.3 and 12.7~$\mu$m, in many dusty environments excited by UV photons
\citep{leg89,all89}. The AIBs are the spectral fingerprint of the excitation 
of vibrations in aromatic C\textendash C and C\textendash H bonds \citep{dul81}. Furthermore, 
PAHs and their cations were also supposed to account for a subset of 
``Diffuse Interstellar Bands'' (DIBs) \citep{leg85,van85,cra85}, more than 
$300$ absorption features observed in the near-UV, visible and near-IR in 
the spectra of reddened stars \citep{her95,ehr00,dra03} in the Milky Way as 
well as in external galaxies \citep{ehr02,sol05}. 

PAHs are believed to play an important role in the physics 
and chemistry of the ISM, showing intermediate properties between gas and 
dust phases, i.e. behaving at the same time both as very small grains and 
large molecules. This has motivated many observational, experimental and 
theoretical efforts in the last two decades, which confirmed PAH\textendash related 
species to be promising candidates to explain AIBs and some of the DIBs, 
accounting for a substantial fraction of the total interstellar carbon budget 
\citep{sal99b}. 
However, despite the large number of studies
and the tentative identification of neutral anthracene and 
pyrene in the Red Rectangle nebula \citep{vij04,vij05,mul06}, no 
\emph{definitive} spectral identification of any \emph{specific} 
individual member in this class exists to date. 

Concerning optical absorption spectra, in fact, while the large number of 
low\textendash temperature matrix isolation spectroscopy studies of PAHs 
are of fundamental importance to assess the link 
between PAHs and DIBs \citep{job95,sal96,sal99a,sal99b,rui02,rui05}, at the 
same time they do not permit an unambiguous identification of any single PAH, 
due to the unpredictable matrix\textendash induced broadening and shift of the absorption
bands. Measurements of the electronic absorption spectrum of cold PAH ions 
in gas\textendash phase \citep{rom99,pin99,bre99,bie03,bie04,suk04,tan05b,tan05a}
are ``the right way'' to identify specific PAHs based 
on their optical absorption spectrum, but each of them still represents a 
demanding experimental task, which can hardly be generalised in a systematic 
way to a representative sample of this vast class of molecules. 

As to IR spectroscopy, ``classical'' AIBs do not permit an 
unambiguous identification of any single species, because 
vibrational transitions in the near and medium IR, are a common
feature of the whole class of PAHs \citep{lan96,sal99b}. These vibrational 
modes just probe specific chemical bonds and not the overall structure 
of the whole molecule. Indeed, the emission in these bands is usually 
explained assuming a whole population of different PAHs to contribute 
to them \citep{shu93,coo98,all99,bak01b,bak01a,pec02}. 

On the other hand, every single such molecule ought to show a unique 
spectral fingerprint in the far\textendash IR spectral region, which contains 
the low\textendash frequency vibrational modes associated with collective oscillations 
of the whole skeletal structure
\citep{zha96,job02,mul03}. Moreover, far\textendash IR bands are expected to be 
emitted preferentially when the excitation energy of the molecule is 
relatively low \citep{job02}; this results in slower internal vibrational 
redistribution of energy and thus smaller lifetime broadening, implying 
that their rotational envelopes might be still discernible, as hinted by 
the gas phase laboratory measurements of \citet{zha96}. 
It might then be possible to resolve the rotational structure with a
high resolution spectrometer, such as the HIFI heterodyne spectrometer on 
board of the forthcoming Herschel Space Observatory (HSO)
mission\footnote{\texttt{www.sron.nl/divisions/lea/hifi/directory.html}},
providing one more crucial identification 
element for interstellar PAHs \citep{job02}.

The far\textendash IR spectrum emitted by a given interstellar PAH will depend 
essentially just on its molecular properties and on the exciting radiation 
field, and can be modelled in detail if all these ``ingredients'' are 
known \citep{job02,mal03c,mul03,mul06}.
Here we make use of a database of theoretical 
spectral properties of a sample of PAHs \citep{mal06}
to derive their expected far\textendash IR emission in a grid of radiation fields (RFs) 
covering some of the environments in which the AIBs, commonly attributed to
PAHs, have been observed. 

In this paper we describe the modelling procedure, we detail the 
approximations used and assess their impact on the resulting calculated 
spectra, both in terms of absolute fluxes and band ratios. Then we 
present the calculated spectra for a large sample of PAHs and their cations.
Our approach can be used also to model molecular rotation, and thus 
the expected rotational envelopes of far\textendash IR emission bands 
\citep{job02} 
This latter calculation is very demanding from a computational point of 
view 
and is the subject of more forthcoming work \citep{mul06}.

In Sect.~\ref{modelling} we summarise the modelling 
procedure we used. We then proceed in Sect.~\ref{validation} to validate 
our modelling approach with respect to the use of 
theoretical UV\textendash visible photo\textendash absorption spectra (Sect.~\ref{validation1}), the
impact of some necessary simplifying assumptions for the molecular relaxation 
processes following electronic excitation (Sect.~\ref{relaxation})
and its comparison with an independently developed model 
(Sect.~\ref{dominique}).
Results are presented in Sect.~\ref{results}, while
Sect.~\ref{discussion} discusses the diagnostic potential and limitations 
of the calculated spectra.

\section{Modelling procedure}\label{modelling}

Our Monte\textendash Carlo (MC) modelling procedure is described in 
detail elsewhere \citep{mul98,job02,mal03c,mul03,mul06}. It requires the
following parameters: \begin{enumerate}
\item \label{itemRF} the exciting radiation field;
\item \label{itemsigma} the UV\textendash visible photo\textendash absorption cross\textendash section up to 
the Lyman limit;
\item \label{itempaths} the dominant de\textendash excitation pathways following the 
absorption of an UV\textendash visible photon;
\item \label{itemvibs} the complete vibrational analysis in the electronic 
ground state (frequencies, symmetry of the modes and intensities of the 
IR\textendash active modes).
\end{enumerate}


These data, along with some general assumptions on the photophysics 
of PAHs, enable one to calculate all the probabilities involved in the 
exchange of photons between the molecule and the radiation field.
Until recently the general applicability of this procedure was hindered by the 
scarcity of complete photo\textendash absorption spectra of 
neutral and cationic PAHs ranging from the visible to the Lyman limit at 
13.6~eV. However, we recently produced theoretical UV\textendash visible 
photo\textendash absorption spectra for a sample of 20 PAHs and their cations,
using the Time\textendash Dependent Density Functional Theory \citep[TD\textendash DFT, e.~g.][]
{mar04} implementation in the \textsc{Octopus} computer code \citep{mar03}, 
and showed them to be in good agreement with experimental data when available 
\citep{job92a,job92b}, validating their use as a quite decent surrogate 
when they are not \citep{mal04}.

We here extend our previous work \citep{mul03}
to the whole sample of molecules studied in \citet{mal04}, using the 
theoretical photo\textendash absorption cross\textendash sections published therein and 
computing the vibrational frequencies with the \textsc{NWChem} code 
\citep{str03}. Following previously published calibration calculations
\citep{lan96,bau97}, we obtained the full vibrational analyses 
using the exchange\textendash correlation functional B3LYP \citep{bec93} and the 
\mbox{4-31G} basis set \citep{fri84}. 
The vibrational frequencies obtained at the B3LYP/4-31G level of theory 
are usually scaled with an empirical factor of $0.958$, which accounts for 
anharmonicity and other minor corrections \citep{lan96,bau97,hud01} and 
brings them into near coincidence with experimental data. Our calculated 
vibrational frequencies are in good agreement with the results previously 
published. Since the aim of the present paper is to 
produce as accurate as possible synthetic spectra for direct comparison 
with observational data, we did scale the computed vibrational frequencies 
with the above empirical factor, differently from what we did in our 
proof\textendash of\textendash concept papers on ovalene \citep{mal03c,mul03}. The only gas\textendash phase 
emission data 
for PAHs in the far\textendash IR currently available are, to the best of our knowledge, 
those by \citet{zha96} and those by \citet{pir04}.
The comparison between our calculated band positions and the experimental 
data is shown in Table~\ref{empirical_factor}. We remark that the differences 
among frequencies may be partly due to the temperature effects, 
since in both experiments PAHs have to be heated. Inspection of 
Table~\ref{empirical_factor} shows that the empirical scale factor calibrated 
on mid\textendash IR data is probably not the best one for far\textendash IR bands. More work is 
needed both from an experimental point of view, to study the temperature 
effect on band position and width, and from a theoretical point of view 
to derive the best empirical 
scaling procedure for this spectral range. For the time being, the calculated 
positions of the bands in this work may therefore be affected by 
a small systematic error due to the scaling adopted, and some caution is in 
order in their interpretation. All the vibrational frequencies (not only the 
IR\textendash active ones) are used within the MC model to compute the density 
of vibrational states as a function of energy \citep[e.~g.][]{mul98,coo98}.
These data are being integrated in an online database of computed 
molecular properties, which is under construction \citep{mal06}.

\begin{table}[t!]
\begin{center}
\caption{Comparison between available far\textendash IR gas\textendash phase laboratory 
measurements \citep[][ first column]{zha96,pir04}, our DFT 
results, both unscaled (second column) and scaled with an empirical factor 
of $\sim0.958$ (third column) and the previously published DFT results
\citep[][ last column]{lan96,mar96}.}
\label{empirical_factor}
\begin{tabular}{cccc}
\hline 
\hline
\noalign{\smallskip}
\multicolumn{4}{c}{Band positions (cm$^{-1}$)} \\
\noalign{\smallskip}
Experimental & Unscaled DFT & Scaled DFT  & Previous DFT \\
\noalign{\smallskip}
\hline
\noalign{\smallskip}
\multicolumn{4}{c}{Naphthalene (C$_{10}$H$_{8}$)}\\
\noalign{\smallskip}
167.0$^{a}$, 165.6$^{b}$ & 178.2 & 170.8 & 171.8$^{c}$, 171$^{d}$ \\
360.6$^{a}$, 358.7$^{b}$ & 376.3 & 360.5 & 357.7$^{c}$, 359$^{d}$ \\
\noalign{\smallskip}
\hline
\noalign{\smallskip}
\multicolumn{4}{c}{Anthracene (C$_{14}$H$_{10}$)}\\
\noalign{\smallskip}
87.2$^{b}$ & 94.7 & 90.7 & 91.0$^{c}$, 90$^{d}$ \\
227.3$^{b}$ & 239.0 & 228.9 & 232.3$^{c}$, 232$^{d}$ \\
\noalign{\smallskip}
\hline
\noalign{\smallskip}
\multicolumn{4}{c}{Phenanthrene (C$_{14}$H$_{10}$)}\\
\noalign{\smallskip}
96.7$^{b}$ & 104.3 & 99.9 & 99$^{d}$ \\
221.1$^{b}$ & 235.8 & 225.9 & 226.8$^{c}$, 226$^{d}$ \\
\noalign{\smallskip}
\hline
\noalign{\smallskip}
\multicolumn{4}{c}{Pyrene (C$_{16}$H$_{10}$)}\\
\noalign{\smallskip}
95.0$^{a}$, 94.6$^{b}$ & 102.7 & 98.4 & 98.6$^{c}$ \\
214.2$^{a}$, 206.7$^{b}$ & 218.4 & 209.2 & 210.0$^{c}$ \\
350.0$^{a}$ & 368.6 & 353.1 & 353.5$^{c}$ \\
\noalign{\smallskip}
\hline
\noalign{\smallskip}
\multicolumn{4}{c}{Chrysene (C$_{18}$H$_{12}$)}\\
\noalign{\smallskip}
169.9$^{b}$ & 192.8 & 184.7 & \textemdash{} \\
231.5$^{a}$, 229.9$^{b}$ & 242.7 & 232.5 & 233.0$^{c}$ \\
\noalign{\smallskip}
\hline
\noalign{\smallskip}
\multicolumn{4}{c}{Perylene (C$_{20}$H$_{12}$)}\\
\noalign{\smallskip}
88.1$^{b}$ & 98.9 & 94.8 & 98.6$^{c}$ \\
 \textemdash{}    & 182.7 & 175.0 & 175.9$^{c}$\\
464.9$^{b}$ & 479.7 & 459.6 & 462.0$^{c}$ \\
\noalign{\smallskip}
\hline
\noalign{\smallskip}
\end{tabular}
\end{center}
\vspace{-0.4cm}
$^{a}$ Gas\textendash phase emission measurements \citep{zha96};\\ 
$^{b}$ Gas\textendash phase emission measurements \citep{pir04};\\
$^{c}$ B3LYP/4-31G calculation \citep{lan96};\\
$^{d}$ B3LYP/cc-pvdz calculation \citep{mar96}.
\vspace{-0.3cm}
\end{table}

\begin{figure}
 
\includegraphics{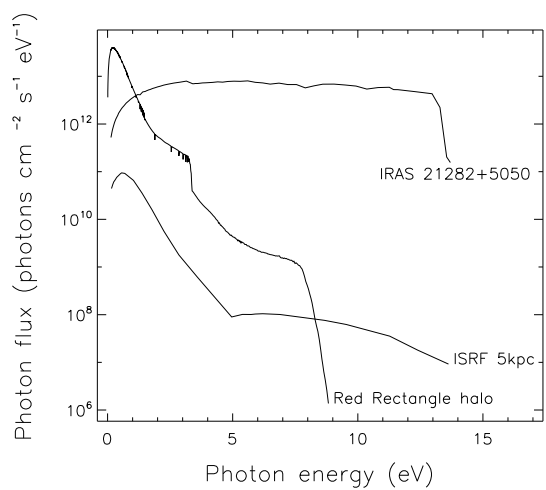}
\caption{The RFs considered, namely the ISRF 
given by \citet{mat83}, the RF of the extended halo 
surrounding the Red Rectangle proto\textendash planetary nebula and the relatively 
UV\textendash rich RF in the PDR of the planetary nebula
IRAS~21282+5050; the latter two were obtained respectively from observational
data \citep{vij04,vij05} and from an appropriate stellar atmospheric model of 
\citet{kur92}.
}
\label{radfields}

\end{figure}

For the present simulations we used the relatively UV\textendash poor RF of the 
extended halo surrounding the Red Rectangle proto\textendash planetary nebula 
\citep{men02,mul06}, the interstellar RF (ISRF) 
in the galactic plane at about 5~kpc from the galactic centre 
\citep[as given by][]{mat83} and the relatively UV\textendash rich 
RF in the photodissociation region (PDR) of the planetary nebula 
IRAS~21282+5050 \citep{pec02}. The RFs considered are shown in 
Fig.~\ref{radfields}; note that the RF of the extended halo of the Red 
Rectangle drops to negligible values above $\sim9$~eV on the scale of this plot. 
This choice of RFs is not meant by any means to represent a choice of the
environments in which we expect the far\textendash IR bands of PAHs to be most easily
detected. They were chosen to span the range of conditions in which
AIBs have been detected. As will be explained later in Sect.~\ref{discussion}, 
this choice is not a limiting one.

\section{Model validation}\label{validation}

\subsection{Use of theoretical photo\textendash absorption spectra}
\label{validation1}

To test and validate the use of the calculated absorption spectra of
\citet{mal04} in the MC simulation we compared them 
with the results obtained applying the same modelling procedure on the 
experimental spectra of anthracene, pyrene, coronene and ovalene 
\citep{job92a,job92b}.
For all of these molecules the agreement between theoretical and laboratory 
data is reasonably good, with typical absolute errors of $\sim$0.3~eV between 
the calculated and measured positions of the absorption bands, as expected for 
TD\textendash DFT calculations \citep[e.~g. ][]{hir99,hir03,mal04}.
In some cases 
these relatively small differences are widely amplified in the low\textendash energy 
part of the estimated photon absorption rates for the Red Rectangle and, to a 
lesser extent, for the ISRF, because these RFs are very 
steep (see Fig.~\ref{radfields}), i.e. the photon flux changes up to about 
an order of magnitude in 0.3~eV.
To make sure that any differences depend \emph{only} on the different 
$\sigma$ adopted, for this comparison we used 
in all cases theoretical frequencies and intensities of the vibrational bands.
%
%

Appendix~\ref{rates} shows in detail the comparison between the synthetic 
($\sigma_\mathrm{th}$) and laboratory ($\sigma_\mathrm{lab}$) photo\textendash absorption cross 
sections of anthracene, pyrene, coronene and ovalene, as well as the impact of 
their differences 
on the estimated 
energy\textendash dependent absorption rates in the three RFs considered. The agreement 
between IR emission fluxes calculated using $\sigma_\mathrm{th}$ and those 
calculated using $\sigma_\mathrm{lab}$ is found to be \emph{quantitatively} very 
good for all of the four molecules, the worst error being a factor of 2 
for pyrene in the Red Rectangle, much better in all other cases. In all cases, 
\emph{relative} intensities among IR bands appear very accurate, with 
differences of the order of, at most, $\sim$20\%.

\subsection{Molecular relaxation processes}\label{relaxation}

The predicted IR emission spectra 
were derived under the assumption that the absorption of a UV\textendash visible 
photon be immediately followed by one or several non\textendash radiative transitions 
leaving the molecule in a highly excited vibrational level of the 
electronic ground\textendash state $E_0$, from which it relaxes by vibrational 
transitions \citep[][]{all89,lea95a,lea95b}. 
Quite generally, upon reaching $E_0$, any PAH very 
quickly redistributes its remaining excitation energy among the accessible 
vibrational modes, to achieve Internal Thermal Equilibrium (ITE). Fast
intramolecular radiationless transitions ensure that ITE be efficiently
established between vibrational transitions. 

Since the molecule is a closed, isolated system in which energy is conserved 
between photon emissions or absorptions, microcanonical statistics 
applies for the distribution of energy among vibrational degrees of 
freedom in this regime. However, the rate of intramolecular transitions is a 
sensitive function of excitation energy, and becomes ineffective below a 
molecule\textendash specific decoupling threshold $E_\mathrm{dec}$. Since energy gets 
emitted in discrete steps during relaxation, the molecule is likely to 
effectively jump from an excitation energy above the threshold, i.e. in ITE, 
to one below the threshold, in which redistribution of energy does not occur
anymore. When this happens, the number of vibrational quanta in each 
energetically accessible mode is likely to be frozen according to 
microcanonical statistics at that energy. In perfect isolation conditions, 
all of these vibrational quanta must be emitted via radiative transitions, 
even if they are electric\textendash dipole forbidden. Since the population of 
vibrational modes upon decoupling depends \emph{only} on the statistics of 
accessible states, very low\textendash energy modes are clearly favoured, in the final 
redistribution of excitation energy immediately before decoupling, 
regardless of their IR\textendash activity. 

The dominant process competing with radiative relaxation is the absorption 
of another UV\textendash visible photon before the molecule reaches the vibrational 
ground state. When this happens, the total available energy rises 
above $E_\mathrm{dec}$ again. 
Emission in the weakest IR bands, towards the low\textendash energy end, essentially 
happens only when the stronger ones are energetically inaccessible;
if vibrational cascades are interrupted, IR\textendash inactive and weak IR\textendash active 
bands are consequently suppressed, in favour of the stronger 
ones, which are emitted much more quickly when the molecule excitation 
energy is high enough. Since no data is available on the spontaneous 
transition rates in IR\textendash inactive vibrational modes of PAHs and their 
calculation is not presently implemented in commonly available quantum 
chemistry packages, we modelled them as IR\textendash active bands with an oscillator 
strength equal to the average of all IR\textendash active bands scaled down by a factor 
10$^4$, which yields the appropriate order of magnitude for electric quadrupole 
transitions. 
As to the decoupling energy $E_\mathrm{dec}$, its only available direct 
measurements 
to date were performed on perylene and anthracene \citep{bou83,fel84}; in our 
model runs we therefore chose the threshold for other PAHs by requiring the 
density of vibrational states to be the same as in anthracene at its measured 
$E_\mathrm{dec}$ ($\sim$1700~cm$^{-1}$).

In all environments considered, collisions are less frequent than
photon absorption events \citep{omo86}, and we neglected them in our 
treatment. They may play a role in populating/depopulating IR\textendash inactive 
modes after decoupling, when timescales between photon emission/absorption 
events are relatively long, especially for ions, which have larger collisional 
cross sections. This can be expected to become relatively more important for 
environments for which the ratio between gas density and UV\textendash visible radiation 
density is higher. However, while not necessarily negligible, this is a higher 
order correction in all cases considered here.

Immediate conversion of all excitation energy to vibrational energy 
in $E_0$
is not necessarily the only relaxation path 
available. 
Photoionisation can be relevant for neutral PAHs in the RFs considered,
hence we took it into account in our modelling calculating
the energy\textendash dependent yield using the analytical expression given by
\citet{lep01} and the experimental ionisation energies 
\citep{lia05}. Since the RFs considered do not contain significant
fluxes of photons capable of doubly ionising the PAHs considered here, 
we neglect ionisation for cations. The probabilities of the main dissociation 
channels have been estimated to be negligible for our purposes 
\citep{all96a,all96b}. However, recent results with the PIRENEA 
experiment \citep{job06} show that the coronene cation loses one H atom at 
an average excitation energy of 10.5~eV. Therefore, a fully detailed model 
would need to follow the photophysical and chemical evolution of a PAH 
population including ionisation, electron recombination, photodissociation 
and reactivity in particular with H atoms. This is outside of the scope 
of the present paper.
In any case, 
as far as the IR emission spectrum is concerned, the effect of
photodissociation and/or photoionisation is a higher order correction. 
Indeed, the overall IR emission spectrum stems from the sum over a large 
number of relaxation cascades, which in this respect can be considered 
essentially independent, and photodissociation, photoionisation etc. happen 
only in a small fraction of them.

Provided they neither ionise nor dissociate, all closed\textendash shell 
(e.~g. neutral, fully hydrogenated) PAHs, upon excitation 
from their ground electronic singlet state
\mbox{$\mathrm{S}_n\gets\mathrm{S}_0$}, will undergo very fast 
internal conversion to a low lying singlet electronic level
$\mathrm{S}_m$, which is almost always $\mathrm{S}_1$ \citep[e.~g.]
[]{lea95a,lea95b}.
Afterwards, the molecule relaxes 
radiatively in the visible and/or in the IR. For a thorough 
discussion see e.~g.~\citet{mul06} where all the relaxation channels 
were modelled in detail. A short resum\'e is given also in 
Appendix~\ref{modelcomp}.


\begin{table*}[ht!]
\caption{Comparison between two sets of model runs for coronene
(C$_{24}$H$_{12}$) in different exciting RFs. The results from the present
model based on \citet{mul98} (disregarding the possibility of ionisation) 
are labelled as model~1, while those from the corrected 
model of \citet{job02} are labelled as model~2. For each model 
run and each band we list the absolute integrated flux and, in parentheses, 
the flux fraction.}
\label{coronene_jobmul_table}
\begin{center}
\begin{tabular}{ccccccc}
\hline \hline 
\noalign{\smallskip}
& \multicolumn{2}{c}{ISRF} & \multicolumn{2}{c}{Red Rectangle} & 
\multicolumn{2}{c}{IRAS~21282+5050}\\
Peak & model~1 & model~2 & model~1 & model~2 & model~1 & model~2 \\
($\mu$m) & (W$ $sr$^{-1}$) (\%) & (W$ $sr$^{-1}$) (\%) & (W$ $sr$^{-1}$) (\%) & 
(W$ $sr$^{-1}$) (\%) & (W$ $sr$^{-1}$) (\%) & (W$ $sr$^{-1}$) (\%) \\
\noalign{\smallskip} \hline \noalign{\smallskip}
3.26 & $3.8 \,\, 10^{-27}$ (37.96) & $3.9 \,\, 10^{-27}$ (38.15) & 
$5.8 \,\, 10^{-26}$ (24.76) & $6.3 \,\, 10^{-26}$ (25.36) & 
$4.3 \,\, 10^{-22}$ (46.68) & $4.2 \,\, 10^{-22}$ (46.50) \\
3.29 & $2.6 \,\, 10^{-28}$ (2.58) & $2.6 \,\, 10^{-28}$ (2.56) & 
$4.0 \,\, 10^{-27}$ (1.72) & $4.2 \,\, 10^{-27}$ (1.70) & 
$2.9 \,\, 10^{-23}$ (3.18) & $2.9 \,\, 10^{-23}$ (3.16) \\
6.24 & $5.2 \,\, 10^{-28}$ (5.15) & $5.4 \,\, 10^{-28}$ (5.21) & 
$1.3 \,\, 10^{-26}$ (5.47) & $1.4 \,\, 10^{-26}$ (5.51) & 
$4.6 \,\, 10^{-23}$ (5.01) & $4.5 \,\, 10^{-23}$ (5.00) \\
6.69 & $9.2 \,\, 10^{-30}$ (0.09) & $9.7 \,\, 10^{-30}$ (0.09) & 
$2.4 \,\, 10^{-28}$ (0.10) & $2.6 \,\, 10^{-28}$ (0.11) & 
$8.0 \,\, 10^{-25}$ (0.09) & $6.9 \,\, 10^{-25}$ (0.08) \\
7.21 & $1.1 \,\, 10^{-28}$ (1.07) & $1.1 \,\, 10^{-28}$ (1.08) & 
$2.8 \,\, 10^{-27}$ (1.22) & $3.1 \,\, 10^{-27}$ (1.24) & 
$9.1 \,\, 10^{-24}$ (0.99) & $8.8 \,\, 10^{-24}$ (0.97) \\
7.63 & $6.6 \,\, 10^{-28}$ (6.59) & $7.0 \,\, 10^{-28}$ (6.77) & 
$1.7 \,\, 10^{-26}$ (7.49) & $1.9 \,\, 10^{-26}$ (7.73) & 
$5.5 \,\, 10^{-23}$ (6.05) & $5.5 \,\, 10^{-23}$ (6.10) \\
8.25 & $3.0 \,\, 10^{-28}$ (2.92) & $3.0 \,\, 10^{-28}$ (2.93) & 
$8.0 \,\, 10^{-27}$ (3.42) & $8.7 \,\, 10^{-27}$ (3.53) & 
$2.4 \,\, 10^{-23}$ (2.63) & $2.4 \,\, 10^{-23}$ (2.64) \\
8.78 & $3.1 \,\, 10^{-28}$ (3.10) & $3.3 \,\, 10^{-28}$ (3.21) & 
$8.8 \,\, 10^{-27}$ (3.79) & $9.5 \,\, 10^{-27}$ (3.82) & 
$8.8 \,\, 10^{-23}$ (2.75) & $2.5 \,\, 10^{-23}$ (2.75) \\
11.57 & $3.4 \,\, 10^{-27}$ (33.25) & $3.5 \,\, 10^{-27}$ (33.89) & 
$9.9 \,\, 10^{-26}$ (42.48) & $1.1 \,\, 10^{-25}$ (42.72) & 
$2.6 \,\, 10^{-22}$ (27.89) & $2.5 \,\, 10^{-22}$ (28.02) \\
12.43 & $1.1 \,\, 10^{-29}$ (0.11) & $1.1 \,\, 10^{-29}$ (0.11) & 
$3.3 \,\, 10^{-28}$ (0.14) & $3.7 \,\, 10^{-28}$ (0.15) & 
$8.0 \,\, 10^{-25}$ (0.09) & $8.3 \,\, 10^{-25}$ (0.09) \\
12.90 & $1.6 \,\, 10^{-28}$ (1.56) & $1.6 \,\, 10^{-28}$ (1.60) & 
$4.7 \,\, 10^{-27}$ (2.03) & $5.3 \,\, 10^{-27}$ (2.12) & 
$1.2 \,\, 10^{-23}$ (1.29) & $1.2 \,\, 10^{-23}$ (1.30) \\
18.21 & $3.9 \,\, 10^{-28}$ (3.91) & $4.1 \,\, 10^{-28}$ (4.01) & 
$1.2 \,\, 10^{-26}$ (5.29) & $1.3 \,\, 10^{-26}$ (5.44) & 
$2.8 \,\, 10^{-23}$ (3.04) & $2.8 \,\, 10^{-23}$ (3.11) \\
26.20 & $2.8 \,\, 10^{-29}$ (0.28) & $2.3 \,\, 10^{-29}$ (0.22) & 
$9.7 \,\, 10^{-28}$ (0.42) & $7.8 \,\, 10^{-27}$ (0.32) & 
$1.8 \,\, 10^{-24}$ (0.19) & $1.5 \,\, 10^{-24}$ (0.16)\\
80.60 & $2.4 \,\, 10^{-29}$ (0.24) & $1.6 \,\, 10^{-29}$ (0.16) & 
$8.4 \,\, 10^{-28}$ (0.36) & $6.0 \,\, 10^{-28}$ (0.24) & 
$1.0 \,\, 10^{-24}$ (0.11) & $7.6 \,\, 10^{-25}$ (0.08)\\
\noalign{\smallskip} \hline
\end{tabular}
\end{center}
\vspace{-0.3cm}
\end{table*}

In open\textendash shell PAHs (e.~g. singly charged ions of fully hydrogenated species 
) usually both the ground $\mathrm{D}_0$ and 
first excited $\mathrm{D}_1$ electronic states are doublets and are 
connected by an electric dipole\textendash permitted electronic transition \citep{bir70}.
In this situation internal conversion (IC) 
\mbox{$\mathrm{D}_n\leadsto\mathrm{D}_0$} followed by vibrational transitions, 
is expected to always be the dominant relaxation channel 
\citep{lea95a,lea95b}, fluorescence and phosphorescence being negligible.
Energy\textendash dependent quantum yields for the main relaxation 
channels of closed\textendash shell PAHs are known for very few molecules 
\citep{bre05} and only IC to $\mathrm{D}_0$
is important for open\textendash shell
species. Therefore, since the present work aims to systematically explore 
the far\textendash IR spectral fingerprint of a large sample of molecules, we chose 
to neglect the relaxation channels via fluorescence and phosphorescence.

To assess the impact of this simplification, the spectra presented here can 
be compared with those we obtained in \citet{mul06}, in which we modelled in 
full detail all relaxation channels. 
In Appendix~\ref{modelcomp} we present such a comparison, which
shows that neglecting fluorescence and phosphorescence 
leads to an overestimation of the absolute vibrational emission fluxes in 
each band by up to a factor $\sim$2 in the worst case, for the high energy bands,
while low energy bands are almost unaffected. This means that the calculated 
spectra, despite the use of $\sigma_\mathrm{th}$
and the neglect of relaxation channels involving fluorescence and 
phosphorescence, 
still retain their full potential to identify specific PAHs, this 
approximation appearing to impact essentially the accuracy of calculated 
column densities. 

\subsection{Comparison with an alternative model}
\label{dominique}

The model used here is an enhanced version of the one described in 
\citet{mul98}.
As an internal consistency test, we also derived the IR emission 
spectra using an independent 
model based on \citet{job02}. Such a comparison 
enabled us to discover and correct some implementation errors in both 
of them.
More specifically, an incorrect distribution of vibrational levels was 
used in \citet{job02} and biased absolute values of IR emission band 
intensities were produced in \citet{mul03}. 

The result of this comparison for neutral coronene is shown in 
Table~\ref{coronene_jobmul_table}, which clearly demonstrate 
that  the two models are very close. For both models we assumed a decoupling 
energy threshold of $\sim$1200~cm$^{-1}$, following the prescription discussed in 
Sect.~\ref{relaxation}. We remark that the model used here, as previously 
explained, takes into account the possibility that absorption of a 
high\textendash energy photon results in an ionisation instead of IR emission. Since 
the model by \citet{job02} does not include this possiblity, for a 
proper comparison we deliberately excluded it in the simulations compared in 
Table~\ref{coronene_jobmul_table}.
This results in a slight overestimation of emission which is larger
for high\textendash energy bands; the latter tend to be preferentially emitted when the
molecule is highly excited, hence after the absorption of harder photons 
which, in turn, are more likely to ionise it.

\begin{figure}[t!]
\includegraphics[width=\hsize]{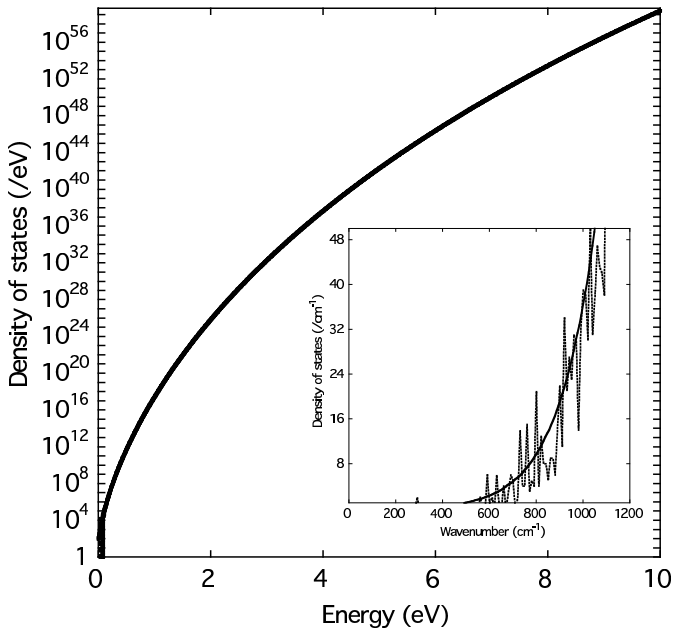}
\caption{Comparison between the DoS of neutral coronene 
computed by the current model, based on \citet{mul98} (continuous line) 
and that of \citet{job02} (dotted line). The DoSs are 
perfectly coincident for energies well above the energy of a single 
vibrational quantum, while they differ for very small energies.}
\label{doscompare}

\end{figure}

The small remaining difference between the two models
is probably due to somewhat different algorithms used to 
estimate the density of vibrational states (DoS), namely the Beyer\textendash Swinehart 
\citep[BS,][]{bey73} method in \citet{job02} and a version of the 
Stein\textendash Rabinovitch \citep[SR,][]{ste73} method in \citet{mul98}, slightly 
modified to ensure its continuity down to low energies. The resulting DoSs 
are plotted in Fig.~\ref{doscompare}, which shows them to be essentially 
coincident except for very small energies, where the modified SR DoS is 
still smooth while the BS one strongly oscillates. 
Such a behaviour depends on the difference between the formal 
definition of DoS and the actual way it is calculated. For a finite 
system, as a large molecule, states are discrete and the DoS is thus not
well\textendash defined as a function, being instead a distribution, i.e. the 
sum of a collection of Dirac deltas, one for each discrete level. On the other 
hand, the integral of the DoS over any finite interval still is a 
well\textendash defined function. Both the BS and the SR algorithms count the number of 
vibrational states in a grid of finite 
energy intervals, and obtain the DoS dividing the resulting 
number by the bin size. This is not, therefore the DoS in its strict 
mathematical sense, but rather its average over a finite energy interval.
As a consequence, the ``DoS'' thus calculated is obviously dependent on the 
chosen bin size, this dependence becoming very extreme for increasingly 
small bin sizes, when the latter becomes comparable to the spacing of the
discrete levels. 

The way out of this apparent formal impasse is to recognise
that what is needed to calculate probabilities in the ergodic regime is 
\emph{not} really the exact DoS, as to its strict mathematical definition,
but rather its average over an energy range given by the Heisenberg
indetermination principle. Moreover, the DoS is \emph{only} used
when the molecule is in ITE, i.e. when intramolecular radiationless 
transitions are fast and, as a consequence, energy indetermination is 
large. Therefore, we slightly modified the SR algorithm to 
make sure that the resulting ``DoS'' is continuous regardless of the 
assumed energy bin size used for calculating the density, and that it 
converges to a well\textendash defined limiting value for vanishingly small bin size. 
Essentially, this is equivalent to calculating the average of the DoS over 
an energy bin of the same size as a single vibrational quantum of the lowest 
frequency mode of the molecule.

We stress that the results of the simulations are fundamentally 
independent of the algorithm used: statistics tend to wash out the effect 
of the strong oscillations in the BS DoS. However, for 
smaller and smaller bin sizes, and correspondingly stronger oscillations in
the DoS, increasingly long simulations are required to obtain numerical 
convergence. When using the explicitly smoothed DoS in the 
simulations, instead, numerical convergence is faster and independent of the
bin size adopted for the DoS calculation.

\section{Results} 
\label{results}

\begin{table*}
\caption{Integrated photon absorption rates $\mathcal{R}$ (s$^{-1}$) and 
average absorbed energy $\mathcal{E}$ (eV), computed using calculated 
photo\textendash absorption cross\textendash sections \citep{mal04} for the whole sample of 20 
neutral PAHs and their cations in the three RFs considered.}
\label{all_rates}
\begin{center}
\begin{tabular}{cccccccc}
\hline 
\hline
\noalign{\smallskip}
& & \multicolumn{6}{c}{Radiation field}\\
\noalign{\smallskip}
Molecule & Charge state & 
\multicolumn{2}{c}{ISRF}& \multicolumn{2}{c}{Red Rectangle} 
& \multicolumn{2}{c}{IRAS~21282+5050}\\
\noalign{\smallskip}
\cline{3-8}
\noalign{\smallskip}
& & $\mathcal{R}$ (s$^{-1}$) & $\mathcal{E}$ (eV) & $\mathcal{R}$ (s$^{-1}$)  
& $\mathcal{E}$ (eV) & $\mathcal{R}$ (s$^{-1}$) & $\mathcal{E}$ (eV)\\
\noalign{\smallskip}
\hline
\noalign{\smallskip}
\multirow{2}*{Naphthalene (C$_{10}$H$_{8}$)} 
& Neutral  
& 3.9 $\,$ 10$^{-8}$ & 7.9 & 5.7 $\,$ 10$^{-7}$ & 5.6 & 3.8 $\,$ 10$^{-3}$ & 8.9\\ 
& Cation 
& 6.6 $\,$ 10$^{-8}$ & 4.7 & 3.1 $\,$ 10$^{-6}$ & 2.8 & 3.1 $\,$ 10$^{-3}$ & 9.0\\
\multirow{2}*{Fluorene (C$_{13}$H$_{10}$)} 
& Neutral  
& 5.2 $\,$ 10$^{-8}$ & 7.8 & 8.9 $\,$ 10$^{-7}$ & 5.2 & 4.9 $\,$ 10$^{-3}$ & 9.2\\
& Cation  
& 1.3 $\,$ 10$^{-7}$ & 4.0 & 6.7 $\,$ 10$^{-6}$ & 2.7 & 4.1 $\,$ 10$^{-3}$ & 9.2\\
\multirow{2}*{Anthracene (C$_{14}$H$_{10}$)}
& Neutral  
& 6.4 $\,$ 10$^{-8}$ & 6.8 & 2.5 $\,$ 10$^{-6}$ & 4.0 & 5.2 $\,$ 10$^{-3}$ & 8.6\\
& Cation  
& 1.7 $\,$ 10$^{-7}$ & 3.4 & 8.5 $\,$ 10$^{-6}$ & 2.5 & 4.4 $\,$ 10$^{-3}$ & 8.6\\
\multirow{2}*{Phenanthrene (C$_{14}$H$_{10}$)}
& Neutral  
& 5.6 $\,$ 10$^{-8}$ & 7.4 & 1.1 $\,$ 10$^{-6}$ & 5.1 & 5.1 $\,$ 10$^{-3}$ & 8.8\\
& Cation 
& 1.7 $\,$ 10$^{-7}$ & 3.2 & 9.1 $\,$ 10$^{-6}$ & 2.3 & 4.3 $\,$ 10$^{-3}$ & 8.8\\
\multirow{2}*{Pyrene (C$_{16}$H$_{10}$)}
& Neutral 
& 8.1 $\,$ 10$^{-8}$ & 6.4 & 3.7 $\,$ 10$^{-6}$ & 3.8 & 5.7 $\,$ 10$^{-3}$ & 8.6\\
& Cation 
& 1.4 $\,$ 10$^{-7}$ & 4.0 & 9.2 $\,$ 10$^{-6}$ & 2.7 & 4.9 $\,$ 10$^{-3}$ & 8.6\\
\multirow{2}*{Fluoranthene (C$_{16}$H$_{10}$)}
& Neutral
& 7.6 $\,$ 10$^{-8}$ & 6.8 & 3.2 $\,$ 10$^{-6}$ & 3.8 & 6.0 $\,$ 10$^{-3}$ & 8.9\\
& Cation
& 2.4 $\,$ 10$^{-7}$ & 3.0 & 1.3 $\,$ 10$^{-5}$ & 2.2 & 5.1 $\,$ 10$^{-3}$ & 8.8\\
\multirow{2}*{Tetracene (C$_{18}$H$_{12}$)}
& Neutral
& 1.3 $\,$ 10$^{-7}$ & 5.4 & 5.8 $\,$ 10$^{-6}$ & 3.5 & 6.8 $\,$ 10$^{-3}$ & 8.4\\
& Cation
& 3.7 $\,$ 10$^{-7}$ & 2.8 & 1.9 $\,$ 10$^{-5}$ & 2.2 & 6.0 $\,$ 10$^{-3}$ & 8.4\\
\multirow{2}*{Chrysene (C$_{18}$H$_{12}$)}
& Neutral
& 8.6 $\,$ 10$^{-8}$ & 6.7 & 2.9 $\,$ 10$^{-6}$ & 4.3 & 6.7 $\,$ 10$^{-3}$ & 8.7\\
& Cation
& 4.6 $\,$ 10$^{-7}$ & 2.2 & 3.0 $\,$ 10$^{-5}$ & 1.6 & 5.9 $\,$ 10$^{-3}$ & 8.7\\
\multirow{2}*{Perylene (C$_{20}$H$_{12}$)}
& Neutral
& 1.7 $\,$ 10$^{-7}$ & 4.6 & 1.0 $\,$ 10$^{-5}$ & 2.9 & 7.1 $\,$ 10$^{-3}$ & 8.7\\
& Cation
& 2.3 $\,$ 10$^{-7}$ & 3.7 & 1.4 $\,$ 10$^{-5}$ & 2.6 & 6.2 $\,$ 10$^{-3}$ & 8.6\\
\multirow{2}*{Pentacene (C$_{22}$H$_{14}$)}
& Neutral
& 1.6 $\,$ 10$^{-7}$ & 5.4 & 7.2 $\,$ 10$^{-6}$ & 3.8 & 8.2 $\,$ 10$^{-3}$ & 8.3\\
& Cation
& 7.2 $\,$ 10$^{-7}$ & 2.3 & 4.3 $\,$ 10$^{-5}$ & 2.0 & 7.5 $\,$ 10$^{-3}$ & 8.2\\
\multirow{2}*{Benzo[g,h,i]perylene (C$_{22}$H$_{12}$)}
& Neutral
& 1.3 $\,$ 10$^{-7}$ & 5.8 & 7.0 $\,$ 10$^{-6}$ & 3.4 & 7.8 $\,$ 10$^{-3}$ & 8.5\\
& Cation
& 2.8 $\,$ 10$^{-7}$ & 3.3 & 1.6 $\,$ 10$^{-5}$ & 2.4 & 6.9 $\,$ 10$^{-3}$ & 8.4\\
\multirow{2}*{Anthanthrene (C$_{22}$H$_{12}$)}
& Neutral
& 1.7 $\,$ 10$^{-7}$ & 4.9 & 1.1 $\,$ 10$^{-5}$ & 3.1 & 7.9 $\,$ 10$^{-3}$ & 8.4\\
& Cation
& 3.2 $\,$ 10$^{-7}$ & 3.1 & 2.0 $\,$ 10$^{-5}$ & 2.3 & 7.1 $\,$ 10$^{-3}$ & 8.4\\
\multirow{2}*{Coronene (C$_{24}$H$_{12}$)}
& Neutral
& 1.3 $\,$ 10$^{-7}$ & 6.2 & 4.5 $\,$ 10$^{-6}$ & 4.1 & 8.5 $\,$ 10$^{-3}$ & 8.5\\
& Cation
& 3.1 $\,$ 10$^{-7}$ & 3.3 & 1.6 $\,$ 10$^{-5}$ & 2.4 & 7.6 $\,$ 10$^{-3}$ & 8.4\\
\multirow{2}*{Dibenzo[cd,lm]perylene (C$_{26}$H$_{14}$)}
& Neutral
& 4.6 $\,$ 10$^{-7}$ & 3.0 & 4.0 $\,$ 10$^{-5}$ & 1.9 & 9.5 $\,$ 10$^{-3}$ & 8.5\\
& Cation
& 4.5 $\,$ 10$^{-7}$ & 3.1 & 2.8 $\,$ 10$^{-5}$ & 2.5 & 8.4 $\,$ 10$^{-3}$ & 8.4\\
\multirow{2}*{Bisanthene (C$_{28}$H$_{14}$)}
& Neutral
& 5.1 $\,$ 10$^{-7}$ & 2.7 & 2.5 $\,$ 10$^{-5}$ & 2.0 & 9.9 $\,$ 10$^{-3}$ & 8.5\\
& Cation
& 5.7 $\,$ 10$^{-7}$ & 2.5 & 3.6 $\,$ 10$^{-5}$ & 1.7 & 9.0 $\,$ 10$^{-3}$ & 8.4\\
\multirow{2}*{Terrylene (C$_{30}$H$_{16}$)}
& Neutral
& 7.7 $\,$ 10$^{-7}$ & 2.7 & 3.9 $\,$ 10$^{-5}$ & 2.1 & 1.1 $\,$ 10$^{-2}$ & 8.6\\
& Cation
& 9.6 $\,$ 10$^{-7}$ & 2.4 & 4.8 $\,$ 10$^{-5}$ & 2.0 & 1.0 $\,$ 10$^{-2}$ & 8.6\\
\multirow{2}*{Ovalene (C$_{32}$H$_{14}$)}
& Neutral
& 2.8 $\,$ 10$^{-7}$ & 4.6 & 2.0 $\,$ 10$^{-5}$ & 3.1 & 1.1 $\,$ 10$^{-2}$ & 8.2\\
& Cation
& 5.5 $\,$ 10$^{-7}$ & 2.9 & 3.8 $\,$ 10$^{-5}$ & 2.2 & 1.1 $\,$ 10$^{-2}$ & 8.2\\
\multirow{2}*{Circumbiphenyl (C$_{38}$H$_{16}$)}
& Neutral
& 3.1 $\,$ 10$^{-7}$ & 4.7 & 2.7 $\,$ 10$^{-5}$ & 3.1 & 1.3 $\,$ 10$^{-2}$ & 8.2\\
& Cation
& 8.7 $\,$ 10$^{-7}$ & 2.4 & 6.9 $\,$ 10$^{-5}$ & 1.9 & 1.2 $\,$ 10$^{-2}$ & 8.2\\
\multirow{2}*{Quaterrylene (C$_{40}$H$_{20}$)}
& Neutral
& 2.2 $\,$ 10$^{-6}$ & 2.0 & 1.1 $\,$ 10$^{-4}$ & 1.7 & 1.5 $\,$ 10$^{-2}$ & 8.6\\
& Cation
& 2.7 $\,$ 10$^{-6}$ & 1.9 & 1.4 $\,$ 10$^{-4}$ & 1.6 & 1.4 $\,$ 10$^{-2}$ & 8.5\\
\multirow{2}*{Dicoronylene (C$_{48}$H$_{20}$)}
& Neutral
& 8.8 $\,$ 10$^{-7}$ & 3.2 & 5.8 $\,$ 10$^{-5}$ & 2.6 & 1.7 $\,$ 10$^{-2}$ & 8.1\\
& Cation
& 2.2 $\,$ 10$^{-6}$ & 2.0 & 1.5 $\,$ 10$^{-4}$ & 1.6 & 1.6 $\,$ 10$^{-2}$ & 8.0\\
\noalign{\smallskip}
\hline
\end{tabular}
\end{center}
\end{table*}

The whole sample of molecules investigated is shown in Fig.~2 of 
\citet{mal04}. Table~\ref{all_rates} summarises the resulting integrated 
photon absorption rates $\mathcal{R}$ and average energies $\mathcal{E}$ of the
absorbed photons. The energy absorption rate is $\mathcal{R}\,\mathcal{E}$. 

Our model runs produce band positions and the corresponding 
integrated fluxes. More specifically, the model outputs the total number of 
photons, and hence the total power $\mathcal{P}$, isotropically radiated in 
each vibrational mode by one molecule embedded in a given RF. 
The dimensions of this quantity are energy per unit time per unit solid angle 
per molecule. The statistical error on the number of photons is Poissonian, 
therefore the purely statistical contribution to errors can be made 
negligible by extending enough the length of the model runs. This is 
tipically achieved simulating $\sim$10$^4$ photon absorption events and the 
following de\textendash excitation cascades. 
The accuracy of our model results 
is limited by that of the level of theory we used to perform the 
vibrational analyses. Our calculated positions of far\textendash IR vibrational bands 
compare rather favourably with the sparse experimental measurements available 
(see Table~\ref{empirical_factor}). As to the calculated Einstein 
coefficients for spontaneous emission in the IR bands, a comparison
between gas\textendash phase measurements of band intensities \citep{job94,job95b} and 
DFT calculations at the level of theory we used here \citep{lan96} show 
differences which range from $\sim$10\% to a factor of 2 for single bands.
The largest discrepancy is found for the high frequency in\textendash plane C\textendash H stretch 
modes, with a generally better agreement for the other modes.
To the best of our knowledge, with the
exception of the works by \citet{zha96} and \citet{pir04}, there are no 
other gas\textendash phase laboratory data available for far\textendash IR bands of our sample
of PAHs. More experimental work is therefore definitely needed for a 
systematic assessment of their accuracy and reliability. 


Tables \ref{naphthalene} to \ref{dicoronylene+} in 
Appendix~\ref{tableappendix} list the calculated far\textendash IR emission spectra of 
all 20 PAHs considered and of their cations. In such tables permitted 
transitions are distinguished between parallel and perpendicular, depending
on the direction of the electric dipole transition moment with respect to
the plane of the molecule. In addition, 
we explicitly marked electric dipole forbidden bands.

Moreover, no data are available, to the best of our knowledge, on the 
intensity of bands which are IR\textendash inactive in the harmonic electric\textendash dipole 
approximation. 
Further experimental and theoretical work on these bands 
is needed for an accurate assessment of the photon absorption rates below 
which each electric\textendash dipole\textendash forbidden band might be emitted efficiently.

We assumed 100\% efficient conversion of absorbed energy to IR emission,
overestimating our calculated fluxes for closed\textendash shell species (see 
Sect.~\ref{validation}), very slightly in the case of far\textendash IR bands.
This affects only marginally the diagnostic capability of the calculated 
spectra, i.e. the possibility to use them to identify specific PAHs
upon comparison with astronomical observations, but instead affects the 
accuracy of column density (limits) derived using them. 
This is not an issue for open\textendash shell species.

Overall, as a conservative estimate, we expect our absolute estimated far\textendash IR 
fluxes to be accurate within a factor of 2 and our band positions within 
better than $\sim$2\%.

\section{Discussion} \label{discussion}
 
The results in Appendix~\ref{tableappendix} show the diagnostic 
potential of far\textendash IR observations for the identification of \emph{specific} 
PAHs. All the molecules in our sample, in the RFs considered, exhibit 
a few relatively strong IR\textendash active bands which dominate its far\textendash IR spectrum. 
These bands, in most cases flopping modes of the whole molecule, strongly 
depend on its size and shape and on the rigidity of its 
bonds. This latter dependence usually makes the far\textendash IR spectrum of each 
cation distinguishable from that of its parent neutral, even if 
similar to it.

For each species the IR emission spectrum expected for the RF of the Red 
Rectangle halo is considerably ``colder'' than the one expected for the 
ISM, which in turn is ``colder'' than that in the PDR of
IRAS~21282+5050, i.e. a larger and larger flux fraction is emitted in 
higher energy bands going from the first to the last. This is just the 
effect of the spectral energy distribution of the exciting RF, which 
in the Monte\textendash Carlo simulations translates to a higher average excitation 
energy $\mathcal{E}$ of the molecules (see Table~\ref{all_rates}).

Another similar trend which can be seen is that the IR emission spectrum
of larger molecules is ``colder'', in the same sense as above, than that
of smaller ones in the same RF. Again, this is expected due to the larger 
number of vibrational degrees of freedom for increasing number of atoms.

An interesting 
result of our simulations is the prediction of non\textendash negligible 
emission in far\textendash IR electric\textendash dipole forbidden vibrational transitions. 
Such emission is 
essentially the excitation energy remaining ``trapped'' in IR\textendash inactive 
modes at the end of the vibrational cascade following UV\textendash visible absorption. 
In perfect isolation conditions, this energy has no other way out but to be 
slowly emitted via forbidden or very weakly permitted vibrational transitions. 
The absorption of an additional UV\textendash visible photon by the molecule before 
it can emit the energy trapped in its IR\textendash inactive modes effectively quenches
the emission of far\textendash IR photons. Hence emission in an IR\textendash inactive mode
only occurs if its time scale is shorter than or at least comparable to 
the rate of UV\textendash visible photon absorption. 
Such conditions are by and large met in the ISRF, in which the 
calculated fluxes in such bands can be comparable to the lowest energy 
permitted ones. In the Red Rectangle halo, on the other hand, the radiation 
density is high enough to make the absorption of another UV\textendash visible photon 
happen on a timescale comparable to that of the emission of photons in 
IR\textendash inactive modes, which is reduced. In the PDR of 
IRAS~21282+5050 the absorption of UV\textendash visible photons completely dominates 
over the emission in all weak IR bands, essentially suppressing them. 
An increase in molecular size produces the same effect, 
described above, of an increase in RF intensity: this happens 
because the photo\textendash absorption cross\textendash sections and, consequently, the photon 
absorption rates $\mathcal{R}$, roughly scale with the number of carbon atoms 
in the molecule \citep{job92a,job92b}, while the total IR emissivity at the 
low\textendash energy end does not scale in the same way for this sample of molecules. 

Quite generally, emission in
far\textendash IR bands can be divided in three different regimes, depending only on the
photon absorption rate: \begin{enumerate}
\item for very dilute RFs, e.~g. in the diffuse ISM,
a molecule always has time to completely relax after each photon absorption,
and consequently the fluxes in \emph{all} far\textendash IR bands scale linearly with 
the photon absorption rate;
\item for intermediate RFs, e.~g. in the Red Rectangle halo,
a molecule always has time to emit all quanta in IR\textendash active modes, but absorbs
another photon before being able to completely relax IR\textendash inactive modes; 
hence, fluxes in IR\textendash active bands still scale linearly with the photon 
absorption rate, while some emission in electric\textendash dipole forbidden 
bands is suppressed;
\item for very strong RFs, e.~g. in the PDR of IRAS~21282+5050, a molecule 
is always vibrationally ``warm'', since photon absorption events are very 
frequent, and it almost never reaches the bottom of its IR emission cascade; 
as a result, \emph{all} weak IR bands tend to be progressively suppressed 
in favour of the stronger ones (i.e. classical AIBs).
\end{enumerate}
The boundaries between these regimes will depend on the molecule, since
different molecules will have different photon absorption rates in the 
same RF. 

Within the same regime, far\textendash IR fluxes scale linearly with the dilution
of the RF adopted, which means that the data we present here cover a 
relatively wide range of environments. 
For example, the ISRF estimated by \citet{dra78} is very similar to the 
interstellar RF we used, with just a different dilution factor; the former 
has an intensity, in Habing units, of $G_1\simeq1.68~G_0$ \citep{wein01} while the 
latter corresponds to $G_2\simeq3.49~G_0$: therefore, the expected far\textendash IR spectrum 
for the Draine ISRF can be obtained simply by multiplying the fluxes we 
calculated by the ratio $G_1/G_2\simeq0.48$. As another example, the RF in the Orion 
Bar is very similar to the one of IRAS~21282+5050, with a $\sim$10\textendash fold relative 
dilution, hence the expected far\textendash IR spectrum for the Orion Bar can be 
obtained, as a first order approximation, dividing the fluxes we calculated 
for IRAS~21282+5050 by a factor of 10.

PAH emission in IR\textendash inactive modes, if detectable, will provide a powerful 
probe both of the absolute intensity of the RF the emitting molecule is 
embedded in and of some poorly known molecular parameters
\citep{job02,mal03c,mul03,mul06}. 
However, as a cautionary remark, we recall that in the absence of 
any information on the intensity of electric\textendash dipole forbidden vibrational
bands, in the model we made the simplification of considering them as 
electric\textendash dipole permitted bands with an oscillator strength $10^4$ times 
smaller than the average IR\textendash active bands.
Therefore, while both
the asymptotic results calculated for the ISRF (perfect isolation) and 
for the PDR of IRAS~21282+5050 (complete suppression) are expected to be 
correct, as well as the general trends predicted, accurate quantitative 
results for
intermediate cases will require either the measurement or the calculation
of IR\textendash inactive band intensities. On the plus side, the impact of this 
uncertainty on the accuracy of fluxes in IR\textendash active bands is by and large 
negligible in all cases.

For comparison with observational data, in the optically thin case, the 
spectra we presented here must be multiplied by the assumed column density 
and the result integrated over the instrumental aperture on the sky for 
extended sources. A detailed radiative transfer analysis is needed for the 
optically thick case. We previously performed such a comparison
for three small, neutral PAHs in the Red Rectangle \citep{mul06}, and
in this case found the expected band intensities to be well within the
sensitivity of available Infrared Space Observatory (ISO) data. This 
example clearly demonstrated the discriminating capabilities of our synthetic 
spectra with presently available ISO observations and even more so for the 
forthcoming HSO mission, which is expected to go about two orders of 
magnitude deeper than ISO in the far\textendash IR.

\subsection{Detectability}\label{detectability}

To assess the possibility to identify specific interstellar PAHs in a 
large population of species, we computed the spectrum of a weighted sum of
all the species in our sample, using the inverse of the number of carbon 
atoms in each molecule as its weight, i.e.,
\begin{equation} \label{sum}
\mathrm{I}_\lambda^\mathrm{tot} = k \sum_\mathrm{PAH} 
\frac{1}{\mathrm{N}_\mathrm{C}(\mathrm{PAH})}\mathrm{I}_\lambda^\mathrm{PAH},
\end{equation}
the summation being over all molecules and charge states in our sample, and
the single spectra calculated for the Red Rectangle halo; k is a normalisation
constant which is defined later.

To assess the problem of detectability, some assumption on band 
shapes had to be made. One major cause of band broadening is the overlap of 
many nearly degenerate vibrational modes due to different occurences of the 
same  chemical bond in the molecule, such as e.g. out\textendash of\textendash plane C\textendash H 
bends. This is automatically taken into account in our model, which calculates 
separate expected fluxes for all bands, regardless of their near degeneracy.

\begin{figure}[b!]
\includegraphics[width=\hsize]{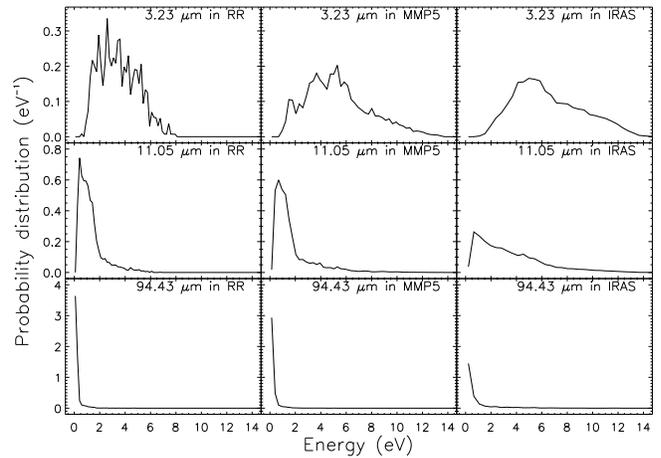}
\caption{Distribution of the excitation energies at which photons are emitted
by neutral bisanthene in three RFs and three vibrational 
bands.}
\label{histo_bisanthene_n}
\vspace{-0.4cm}
\end{figure}

\begin{figure}[b!]
\includegraphics[width=\hsize]{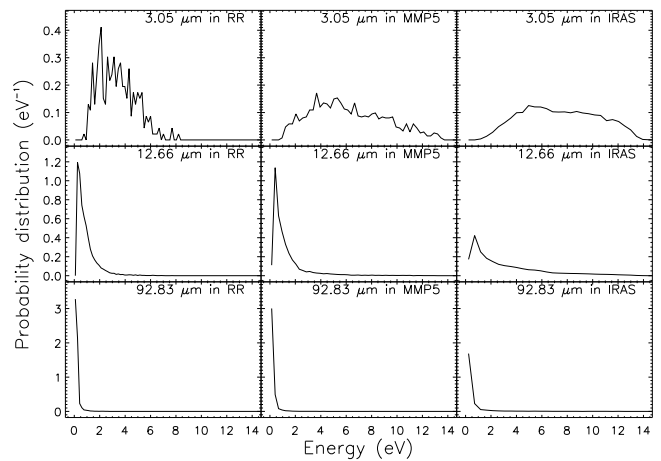}
\caption{As Fig.~\ref{histo_bisanthene_n} for the bisanthene cation.}
\label{histo_bisanthene_c}
\vspace{-0.4cm}
\end{figure}

Another cause of broadening is anharmonicity: almost all emitted 
photons are not due to a transition from a state in which one single 
vibrational mode was singly excited to the ground state, but from 
highly excited states. 
As an example, Figs.~\ref{histo_bisanthene_n} and 
\ref{histo_bisanthene_c}
show the distribution of the number of photons emitted in three 
vibrational modes as a function of the excitation energy of neutral and 
cationic bisanthene (C$_{28}$H$_{14}$), a medium\textendash sized molecule of our sample, 
in the three modelled RFs. 

\begin{figure}[t!]
\includegraphics[width=\hsize]{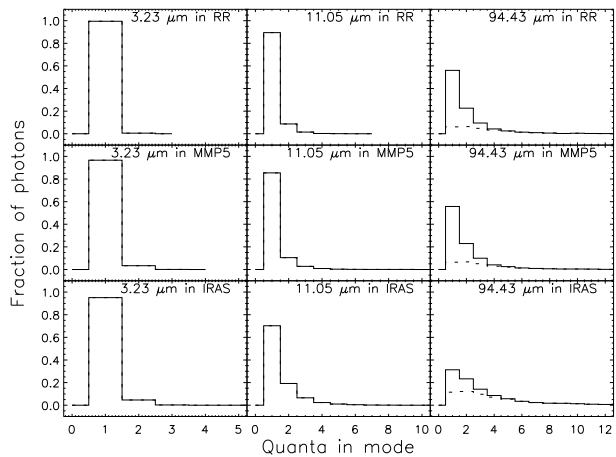}
\caption{Distribution of the photons emitted by neutral bisanthene 
in three RFs and three vibrational bands as a function of the 
vibrational excitation quanta in the mode.}
\label{quantahisto_bisanthene_n}

\end{figure}

\begin{figure}[t!]
\includegraphics[width=\hsize]{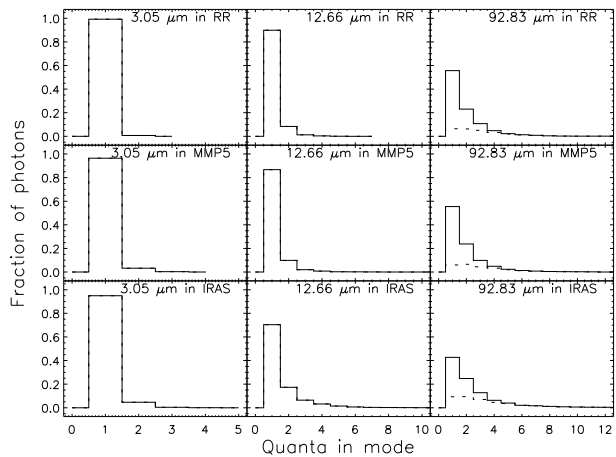}
\caption{As Fig.~\ref{quantahisto_bisanthene_n} for the bisanthene cation.}
\label{quantahisto_bisanthene_c}
\end{figure}

Figures~\ref{quantahisto_bisanthene_n} and \ref{quantahisto_bisanthene_c} 
show the distribution of the number of photons emitted in the same three 
specific vibrational modes as a function of the excitation energy in that 
specific vibrational mode. As described in detail by \citet{oom03}, 
the peak position of the bands is a function of vibrational energy
(i.~e. cross\textendash anharmonic shifts) and quantum number 
(i.~e. anharmonic shifts) of the emitting vibrational mode.
Knowledge of these molecular parameters, together with the distributions
in energy and quantum number (i.~e. Figs.~\ref{histo_bisanthene_n} to 
\ref{quantahisto_bisanthene_c}) would enable one to derive the 
effect of anharmonicity on band shape for each band.

Furthermore, intramolecular energy redistribution is extremely 
effective at high excitation energies and causes strong lifetime broadening. 
The rate of radiationless transitions causing intramolecular energy 
redistribution is known to be a steep function of molecular vibrational 
excitation energy. Thus again, in principle, if this function were known this 
broadening might be estimated from the distribution of energies of the 
emitting molecules. 
In the case of the lowest frequency modes, lifetime broadening 
becomes negligible, since they are emitted preferentially below or only 
slightly above the decoupling energy $E_\mathrm{dec}$.

We performed a completely detailed calculation of anharmonic and
lifetime broadening effects on band shape for 
the C\textendash H stretching modes near 3.3~$\mu$m for neutral anthracene, 
phenanthrene and pyrene in the Red Rectangle \citep{mul06}, using 
experimental data \citep{job95b,pec02}.
Since the molecular parameters required for this are only 
available for few bands of few species, we did not include this 
effect in our present calculations.


Finally, far\textendash IR bands are expected to show 
rotational profiles. The population of rotational levels, for isolated
PAHs in space, is probably far from thermal equilibrium 
\citep{rou92,lec98,lec99}. While our modelling approach can be extended
to estimate this precisely, this implies a huge increase in computational
costs, with a factor of the order of the number of rotational levels to
be traced (i.~e. $\sim10^5$). It also requires the detailed knowledge of 
more molecular parameters \citep[e.~g. the matrix of vibration\textendash rotation
constants and anharmonic vibrational parameters,][]{bar05}, which are not 
readily available for our sample and far from trivial to calculate. We 
performed such a complete calculation for two of the smallest neutral 
PAHs in our sample, namely naphthalene and anthracene, as a test case 
\citep{mul06b}.
For the present paper, we simulated
rotational profiles by including, for each band, two relatively broad
gaussians, displaced by their width $\sigma_\mathrm{PR}$ on either side of the 
calculated band origin, representing P and R branches and, for 
perpendicular bands, a narrower gaussian of width $\sigma_\mathrm{Q}$
representing the central Q branch. A fraction of $\sim20\%$ of the flux
in the band was assigned to the Q branch, when present, the remaining
split evenly between the P and R branches. Real values will deviate possibly 
by a factor of $\sim2$ from this. 
The population of levels between 
different values of the quantum number $J$ of the total angular momentum
is qualitatively expected to follow a pseudo\textendash thermal distribution
\citep{rou92}, its pseudo\textendash temperature scaling with the average energy of
the emitted IR photons. The parameter $\sigma_\mathrm{PR}$ in our approximate
representation of the P and R branches scales with the 
square root of this pseudo\textendash temperature, since they involve transitions
between states of different $J$ \citep[see e.~g.][]{cam04}.

A c\textendash type\footnote{IR\textendash permitted bands are labelled according to the 
direction of the change of the electric dipole moment in the involved 
vibrational mode, in relation to the axes of the principal momenta of inertia 
(I$_i$, with $i=x,y,z$) of the molecule. These axes, and the corresponding 
rotational constants and transition types, are labelled in order of 
increasing I$_i$. A c\textendash type band therefore corresponds to a transition in which
the electric dipole moment changes along the direction of maximum momentum of 
inertia, which for planar molecules is always perpendicular to the symmetry
plane \citep{her91b}.} band of a planar molecule such as a PAH will generally 
exhibit a sharp central Q branch, since in the rigid rotor
approximation all the transitions composing it are coincident in energy
if the rotational constants do not change between the two vibrational 
states involved \citep{her91b}. For this class of molecules, the
fractional variation in rotational constants due to vibrational excitation
is usually of the order of $\sim10^{-4}$, arriving at most to $\sim10^{-2}$ in the
infrequent case of accidental near degeneracy, giving rise to very strong
Coriolis perturbations \citep{her91b,mul06b}. The transitions in the 
Q branch, by definition, connect states with the same $J$.
The relative population of levels with
the same $J$, due to Internal Vibration\textendash Rotation Energy Transfer
\citep[IVRET][]{rou92} remains in thermal equilibrium with the
bath of vibrational states down to the decoupling energy. This means that
the relative population of levels with the same $J$ will reflect the 
excitation of the molecule when each photon gets emitted. Since low\textendash energy
photons are emitted near the end of the relaxation cascades, this latter
distribution will be very ``cold'' for them.
This ``cold'' relative population of levels and the almost unchanged 
rotational constants in the transition conspire to produce Q branches
very much sharper than the corresponding P and R branches. 

In the absence of reliable estimates in this respect, we had to assume a 
starting point from which to apply the above scaling relations. We
took $0.7~\mathrm{cm^{-1}} \leq \sigma_\mathrm{PR} \leq 5.5 ~\mathrm{cm^{-1}}$ for neutral
naphthalene, which amounts to pseudo\textendash temperatures in $J$, calculated 
according to the recipe in \citet{cam04}, between $\sim$10~K 
(highly subthermal) and $\sim$500~K (highly suprathermal). Any realistic value 
for the environments considered ought to fall in such a wide range.

For $\sigma_\mathrm{Q}$, the profile of Q branches will be blurred by anharmonic
effects, which are expected to perturb the energy of the transition
by a fractional amount of less than 1\%, essentially of the same order as
the empirical correction factors of (i~.e. $\sim$0.96 in this case) commonly 
used to account for anharmonicity, bringing harmonic frequencies into near 
coincidence with experimental values.

\begin{figure}
 
\includegraphics[width=\hsize]{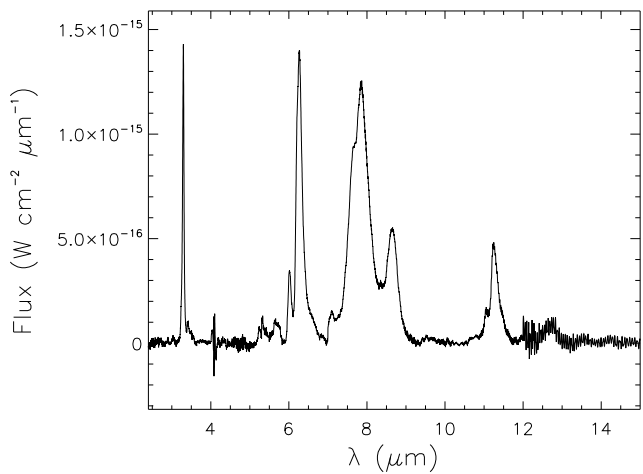}
\caption{Continuum-subtracted AIBs spectrum of the Red Rectangle, from
the online ISO spectral database.}
\label{rraibs}

\end{figure}

\begin{figure*}
 
\includegraphics[width=\hsize]{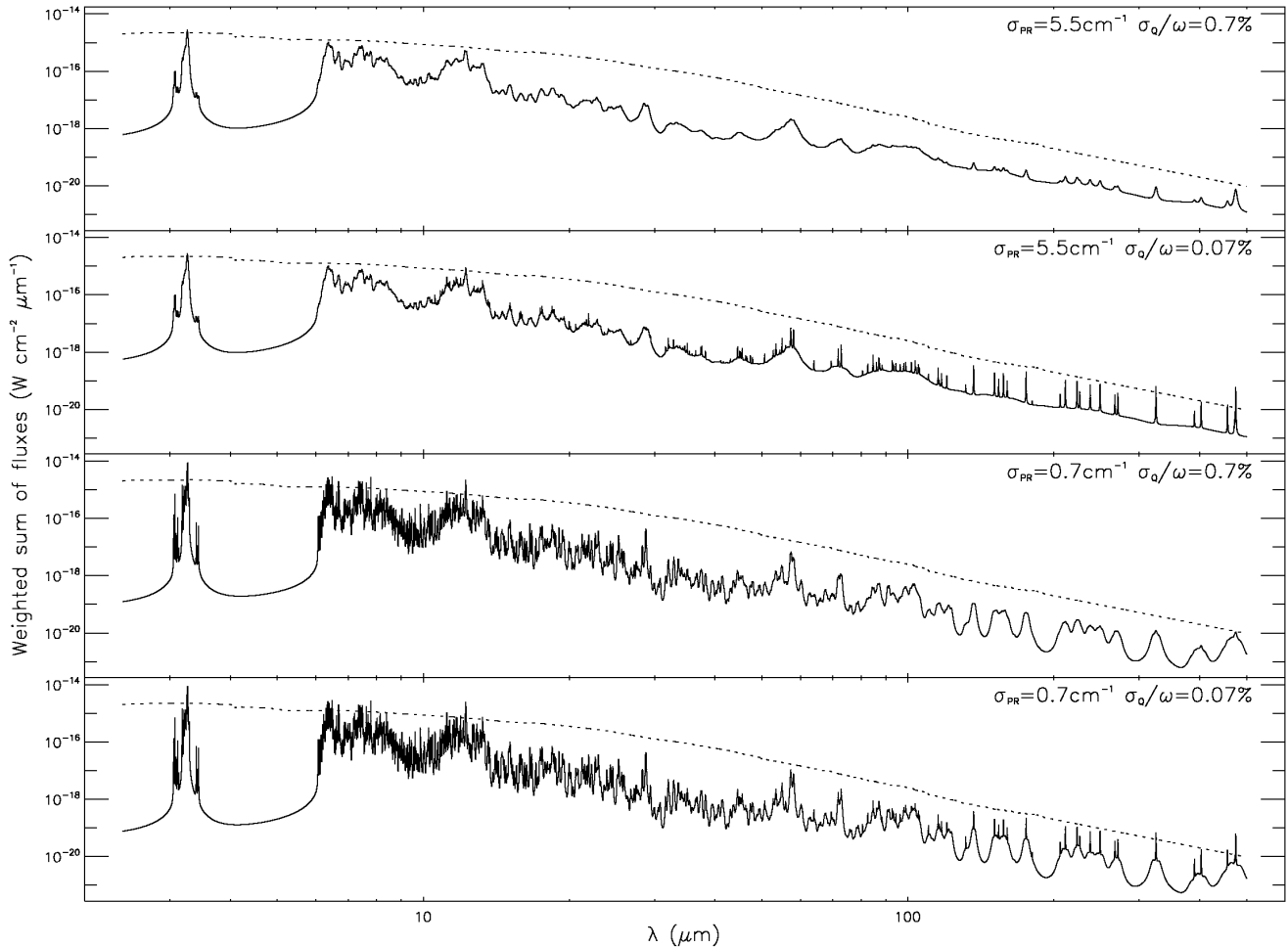}
\caption{Weighted sum of the spectra of all PAHs in our sample, calculated for 
the Red Rectangle halo (see text for its exact definition). The dotted line 
shows, for comparison, the estimated dust continuum in the same source. 
Different panels correspond to different assumptions of rotational 
excitations and anharmonic widths of the bands. The axes are in logarithmic 
scale, wavelength in abscissa and flux in ordinate.}
\label{allsums}

\end{figure*}

\begin{figure}
 
\includegraphics[width=\hsize]{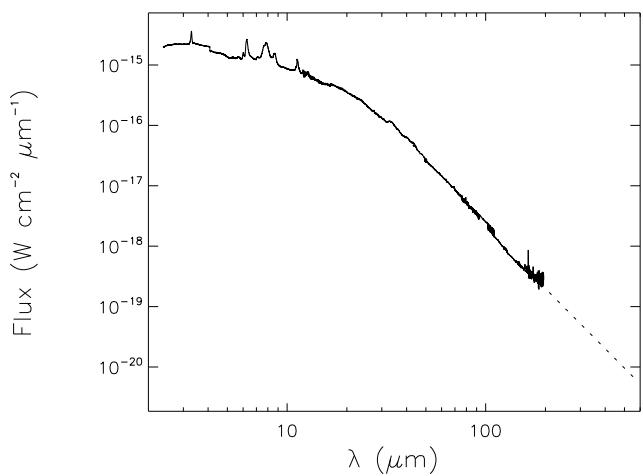}
\caption{ISO SWS and LWS spectrum of the Red Rectangle, showing the
extrapolated dust continuum to long wavelengths (dotted line).}
\label{rrcont}

\end{figure}

For a direct comparison with present and forthcoming observations, 
we chose the normalisation factor $k$ in the definition of 
$\mathrm{I}_\lambda^\mathrm{tot}$ (see Eq.~\ref{sum}) by requiring 
that its integral between 3 and 14~$\mu$m be equal to the integral, over the 
same wavelength interval, of the observed ISO\textendash SWS spectrum of the AIBs in 
the Red Rectangle. The latter spectrum, taken from the online ISO database, 
is reproduced after continuum subtraction in Fig~\ref{rraibs} for reference. 
Our synthetic summed spectrum is shown in Fig.~\ref{allsums}. 
The same figure also shows, overplotted in dotted line 
for comparison, the estimated dust continuum in the Red Rectangle, obtained by 
extrapolating the long wavelength tail of the observed continuum in 
available ISO\textendash LWS observations of the same source with a grey body spectrum
in the Rayleigh\textendash Jeans approximation (see Fig.~\ref{rrcont}), given 
(in W~cm$^{-2}$~$\mu$m$^{-1}$) by $\mathrm{I}_\lambda^\mathrm{dust}~=~A\,\lambda^{-\beta}$,
with $A=8.14~10^{-12}$, $\beta=3.31$ and $\lambda$ in $\mu$m. 
This extrapolation is consistent with the spectral energy distribution 
previously obtained by detailed modelling in \citet{men02}.

The molecules in our sample are likely not to be the interstellar 
population, as the mid\textendash IR spectra of the mixtures show some difference 
with the interstellar spectra. Nonetheless, this combined spectrum 
serves the intended purpose to get a visual estimate of the density of bands, 
and hence of probable spectral congestion, as a function of wavelength.

In almost all assumed conditions, P and R branches tend to blend in a 
structureless continuum, from which Q branches of perpendicular bands 
stand out more or less clearly depending on their assumed width.
Indeed, the main obstacle for the detection and identification of these bands 
is probably not posed by their absolute fluxes, but rather by spectral 
congestion and spectral contrast against a strong background continuum.

We conclude that a\textendash type and b\textendash type bands of PAHs, which display no 
sharp Q branches, will probably remain unobservable for the foreseeable 
future. On the other hand, the Q branches of some c\textendash type bands, for 
some specific molecules, might turn out to be observable, if molecular 
parameters conspire to make them narrow enough.
More theoretical and experimental work will be needed to find the most
promising candidates in the zoo of possible species.

The density of bands decreases with the wavelength. The search for 
such bands appears therefore favoured at lambda $>$100\textendash200~$\mu$m. 
The decrease of the dust continuum towards the millimeter 
range will make the detection of the bands easier. Indeed, Fig.~\ref{allsums} 
shows that, under the most favourable 
combination of molecular parameters, most of the bands in our sample at 
wavelengths beyond $\sim$120~$\mu$m are calculated to have peak intensities over 
10\% of the estimated dust continuum, a few of them even exceeding it. Since 
our sample does not reproduce the actual population of interstallar PAHs,
chemical diversity will dilute the peak intensity of individual bands
by an unknown factor, especially if this population is evenly spread over
a vast number of different species. Chemical diversity is produced by many
concurrent causes, e.~g. different carbon skeletons, different 
hydrogenation and/or charge states, inclusion of heteroatoms, substitution
of hydrogen atoms at the edges by different functional groups, isotopic
substitutions. \citet{lep01,lep03} showed that the population of a given 
PAH in any given environment is expected to be dominated by one or two 
ionisation and hydrogenation states, depending on its size and physical 
conditions. \citet{mul03} showed the effect of the most likely isotopic 
substitutions on low\textendash energy bands to be extremely small in the test case
of ovalene, which makes it negligible in the present context. As to the 
other causes of diversity we mentioned above, chemical selection effects 
will favour the most stable species over the others, so that the most 
abundant of them \emph{may} be detectable in the far\textendash IR.
Dedicated laboratory experiments, based on tentative identifications, 
will be needed to say the final word. 

HSO could be the ideal tool for this difficult quest, with its 
combined sensitivity and spectral resolving power in the far\textendash IR. The 
target of choice for such a search would be a source with the following
properties: 
\begin{itemize} 
\item it should be a known powerful source of the AIBs; 
\item it should be an environment with mild excitation conditions,
i.e. with a not too strong far\textendash UV RF (to avoid a ``hot'' dust continuum)
but with enough near\textendash UV photons to effectively excite PAH emission.
\end{itemize} 
The above description clearly fits carbon\textendash rich circumstellar environments 
such as the Red Rectangle as well as some relatively bright photon\textendash dominated 
regions (e.g. the Orion Bar). On the other hand, neither the diffuse ISM 
(too weak AIBs) nor Planetary Nebulae (too strong far\textendash UV flux) would be among 
the best targets.

\begin{acknowledgements}
G.~Malloci acknowledges the ``Minist\`ere de la Recherche'' and G.~Mulas 
the CNRS for financial support during their stay at CESR in Toulouse. 
We are thankful to the authors of \textsc{Octopus} for making their code 
available under a free license. We acknowledge the 
High Performance Computational Chemistry Group for using their code: 
``NWChem, A Computational Chemistry Package for Parallel Computers, 
version 4.7'' (2005), PNNL, Richland, 
Washington, USA. Part of the calculations used here were performed 
using the CINECA supercomputing facility. We thank the referee for his 
comments, which made us improve the impact that this paper may have.
\end{acknowledgements}


\appendix

\section{Quantitative impact of the use of theoretical photo\textendash absorption 
cross\textendash section}
\label{rates}

Table~\ref{rates_comparison} summarises the integrated photon 
absorption rates $\mathcal{R}$ and the average energy $\mathcal{E}$ of 
the absorbed photons computed using theoretical and laboratory 
photo\textendash absorption cross sections. The overall agreement is good for all 
four molecules in the case of the photodissociation region of 
the planetary nebula IRAS~21282+5050, whose radiation field (RF) 
is rather flat and thus most insensitive to the small differences between 
laboratory and theoretical absorption spectra (see Fig.~\ref{radfields}). 
The largest differences, for opposite reasons, can be seen for the RF of 
the Red Rectangle, mostly for the two smallest molecules, namely anthracene 
and pyrene. In the worst case, the $\mathcal{R}$ value calculated using 
$\sigma_\mathrm{th}$ is overestimated by about a factor of 
$\sim$3, while $\mathcal{E}$ is underestimated by about $\sim$40\%. These two 
errors are systematically in opposite directions, since they both stem from 
the relatively small differences in the absorption band theoretical positions 
at low energies and the steepness of the RFs. Hence the energy absorption 
rate, i~.e. the energy absorbed per unit time by a single molecule 
embedded in a given RF, obviously given by $\mathcal{R} \cdot \mathcal{E}$, is 
more accurate than $\mathcal{R}$. In any standard PAH model the energy 
absorption rate $\mathcal{R} \, \mathcal{E}$ of a molecule equals by and 
large the total power it radiates in all of its vibrational bands, while 
$\mathcal{E}$ governs their relative intensities.

Tables \ref{anthracene_comparison2} to \ref{ovalene_comparison2} compare 
the calculated IR emission spectra resulting from synthetic 
($\sigma_\mathrm{th}$) and laboratory ($\sigma_\mathrm{lab}$) photo\textendash absorption 
cross sections. The agreement is \emph{quantitatively} very good and the 
\emph{relative} intensities among IR bands is very accurate, with 
differences of the order of, at most, $\sim$20\%.

\begin{figure}[t!]

\includegraphics[width=\hsize]{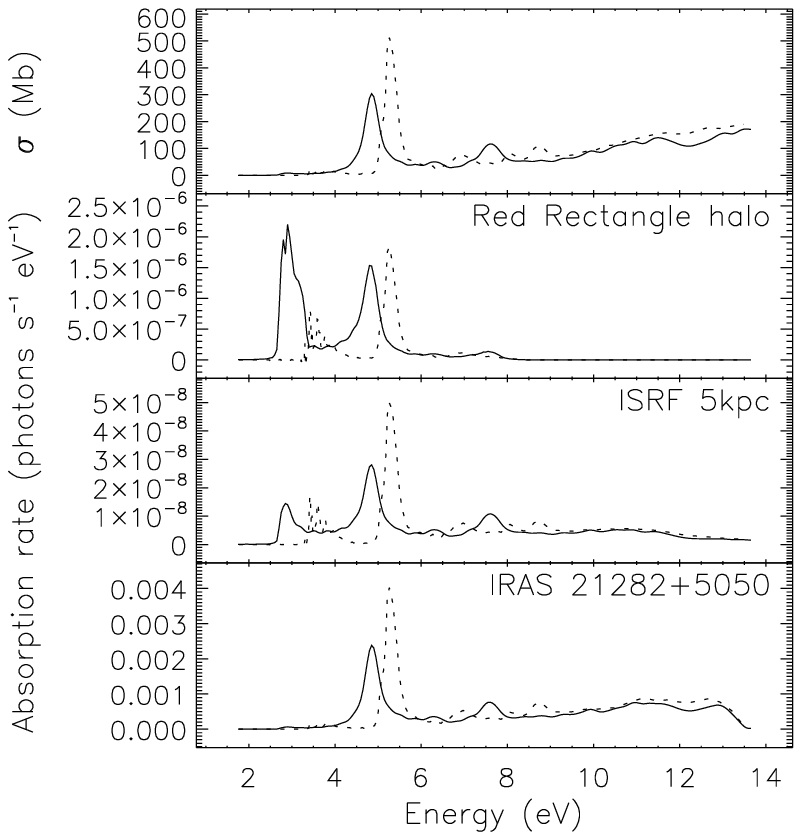}
\caption{The top panel shows the comparison between the measured 
\citep[dotted line;][]{job92a,job92b} and the theoretical photo\textendash absorption 
cross section up to 13.6~eV \citep[continuous line;][]{mal04} for anthracene. 
The cross\textendash sections are expressed in Megabarns (1~Mb=$10^{-18}$cm$^{2}$). The
three panels below show the comparison between the resulting energy\textendash dependent 
photon absorption rates for a molecule embedded in the three RFs considered.}
\label{anthracene_rates}
\end{figure}

\begin{figure}[b!]

\includegraphics[width=\hsize]{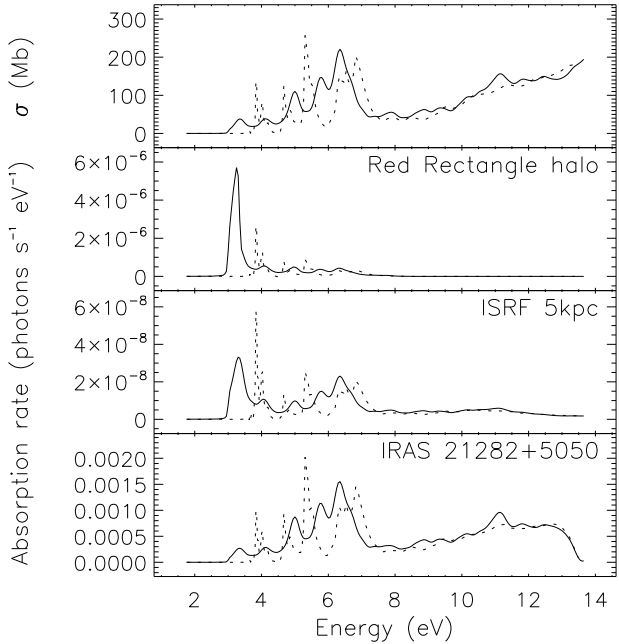}
\caption{Same as Fig.~\ref{anthracene_rates} for pyrene (C$_{16}$H$_{10}$).}
\label{pyrene_rates}
\end{figure}

\begin{figure}[h!]

\includegraphics[width=\hsize]{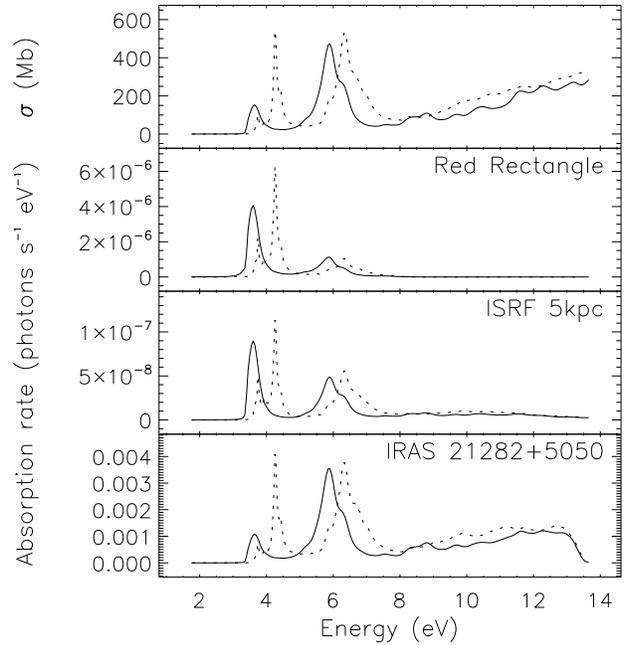}
\caption{Same as Fig.~\ref{anthracene_rates} for coronene (C$_{24}$H$_{12}$).}
\label{coronene_rates}
\end{figure}

\begin{figure}[h!]

\includegraphics[width=\hsize]{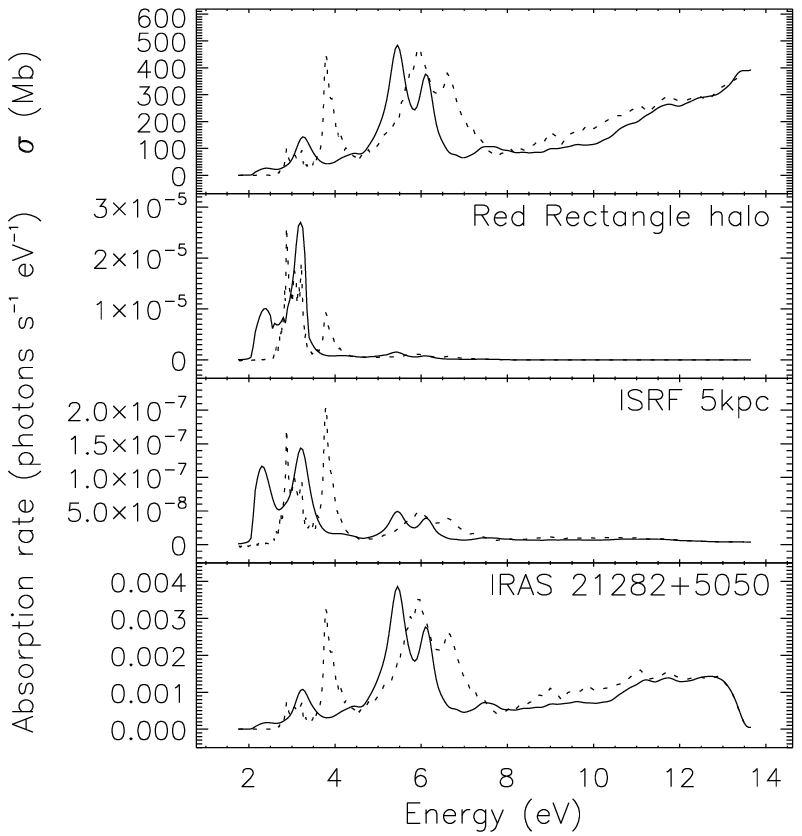}
\caption{Same as Fig.~\ref{anthracene_rates} for ovalene (C$_{32}$H$_{14}$).}
\label{ovalene_rates}
\end{figure}

\begin{table}[htb!]
\caption{Comparison between the integrated photon absorption rates 
$\mathcal{R}$ (in s$^{-1}$) and the average absorbed energy 
$\mathcal{E}$ (in eV) computed using theoretical 
\citep[$\sigma_\mathrm{th}$,][]{mal04} and laboratory 
\citep[$\sigma_\mathrm{lab}$,][]{job92a,job92b} photo\textendash absorption cross\textendash sections for
anthracene, pyrene, coronene and ovalene in the three RFs considered.}
\label{rates_comparison}
\begin{tabular}{ccccccc}
\hline 
\hline
\noalign{\smallskip}
& \multicolumn{6}{c}{Radiation field}\\
\noalign{\smallskip}
& \multicolumn{2}{c}{ISRF}& \multicolumn{2}{c}{Red Rectangle} 
& \multicolumn{2}{c}{IRAS~21282+5050}\\
\noalign{\smallskip}
\hline
\noalign{\smallskip}
& $\mathcal{R}$ & $\mathcal{E}$ & $\mathcal{R}$ & $\mathcal{E}$ & 
$\mathcal{R}$ & $\mathcal{E}$\\
& (s$^{-1}$) & (eV) & (s$^{-1}$) & (eV) & (s$^{-1}$) & (eV) \\
\noalign{\smallskip}
\hline
\noalign{\smallskip}
\multicolumn{7}{c}{Anthracene (C$_{14}$H$_{10}$)}\\
$\sigma_\mathrm{lab}$ & 6.0~10$^{-8}$ & 7.6 & 1.1~10$^{-6}$ & 5.1 & 5.9~10$^{-3}$& 8.9 \\
$\sigma_\mathrm{th}$ & 6.4~10$^{-8}$ & 6.8 & 2.5~10$^{-6}$ & 4.0 & 5.2~10$^{-3}$& 8.6\\
\hline
\noalign{\smallskip}
\multicolumn{7}{c}{Pyrene (C$_{16}$H$_{10}$)}\\
$\sigma_\mathrm{lab}$ & 5.6~10$^{-8}$ & 7.4 & 1.1~10$^{-6}$ & 5.2 & 5.1~10$^{-3}$ & 8.9\\
$\sigma_\mathrm{th}$ & 8.1~10$^{-8}$& 6.4 & 3.7~10$^{-6}$ & 3.8 & 5.7~10$^{-3}$& 8.6\\ 
\hline
\multicolumn{7}{c}{Coronene (C$_{24}$H$_{12}$)}\\
\noalign{\smallskip}
$\sigma_\mathrm{lab}$ & 1.4~10$^{-7}$ & 6.8 & 3.9~10$^{-6}$ & 4.5 & 1.1$ $ 10$^{-2}$ & 8.6\\ 
$\sigma_\mathrm{th}$ & 1.3 ~10$^{-7}$& 6.2 & 4.5~10$^{-6}$ & 4.1 & 8.5~10$^{-3}$ & 8.5\\
\hline
\multicolumn{7}{c}{Ovalene (C$_{32}$H$_{14}$)}\\
$\sigma_\mathrm{lab}$ & 2.4~10$^{-7}$ & 5.6 & 1.4~10$^{-5}$ & 3.5 & 1.3~10$^{-2}$ & 8.2\\
$\sigma_\mathrm{th}$ & 2.8~10$^{-7}$ & 4.6 & 2.0~10$^{-5}$ & 3.1 & 1.1~10$^{-2}$ & 8.2\\
\noalign{\smallskip}
\hline
\end{tabular}
\vspace{-0.4cm}
\end{table}
Figures \ref{anthracene_rates} to \ref{ovalene_rates} show the comparison
between $\sigma_\mathrm{th}$ and $\sigma_\mathrm{lab}$ and the impact of their differences 
on the estimated energy\textendash dependent absorption rates in the three RFs 
considered for anthracene, pyrene, coronene and ovalene.

\begin{table*}
\begin{center}
\caption{Comparison between various model runs for neutral anthracene 
(C$_{14}$H$_{10}$), using either the experimental 
\citep[$\sigma_\mathrm{lab}$,][]{job92a,job92b} or the theoretical 
\citep[$\sigma_\mathrm{th}$,][]{mal04} photo\textendash absorption spectrum, in different RFs. 
Bands which are electric\textendash dipole forbidden are enclosed in square brackets in 
the first column. In each of the other columns, we list the absolute flux 
emitted by one molecule in a given band and RF and, in parentheses, the flux
fraction in that band. Bands whose calculated flux fraction in the 
ISRF is less than 0.05\% are omitted.}
\label{anthracene_comparison2}
\begin{tabular}{ccccccc}
\hline \hline \noalign{\smallskip} 
& \multicolumn{6}{c}{Integrated flux} \\
\noalign{\smallskip} \cline{2-7} \noalign{\smallskip}
 & \multicolumn{2}{c}{ISRF} & \multicolumn{2}{c}{Red Rectangle}
& \multicolumn{2}{c}{IRAS~21282+5050}\\
Peak &  $\sigma_\mathrm{lab}$ & $\sigma_\mathrm{th}$ & $\sigma_\mathrm{lab}$ & $\sigma_\mathrm{th}$ & $\sigma_\mathrm{lab}$ & $\sigma_\mathrm{th}$ \\
($\mu$m) & (W~sr$^{-1}$) (\%) & (W~sr$^{-1}$) (\%) & (W~sr$^{-1}$) (\%) & (W~sr$^{-1}$) (\%) & (W~sr$^{-1}$) (\%) & (W~sr$^{-1}$) (\%) \\
\noalign{\smallskip} \hline \noalign{\smallskip}
3.25 & $1.2 \,\, 10^{-27}$ (27.06) & $1.1 \,\, 10^{-27}$ (25.74) & $1.6 \,\, 10^{-26}$ (22.93) & $2.4 \,\, 10^{-26}$ (19.41) & $1.2 \,\, 10^{-22}$ (29.33) & $9.9 \,\, 10^{-23}$ (28.65) \\
3.26 & $7.5 \,\, 10^{-28}$ (17.55) & $7.0 \,\, 10^{-28}$ (16.69) & $1.1 \,\, 10^{-26}$ (15.09) & $1.6 \,\, 10^{-26}$ (12.80) & $7.6 \,\, 10^{-23}$ (18.96) & $6.5 \,\, 10^{-23}$ (18.89) \\
3.28 & $2.4 \,\, 10^{-30}$ (0.06) & $2.3 \,\, 10^{-30}$ (0.05) & $3.4 \,\, 10^{-29}$ (0.05) & $5.1 \,\, 10^{-29}$ (0.04) & $2.2 \,\, 10^{-25}$ (0.05) & $2.1 \,\, 10^{-25}$ (0.06) \\
3.28 & $2.0 \,\, 10^{-28}$ (4.56) & $1.8 \,\, 10^{-28}$ (4.41) & $2.9 \,\, 10^{-27}$ (3.97) & $4.3 \,\, 10^{-27}$ (3.41) & $2.0 \,\, 10^{-23}$ (4.89) & $1.7 \,\, 10^{-23}$ (4.95) \\
3.29 & $1.1 \,\, 10^{-28}$ (2.53) & $1.0 \,\, 10^{-28}$ (2.43) & $1.6 \,\, 10^{-27}$ (2.17) & $2.3 \,\, 10^{-27}$ (1.84) & $1.1 \,\, 10^{-23}$ (2.78) & $9.2 \,\, 10^{-24}$ (2.68) \\
6.16 & $1.4 \,\, 10^{-28}$ (3.22) & $1.3 \,\, 10^{-28}$ (3.23) & $2.4 \,\, 10^{-27}$ (3.37) & $4.2 \,\, 10^{-27}$ (3.39) & $1.3 \,\, 10^{-23}$ (3.13) & $1.1 \,\, 10^{-23}$ (3.18) \\
6.51 & $3.9 \,\, 10^{-29}$ (0.92) & $3.9 \,\, 10^{-29}$ (0.93) & $6.9 \,\, 10^{-28}$ (0.96) & $1.3 \,\, 10^{-27}$ (1.02) & $3.7 \,\, 10^{-24}$ (0.91) & $3.1 \,\, 10^{-24}$ (0.89) \\
6.86 & $7.8 \,\, 10^{-29}$ (1.82) & $7.6 \,\, 10^{-29}$ (1.81) & $1.4 \,\, 10^{-27}$ (1.88) & $2.5 \,\, 10^{-27}$ (2.03) & $7.0 \,\, 10^{-24}$ (1.74) & $6.1 \,\, 10^{-24}$ (1.75) \\
6.86 & $3.6 \,\, 10^{-29}$ (0.83) & $3.6 \,\, 10^{-29}$ (0.85) & $6.6 \,\, 10^{-28}$ (0.92) & $1.2 \,\, 10^{-27}$ (0.96) & $3.2 \,\, 10^{-24}$ (0.79) & $2.9 \,\, 10^{-24}$ (0.83) \\
7.22 & $2.7 \,\, 10^{-30}$ (0.06) & $2.7 \,\, 10^{-30}$ (0.06) & $5.8 \,\, 10^{-29}$ (0.08) & $9.1 \,\, 10^{-29}$ (0.07) & $2.8 \,\, 10^{-25}$ (0.07) & $2.0 \,\, 10^{-25}$ (0.06) \\
7.43 & $4.7 \,\, 10^{-29}$ (1.10) & $4.7 \,\, 10^{-29}$ (1.13) & $9.0 \,\, 10^{-28}$ (1.25) & $1.7 \,\, 10^{-27}$ (1.33) & $4.4 \,\, 10^{-24}$ (1.10) & $3.8 \,\, 10^{-24}$ (1.09) \\
7.60 & $7.8 \,\, 10^{-29}$ (1.83) & $7.8 \,\, 10^{-29}$ (1.86) & $1.5 \,\, 10^{-27}$ (2.04) & $2.7 \,\, 10^{-27}$ (2.13) & $6.9 \,\, 10^{-24}$ (1.71) & $6.1 \,\, 10^{-24}$ (1.78) \\
7.85 & $9.8 \,\, 10^{-29}$ (2.29) & $9.9 \,\, 10^{-29}$ (2.37) & $1.8 \,\, 10^{-27}$ (2.53) & $3.5 \,\, 10^{-27}$ (2.76) & $8.8 \,\, 10^{-24}$ (2.20) & $7.7 \,\, 10^{-24}$ (2.24) \\
8.55 & $1.3 \,\, 10^{-29}$ (0.31) & $1.4 \,\, 10^{-29}$ (0.33) & $2.5 \,\, 10^{-28}$ (0.35) & $4.9 \,\, 10^{-28}$ (0.39) & $1.2 \,\, 10^{-24}$ (0.30) & $1.0 \,\, 10^{-24}$ (0.30) \\
8.62 & $3.4 \,\, 10^{-29}$ (0.80) & $3.4 \,\, 10^{-29}$ (0.82) & $6.5 \,\, 10^{-28}$ (0.91) & $1.2 \,\, 10^{-27}$ (1.00) & $3.1 \,\, 10^{-24}$ (0.76) & $2.7 \,\, 10^{-24}$ (0.79) \\
8.67 & $9.0 \,\, 10^{-29}$ (2.11) & $9.2 \,\, 10^{-29}$ (2.19) & $1.7 \,\, 10^{-27}$ (2.39) & $3.2 \,\, 10^{-27}$ (2.58) & $8.1 \,\, 10^{-24}$ (2.01) & $6.9 \,\, 10^{-24}$ (2.01) \\
9.95 & $4.2 \,\, 10^{-29}$ (0.98) & $4.2 \,\, 10^{-29}$ (1.01) & $8.1 \,\, 10^{-28}$ (1.13) & $1.6 \,\, 10^{-27}$ (1.24) & $3.8 \,\, 10^{-24}$ (0.94) & $3.2 \,\, 10^{-24}$ (0.94) \\
10.41 & $7.4 \,\, 10^{-29}$ (1.73) & $7.5 \,\, 10^{-29}$ (1.80) & $1.5 \,\, 10^{-27}$ (2.02) & $2.8 \,\, 10^{-27}$ (2.22) & $6.7 \,\, 10^{-24}$ (1.66) & $5.7 \,\, 10^{-24}$ (1.66) \\
11.00 & $1.6 \,\, 10^{-29}$ (0.36) & $1.7 \,\, 10^{-29}$ (0.41) & $3.1 \,\, 10^{-28}$ (0.43) & $6.2 \,\, 10^{-28}$ (0.50) & $1.4 \,\, 10^{-24}$ (0.35) & $1.2 \,\, 10^{-24}$ (0.34) \\
11.32 & $5.4 \,\, 10^{-28}$ (12.60) & $5.5 \,\, 10^{-28}$ (13.25) & $1.1 \,\, 10^{-26}$ (14.72) & $2.1 \,\, 10^{-26}$ (16.66) & $4.8 \,\, 10^{-23}$ (11.92) & $4.2 \,\, 10^{-23}$ (12.14) \\
13.71 & $5.3 \,\, 10^{-28}$ (12.29) & $5.5 \,\, 10^{-28}$ (13.12) & $1.1 \,\, 10^{-26}$ (14.64) & $2.1 \,\, 10^{-26}$ (16.85) & $4.6 \,\, 10^{-23}$ (11.49) & $4.1 \,\, 10^{-23}$ (11.77) \\
15.33 & $9.2 \,\, 10^{-30}$ (0.21) & $9.6 \,\, 10^{-30}$ (0.23) & $2.1 \,\, 10^{-28}$ (0.29) & $3.9 \,\, 10^{-28}$ (0.31) & $7.9 \,\, 10^{-25}$ (0.20) & $7.0 \,\, 10^{-25}$ (0.20) \\
$[15.71]$ & $1.7 \,\, 10^{-30}$ (0.04) & $2.2 \,\, 10^{-30}$ (0.05) & $4.1 \,\, 10^{-29}$ (0.06) & $9.9 \,\, 10^{-29}$ (0.08) & \textemdash{} & \textemdash{} \\
16.34 & $4.2 \,\, 10^{-29}$ (0.99) & $4.4 \,\, 10^{-29}$ (1.04) & $8.5 \,\, 10^{-28}$ (1.18) & $1.7 \,\, 10^{-27}$ (1.36) & $3.6 \,\, 10^{-24}$ (0.90) & $3.2 \,\, 10^{-24}$ (0.93) \\
$[17.11]$ & $2.3 \,\, 10^{-30}$ (0.05) & $2.6 \,\, 10^{-30}$ (0.06) & $5.0 \,\, 10^{-29}$ (0.07) & $1.1 \,\, 10^{-28}$ (0.09) & \textemdash{} & \textemdash{} \\
$[18.71]$ & $2.8 \,\, 10^{-30}$ (0.06) & $3.3 \,\, 10^{-30}$ (0.08) & $6.4 \,\, 10^{-29}$ (0.09) & $1.4 \,\, 10^{-28}$ (0.11) & \textemdash{} & \textemdash{} \\
$[20.04]$ & $3.4 \,\, 10^{-30}$ (0.08) & $3.6 \,\, 10^{-30}$ (0.09) & $7.3 \,\, 10^{-29}$ (0.10) & $1.6 \,\, 10^{-28}$ (0.13) & \textemdash{} & \textemdash{} \\
$[21.02]$ & $3.4 \,\, 10^{-30}$ (0.08) & $3.9 \,\, 10^{-30}$ (0.09) & $8.1 \,\, 10^{-29}$ (0.11) & $1.8 \,\, 10^{-28}$ (0.14) & \textemdash{} & \textemdash{} \\
21.26 & $7.3 \,\, 10^{-29}$ (1.71) & $7.7 \,\, 10^{-29}$ (1.85) & $1.5 \,\, 10^{-27}$ (2.09) & $3.1 \,\, 10^{-27}$ (2.49) & $6.3 \,\, 10^{-24}$ (1.56) & $5.6 \,\, 10^{-24}$ (1.61) \\
$[25.45]$ & $5.0 \,\, 10^{-30}$ (0.12) & $5.5 \,\, 10^{-30}$ (0.13) & $1.0 \,\, 10^{-28}$ (0.14) & $2.2 \,\, 10^{-28}$ (0.18) & \textemdash{} & \textemdash{} \\
$[25.62]$ & $4.7 \,\, 10^{-30}$ (0.11) & $5.4 \,\, 10^{-30}$ (0.13) & $1.1 \,\, 10^{-28}$ (0.15) & $2.3 \,\, 10^{-28}$ (0.18) & \textemdash{} & \textemdash{} \\
26.35 & $5.3 \,\, 10^{-30}$ (0.12) & $5.9 \,\, 10^{-30}$ (0.14) & $1.1 \,\, 10^{-28}$ (0.16) & $2.6 \,\, 10^{-28}$ (0.21) & $6 \,\, 10^{-26}$ (0.02) & $6.5 \,\, 10^{-26}$ (0.02) \\
$[37.41]$ & $7.4 \,\, 10^{-30}$ (0.17) & $8.2 \,\, 10^{-30}$ (0.20) & $1.5 \,\, 10^{-28}$ (0.20) & $3.2 \,\, 10^{-28}$ (0.26) & \textemdash{} & \textemdash{} \\
$[43.01]$ & $8.5 \,\, 10^{-30}$ (0.20) & $8.8 \,\, 10^{-30}$ (0.21) & $1.7 \,\, 10^{-28}$ (0.23) & $3.4 \,\, 10^{-28}$ (0.27) & \textemdash{} & \textemdash{} \\
43.68 & $1.1 \,\, 10^{-29}$ (0.26) & $1.2 \,\, 10^{-29}$ (0.29) & $2.4 \,\, 10^{-28}$ (0.33) & $5.3 \,\, 10^{-28}$ (0.43) & $6.7 \,\, 10^{-25}$ (0.17) & $6.1 \,\, 10^{-25}$ (0.18) \\
$[82.09]$ & $1.1 \,\, 10^{-29}$ (0.25) & $1.3 \,\, 10^{-29}$ (0.30) & $1.8 \,\, 10^{-28}$ (0.25) & $3.0 \,\, 10^{-28}$ (0.24) & \textemdash{} & \textemdash{} \\
110.23 & $1.3 \,\, 10^{-29}$ (0.30) & $1.4 \,\, 10^{-29}$ (0.33) & $2.8 \,\, 10^{-28}$ (0.39) & $6.4 \,\, 10^{-28}$ (0.51) & $1.8 \,\, 10^{-25}$ (0.04) & $1.8 \,\, 10^{-25}$ (0.05) \\
\noalign{\smallskip} \hline
\end{tabular}
\end{center} 
\end{table*}

\begin{table*}
\begin{center}
\caption{As Table~\ref{anthracene_comparison2} for neutral pyrene 
(C$_{16}$H$_{10}$).}
\label{pyrene_comparison2}
\begin{tabular}{ccccccc}
\hline \hline \noalign{\smallskip} 
& \multicolumn{6}{c}{Integrated flux} \\
\noalign{\smallskip} \cline{2-7} \noalign{\smallskip}
& \multicolumn{2}{c}{ISRF} & \multicolumn{2}{c}{Red Rectangle}
& \multicolumn{2}{c}{IRAS~21282+5050}\\
Peak & $\sigma_\mathrm{lab}$ & $\sigma_\mathrm{th}$ & $\sigma_\mathrm{lab}$ & $\sigma_\mathrm{th}$ & $\sigma_\mathrm{lab}$ & $\sigma_\mathrm{th}$ \\
($\mu$m) & (W~sr$^{-1}$) (\%) & (W~sr$^{-1}$) (\%) & (W~sr$^{-1}$) (\%) & (W~sr$^{-1}$) (\%) & (W~sr$^{-1}$) (\%) & (W~sr$^{-1}$) (\%) \\
\noalign{\smallskip} \hline \noalign{\smallskip}
3.25 & $8.9 \,\, 10^{-28}$ (21.97) & $1.1 \,\, 10^{-27}$ (20.58) & $1.4 \,\, 10^{-26}$ (19.15) & $2.7 \,\, 10^{-26}$ (15.10) & $8.2 \,\, 10^{-23}$ (24.28) & $9.0 \,\, 10^{-23}$ (23.97) \\
3.26 & $8.8 \,\, 10^{-28}$ (21.56) & $1.0 \,\, 10^{-27}$ (19.86) & $1.3 \,\, 10^{-26}$ (18.61) & $2.6 \,\, 10^{-26}$ (15.05) & $7.9 \,\, 10^{-23}$ (23.25) & $8.7 \,\, 10^{-23}$ (23.16) \\
3.27 & $2.4 \,\, 10^{-28}$ (5.91) & $2.8 \,\, 10^{-28}$ (5.45) & $3.6 \,\, 10^{-27}$ (5.05) & $6.8 \,\, 10^{-27}$ (3.87) & $2.2 \,\, 10^{-23}$ (6.42) & $2.4 \,\, 10^{-23}$ (6.43) \\
3.29 & $3.3 \,\, 10^{-29}$ (0.81) & $4.0 \,\, 10^{-29}$ (0.77) & $4.9 \,\, 10^{-28}$ (0.68) & $1.0 \,\, 10^{-27}$ (0.59) & $3.2 \,\, 10^{-24}$ (0.94) & $3.5 \,\, 10^{-24}$ (0.92) \\
6.26 & $5.9 \,\, 10^{-29}$ (1.45) & $7.7 \,\, 10^{-29}$ (1.49) & $1.1 \,\, 10^{-27}$ (1.60) & $2.8 \,\, 10^{-27}$ (1.57) & $4.8 \,\, 10^{-24}$ (1.41) & $5.5 \,\, 10^{-24}$ (1.46) \\
6.31 & $1.4 \,\, 10^{-28}$ (3.47) & $1.8 \,\, 10^{-28}$ (3.40) & $2.6 \,\, 10^{-27}$ (3.58) & $6.6 \,\, 10^{-27}$ (3.78) & $1.1 \,\, 10^{-23}$ (3.32) & $1.2 \,\, 10^{-23}$ (3.29) \\
6.78 & $4.2 \,\, 10^{-29}$ (1.03) & $5.4 \,\, 10^{-29}$ (1.05) & $7.9 \,\, 10^{-28}$ (1.10) & $2.1 \,\, 10^{-27}$ (1.18) & $3.4 \,\, 10^{-24}$ (0.99) & $3.8 \,\, 10^{-24}$ (1.02) \\
6.92 & $5.8 \,\, 10^{-30}$ (0.14) & $6.5 \,\, 10^{-30}$ (0.13) & $1.1 \,\, 10^{-28}$ (0.15) & $2.7 \,\, 10^{-28}$ (0.15) & $4.2 \,\, 10^{-25}$ (0.12) & $4.7 \,\, 10^{-25}$ (0.12) \\
7.01 & $1.2 \,\, 10^{-29}$ (0.30) & $1.7 \,\, 10^{-29}$ (0.34) & $2.5 \,\, 10^{-28}$ (0.34) & $5.9 \,\, 10^{-28}$ (0.34) & $1.0 \,\, 10^{-24}$ (0.30) & $1.1 \,\, 10^{-24}$ (0.30) \\
7.01 & $1.2 \,\, 10^{-28}$ (3.06) & $1.6 \,\, 10^{-28}$ (3.14) & $2.3 \,\, 10^{-27}$ (3.26) & $6.3 \,\, 10^{-27}$ (3.58) & $1.0 \,\, 10^{-23}$ (2.97) & $1.1 \,\, 10^{-23}$ (2.93) \\
7.61 & $6.6 \,\, 10^{-29}$ (1.62) & $8.5 \,\, 10^{-29}$ (1.65) & $1.3 \,\, 10^{-27}$ (1.77) & $3.4 \,\, 10^{-27}$ (1.96) & $5.3 \,\, 10^{-24}$ (1.55) & $5.8 \,\, 10^{-24}$ (1.55) \\
7.98 & $3.7 \,\, 10^{-29}$ (0.90) & $4.8 \,\, 10^{-29}$ (0.93) & $7.0 \,\, 10^{-28}$ (0.98) & $2.0 \,\, 10^{-27}$ (1.12) & $2.8 \,\, 10^{-24}$ (0.84) & $3.2 \,\, 10^{-24}$ (0.85) \\
8.42 & $9.9 \,\, 10^{-29}$ (2.44) & $1.3 \,\, 10^{-28}$ (2.59) & $2.0 \,\, 10^{-27}$ (2.74) & $5.4 \,\, 10^{-27}$ (3.08) & $7.9 \,\, 10^{-24}$ (2.32) & $8.8 \,\, 10^{-24}$ (2.34) \\
8.62 & $1.6 \,\, 10^{-29}$ (0.39) & $2.1 \,\, 10^{-29}$ (0.41) & $3.0 \,\, 10^{-28}$ (0.41) & $7.9 \,\, 10^{-28}$ (0.45) & $1.2 \,\, 10^{-24}$ (0.35) & $1.3 \,\, 10^{-24}$ (0.35) \\
9.16 & $4.5 \,\, 10^{-29}$ (1.10) & $5.9 \,\, 10^{-29}$ (1.15) & $8.8 \,\, 10^{-28}$ (1.23) & $2.4 \,\, 10^{-27}$ (1.36) & $3.6 \,\, 10^{-24}$ (1.05) & $3.8 \,\, 10^{-24}$ (1.00) \\
10.04 & $4.7 \,\, 10^{-30}$ (0.12) & $6.7 \,\, 10^{-30}$ (0.13) & $1.1 \,\, 10^{-28}$ (0.15) & $2.8 \,\, 10^{-28}$ (0.16) & $4.0 \,\, 10^{-25}$ (0.12) & $4.6 \,\, 10^{-25}$ (0.12) \\
10.25 & $2.2 \,\, 10^{-29}$ (0.55) & $3.2 \,\, 10^{-29}$ (0.62) & $4.9 \,\, 10^{-28}$ (0.68) & $1.3 \,\, 10^{-27}$ (0.72) & $1.8 \,\, 10^{-24}$ (0.54) & $2.0 \,\, 10^{-24}$ (0.52) \\
11.79 & $9.2 \,\, 10^{-28}$ (22.56) & $1.3 \,\, 10^{-27}$ (24.52) & $1.9 \,\, 10^{-26}$ (25.96) & $5.4 \,\, 10^{-26}$ (30.61) & $7.1 \,\, 10^{-23}$ (20.98) & $8.0 \,\, 10^{-23}$ (21.29)\\
12.20 & $2.4 \,\, 10^{-29}$ (0.59) & $3.2 \,\, 10^{-29}$ (0.63) & $4.9 \,\, 10^{-28}$ (0.69) & $1.4 \,\, 10^{-27}$ (0.81) & $1.8 \,\, 10^{-24}$ (0.54) & $2.1 \,\, 10^{-24}$ (0.55) \\
13.40 & $6.9 \,\, 10^{-29}$ (1.69) & $9.4 \,\, 10^{-29}$ (1.83) & $1.4 \,\, 10^{-27}$ (1.93) & $4.1 \,\, 10^{-27}$ (2.34) & $5.3 \,\, 10^{-24}$ (1.56) & $5.9 \,\, 10^{-24}$ (1.55) \\
14.06 & $2.1 \,\, 10^{-28}$ (5.14) & $2.9 \,\, 10^{-28}$ (5.59) & $4.3 \,\, 10^{-27}$ (6.03) & $1.3 \,\, 10^{-26}$ (7.19) & $1.6 \,\, 10^{-23}$ (4.74) & $1.8 \,\, 10^{-23}$ (4.82) \\
14.43 & $1.9 \,\, 10^{-30}$ (0.05) & $2.9 \,\, 10^{-30}$ (0.06) & $4.2 \,\, 10^{-29}$ (0.06) & $1.5 \,\, 10^{-28}$ (0.08) & $1.1 \,\, 10^{-25}$ (0.03) & $1.3 \,\, 10^{-25}$ (0.03) \\
$[17.26]$ & $2.0 \,\, 10^{-30}$ (0.05) & $3.2 \,\, 10^{-30}$ (0.06) & $4.5 \,\, 10^{-29}$ (0.06) & $1.6 \,\, 10^{-28}$ (0.09) & \textemdash{} & \textemdash{} \\
$[17.34]$ & $2.0 \,\, 10^{-30}$ (0.05) & $3.3 \,\, 10^{-30}$ (0.06) & $4.7 \,\, 10^{-29}$ (0.07) & $1.6 \,\, 10^{-28}$ (0.09) & \textemdash{} & \textemdash{} \,\, \\
18.20 & $1.4 \,\, 10^{-29}$ (0.35) & $2.0 \,\, 10^{-29}$ (0.40) & $3.1 \,\, 10^{-28}$ (0.43) & $9.4 \,\, 10^{-28}$ (0.53) & $1.1 \,\, 10^{-24}$ (0.32) & $1.2 \,\, 10^{-24}$ (0.32) \\
$[18.98]$ & $2.3 \,\, 10^{-30}$ (0.06) & $3.5 \,\, 10^{-30}$ (0.07) & $5.6 \,\, 10^{-29}$ (0.08) & $1.8 \,\, 10^{-28}$ (0.10) & \textemdash{} & \textemdash{} \\
$[19.82]$ & $2.8 \,\, 10^{-30}$ (0.07) & $4.3 \,\, 10^{-30}$ (0.08) & $6.2 \,\, 10^{-29}$ (0.09) & $1.9 \,\, 10^{-28}$ (0.11) & \textemdash{} & \textemdash{} \\
$[19.93]$ & $2.8 \,\, 10^{-30}$ (0.07) & $4.0 \,\, 10^{-30}$ (0.08) & $5.8 \,\, 10^{-29}$ (0.08) & $1.9 \,\, 10^{-28}$ (0.11) & \textemdash{} & \textemdash{} \\
20.00 & $1.3 \,\, 10^{-29}$ (0.33) & $1.9 \,\, 10^{-29}$ (0.38) & $2.9 \,\, 10^{-28}$ (0.40) & $9.0 \,\, 10^{-28}$ (0.51) & $9.9 \,\, 10^{-25}$ (0.29) & $1.1 \,\, 10^{-24}$ (0.30) \\
20.38 & $9.2 \,\, 10^{-30}$ (0.23) & $1.3 \,\, 10^{-29}$ (0.26) & $2.0 \,\, 10^{-28}$ (0.27) & $6.3 \,\, 10^{-28}$ (0.36) & $6.7 \,\, 10^{-25}$ (0.20) & $7.8 \,\, 10^{-25}$ (0.21) \\
$[21.96]$ & $3.4 \,\, 10^{-30}$ (0.08) & $4.7 \,\, 10^{-30}$ (0.09) & $7.5 \,\, 10^{-29}$ (0.10) & $2.3 \,\, 10^{-28}$ (0.13) & \textemdash{} & \textemdash{} \\
$[24.64]$ & $4.1 \,\, 10^{-30}$ (0.10) & $6.2 \,\, 10^{-30}$ (0.12) & $8.7 \,\, 10^{-29}$ (0.12) & $2.7 \,\, 10^{-28}$ (0.16) & \textemdash{} & \textemdash{} \\
$[25.27]$ & $4.4 \,\, 10^{-30}$ (0.11) & $6.2 \,\, 10^{-30}$ (0.12) & $8.8 \,\, 10^{-29}$ (0.12) & $2.9 \,\, 10^{-28}$ (0.17) & \textemdash{} & \textemdash{} \\
28.32 & $8.6 \,\, 10^{-30}$ (0.21) & $1.2 \,\, 10^{-29}$ (0.24) & $1.8 \,\, 10^{-28}$ (0.26) & $6.0 \,\, 10^{-28}$ (0.34) & $5.5 \,\, 10^{-25}$ (0.16) & $6.1 \,\, 10^{-25}$ (0.16) \\
$[38.63]$ & $6.6 \,\, 10^{-30}$ (0.16) & $1.0 \,\, 10^{-29}$ (0.19) & $1.4 \,\, 10^{-28}$ (0.20) & $4.1 \,\, 10^{-28}$ (0.23) & \textemdash{} & \textemdash{} \\
$[40.66]$ & $7.0 \,\, 10^{-30}$ (0.17) & $1.0 \,\, 10^{-29}$ (0.20) & $1.5 \,\, 10^{-28}$ (0.21) & $4.2 \,\, 10^{-28}$ (0.24) & \textemdash{} & \textemdash{} \\
47.80 & $1.6 \,\, 10^{-29}$ (0.39) & $2.4 \,\, 10^{-29}$ (0.46) & $3.6 \,\, 10^{-28}$ (0.50) & $1.2 \,\, 10^{-27}$ (0.66) & $1.1 \,\, 10^{-24}$ (0.33) & $1.3 \,\, 10^{-24}$ (0.33) \\
$[66.26]$ & $9.1 \,\, 10^{-30}$ (0.22) & $1.4 \,\, 10^{-29}$ (0.27) & $1.6 \,\, 10^{-28}$ (0.23) & $3.9 \,\, 10^{-28}$ (0.22) & \textemdash{} & \textemdash{} \\
101.60 & $1.1 \,\, 10^{-29}$ (0.27) & $1.7 \,\, 10^{-29}$ (0.32) & $2.5 \,\, 10^{-28}$ (0.35) & $8.6 \,\, 10^{-28}$ (0.49) & $8.1 \,\, 10^{-26}$ (0.02) & $9.0 \,\, 10^{-26}$ (0.02) \\
\noalign{\smallskip} \hline 
\end{tabular}
\end{center} 
\end{table*}

\begin{table*}
\begin{center}
\caption{As Table~\ref{anthracene_comparison2} for neutral coronene 
(C$_{24}$H$_{12}$).}
\label{coronene_comparison2}
\begin{tabular}{ccccccc}
\hline \hline \noalign{\smallskip} 
& \multicolumn{6}{c}{Integrated flux} \\
\noalign{\smallskip} \cline{2-7} \noalign{\smallskip}
& \multicolumn{2}{c}{ISRF} & \multicolumn{2}{c}{Red Rectangle}
& \multicolumn{2}{c}{IRAS~21282+5050}\\
Peak & $\sigma_\mathrm{lab}$ & $\sigma_\mathrm{th}$ & $\sigma_\mathrm{lab}$ & $\sigma_\mathrm{th}$ & $\sigma_\mathrm{lab}$ & $\sigma_\mathrm{th}$ \\
($\mu$m) & (W~sr$^{-1}$) (\%) & (W~sr$^{-1}$) (\%) & (W~sr$^{-1}$) (\%) & (W~sr$^{-1}$) (\%) & (W~sr$^{-1}$) (\%) & (W~sr$^{-1}$) (\%) \\
\noalign{\smallskip} \hline \noalign{\smallskip}
3.26 & $3.4 \,\, 10^{-27}$ (36.85) & $2.8 \,\, 10^{-27}$ (34.47) & $6.1 \,\, 10^{-26}$ (27.41) & $5.8 \,\, 10^{-26}$ (24.75) & $3.1 \,\, 10^{-22}$ (43.24) & $2.3 \,\, 10^{-22}$ (42.58) \\
3.29 & $2.4 \,\, 10^{-28}$ (2.56) & $1.9 \,\, 10^{-28}$ (2.33) & $4.1 \,\, 10^{-27}$ (1.86) & $4.0 \,\, 10^{-27}$ (1.72) & $2.1 \,\, 10^{-23}$ (2.99) & $1.6 \,\, 10^{-23}$ (2.88) \\
6.24 & $4.9 \,\, 10^{-28}$ (5.24) & $4.2 \,\, 10^{-28}$ (5.27) & $1.2 \,\, 10^{-26}$ (5.45) & $1.3 \,\, 10^{-26}$ (5.47) & $3.7 \,\, 10^{-23}$ (5.16) & $2.9 \,\, 10^{-23}$ (5.21) \\
6.69 & $8.4 \,\, 10^{-30}$ (0.09) & $7.5 \,\, 10^{-30}$ (0.09) & $2.0 \,\, 10^{-28}$ (0.09) & $2.4 \,\, 10^{-28}$ (0.10) & $5.9 \,\, 10^{-25}$ (0.08) & $5.1 \,\, 10^{-25}$ (0.09) \\
7.21 & $1.0 \,\, 10^{-28}$ (1.07) & $8.9 \,\, 10^{-29}$ (1.11) & $2.6 \,\, 10^{-27}$ (1.17) & $2.8 \,\, 10^{-27}$ (1.22) & $7.4 \,\, 10^{-24}$ (1.04) & $5.8 \,\, 10^{-24}$ (1.06) \\
7.63 & $6.3 \,\, 10^{-28}$ (6.76) & $5.5 \,\, 10^{-28}$ (6.88) & $1.7 \,\, 10^{-26}$ (7.47) & $1.7 \,\, 10^{-26}$ (7.49) & $4.5 \,\, 10^{-23}$ (6.31) & $3.5 \,\, 10^{-23}$ (6.42) \\
8.25 & $2.8 \,\, 10^{-28}$ (2.99) & $2.5 \,\, 10^{-28}$ (3.08) & $7.3 \,\, 10^{-27}$ (3.30) & $8.0 \,\, 10^{-27}$ (3.42) & $2.0 \,\, 10^{-23}$ (2.80) & $1.6 \,\, 10^{-23}$ (2.82) \\
8.78 & $2.9 \,\, 10^{-28}$ (3.16) & $2.6 \,\, 10^{-28}$ (3.26) & $8.1 \,\, 10^{-27}$ (3.64) & $8.8 \,\, 10^{-27}$ (3.79) & $2.1 \,\, 10^{-23}$ (2.98) & $1.6 \,\, 10^{-23}$ (2.96) \\
11.57 & $3.2 \,\, 10^{-27}$ (33.86) & $2.8 \,\, 10^{-27}$ (35.56) & $9.0 \,\, 10^{-26}$ (40.58) & $9.9 \,\, 10^{-26}$ (42.48) & $2.2 \,\, 10^{-22}$ (30.21) & $1.7 \,\, 10^{-22}$ (30.70) \\
12.43 & $1.0 \,\, 10^{-29}$ (0.11) & $9.3 \,\, 10^{-30}$ (0.12) & $3.0 \,\, 10^{-28}$ (0.13) & $3.3 \,\, 10^{-28}$ (0.14) & $6.8 \,\, 10^{-25}$ (0.09) & $5.2 \,\, 10^{-25}$ (0.09) \\
12.90 & $1.5 \,\, 10^{-28}$ (1.59) & $1.3 \,\, 10^{-28}$ (1.69) & $4.4 \,\, 10^{-27}$ (1.99) & $4.7 \,\, 10^{-27}$ (2.03) & $1.0 \,\, 10^{-23}$ (1.41) & $7.8 \,\, 10^{-24}$ (1.42) \\
18.21 & $3.7 \,\, 10^{-28}$ (3.99) & $3.4 \,\, 10^{-28}$ (4.24) & $1.1 \,\, 10^{-26}$ (4.99) & $1.2 \,\, 10^{-26}$ (5.29) & $2.4 \,\, 10^{-23}$ (3.35) & $1.9 \,\, 10^{-23}$ (3.41) \\
$[22.28]$ & $4.9 \,\, 10^{-30}$ (0.05) & $5.3 \,\, 10^{-30}$ (0.07) & $1.6 \,\, 10^{-28}$ (0.07) & $1.8 \,\, 10^{-28}$ (0.08) & \textemdash{} & \textemdash{} \\
26.20 & $2.7 \,\, 10^{-29}$ (0.29) & $2.5 \,\, 10^{-29}$ (0.31) & $8.5 \,\, 10^{-28}$ (0.38) & $9.7 \,\, 10^{-28}$ (0.42) & $1.5 \,\, 10^{-24}$ (0.21) & $1.2 \,\, 10^{-24}$ (0.22) \\
$[27.34]$ & $8.8 \,\, 10^{-30}$ (0.09) & $8.3 \,\, 10^{-30}$ (0.10) & $2.6 \,\, 10^{-28}$ (0.12) & $3.1 \,\, 10^{-28}$ (0.13) & \textemdash{} & \textemdash{} \\
$[33.59]$ & $1.3 \,\, 10^{-29}$ (0.14) & $1.2 \,\, 10^{-29}$ (0.15) & $3.8 \,\, 10^{-28}$ (0.17) & $4.1 \,\, 10^{-28}$ (0.18) & \textemdash{} & \textemdash{} \\
$[34.34]$ & $1.3 \,\, 10^{-29}$ (0.14) & $1.3 \,\, 10^{-29}$ (0.17) & $3.8 \,\, 10^{-28}$ (0.17) & $4.4 \,\, 10^{-28}$ (0.19) & \textemdash{} & \textemdash{} \\
$[44.57]$ & $9.6 \,\, 10^{-30}$ (0.10) & $8.8 \,\, 10^{-30}$ (0.11) & $2.4 \,\, 10^{-28}$ (0.11) & $2.9 \,\, 10^{-28}$ (0.12) & \textemdash{} & \textemdash{} \\
$[61.10]$ & $1.2 \,\, 10^{-29}$ (0.13) & $1.2 \,\, 10^{-29}$ (0.15) & $2.8 \,\, 10^{-28}$ (0.13) & $3.1 \,\, 10^{-28}$ (0.13) & \textemdash{} & \textemdash{} \\
80.60 & $2.3 \,\, 10^{-29}$ (0.24) & $2.2 \,\, 10^{-29}$ (0.27) & $7.4 \,\, 10^{-28}$ (0.33) & $8.4 \,\, 10^{-28}$ (0.36) & $8.4 \,\, 10^{-25}$ (0.12) & $7.2 \,\, 10^{-25}$ (0.13) \\
$[113.37]$ & $3.0 \,\, 10^{-29}$ (0.33) & $2.8 \,\, 10^{-29}$ (0.36) & $3.9 \,\, 10^{-28}$ (0.18) & $4.2 \,\, 10^{-28}$ (0.18) & \textemdash{} & \textemdash{} \\
\noalign{\smallskip} \hline 
\end{tabular}
\end{center} 
\end{table*}

\begin{table*}
\begin{center}
\caption{As Table~\ref{anthracene_comparison2} for neutral ovalene 
(C$_{32}$H$_{14}$).}
\label{ovalene_comparison2}
\begin{tabular}{ccccccc}
\hline \hline \noalign{\smallskip} 
& \multicolumn{6}{c}{Integrated flux} \\
\noalign{\smallskip} \cline{2-7} \noalign{\smallskip}
& \multicolumn{2}{c}{ISRF} & \multicolumn{2}{c}{Red Rectangle}
& \multicolumn{2}{c}{IRAS~21282+5050}\\
Peak & $\sigma_\mathrm{lab}$ & $\sigma_\mathrm{th}$ & $\sigma_\mathrm{lab}$ & $\sigma_\mathrm{th}$ & $\sigma_\mathrm{lab}$ & $\sigma_\mathrm{th}$ \\
($\mu$m) & (W~sr$^{-1}$) (\%) & (W~sr$^{-1}$) (\%) & (W~sr$^{-1}$) (\%) & (W~sr$^{-1}$) (\%) & (W~sr$^{-1}$) (\%) & (W~sr$^{-1}$) (\%) \\
\noalign{\smallskip} \hline \noalign{\smallskip}
3.26 & $1.4 \,\, 10^{-27}$ (10.25) & $1.2 \,\, 10^{-27}$ (8.79) & $3.6 \,\, 10^{-26}$ (5.92) & $3.7 \,\, 10^{-26}$ (4.84) & $1.3 \,\, 10^{-22}$ (15.06) & $1.0 \,\, 10^{-22}$ (14.81) \\
3.26 & $1.1 \,\, 10^{-27}$ (8.10) & $9.1 \,\, 10^{-28}$ (6.88) & $2.8 \,\, 10^{-26}$ (4.59) & $2.8 \,\, 10^{-26}$ (3.62) & $1.0 \,\, 10^{-22}$ (11.62) & $7.8 \,\, 10^{-23}$ (11.43) \\
3.27 & $9.5 \,\, 10^{-29}$ (0.68) & $8.1 \,\, 10^{-29}$ (0.61) & $2.4 \,\, 10^{-27}$ (0.39) & $2.7 \,\, 10^{-27}$ (0.35) & $8.9 \,\, 10^{-24}$ (1.03) & $6.8 \,\, 10^{-24}$ (0.99) \\
3.29 & $6.1 \,\, 10^{-29}$ (0.44) & $4.8 \,\, 10^{-29}$ (0.36) & $1.6 \,\, 10^{-27}$ (0.26) & $1.5 \,\, 10^{-27}$ (0.20) & $5.4 \,\, 10^{-24}$ (0.62) & $4.5 \,\, 10^{-24}$ (0.65) \\
3.29 & $1.2 \,\, 10^{-28}$ (0.87) & $9.9 \,\, 10^{-29}$ (0.75) & $3.0 \,\, 10^{-27}$ (0.50) & $3.1 \,\, 10^{-27}$ (0.40) & $1.1 \,\, 10^{-23}$ (1.28) & $8.2 \,\, 10^{-24}$ (1.20) \\
3.29 & $5.3 \,\, 10^{-29}$ (0.38) & $4.8 \,\, 10^{-29}$ (0.36) & $1.3 \,\, 10^{-27}$ (0.21) & $1.5 \,\, 10^{-27}$ (0.20) & $5.2 \,\, 10^{-24}$ (0.60) & $3.8 \,\, 10^{-24}$ (0.56) \\
6.19 & $5.9 \,\, 10^{-28}$ (4.25) & $5.3 \,\, 10^{-28}$ (4.01) & $2.4 \,\, 10^{-26}$ (4.01) & $3.0 \,\, 10^{-26}$ (3.84) & $3.9 \,\, 10^{-23}$ (4.53) & $3.0 \,\, 10^{-23}$ (4.44) \\
6.26 & $3.4 \,\, 10^{-28}$ (2.43) & $3.1 \,\, 10^{-28}$ (2.34) & $1.4 \,\, 10^{-26}$ (2.36) & $1.7 \,\, 10^{-26}$ (2.15) & $2.3 \,\, 10^{-23}$ (2.63) & $1.7 \,\, 10^{-23}$ (2.56) \\
6.35 & $1.5 \,\, 10^{-29}$ (0.11) & $1.4 \,\, 10^{-29}$ (0.10) & $7.0 \,\, 10^{-28}$ (0.11) & $6.4 \,\, 10^{-28}$ (0.08) & $9.1 \,\, 10^{-25}$ (0.11) & $7.8 \,\, 10^{-25}$ (0.11) \\
6.49 & $1.2 \,\, 10^{-29}$ (0.09) & $1.3 \,\, 10^{-29}$ (0.10) & $5.8 \,\, 10^{-28}$ (0.09) & $6.6 \,\, 10^{-28}$ (0.09) & $8.0 \,\, 10^{-25}$ (0.09) & $6.2 \,\, 10^{-25}$ (0.09) \\
6.57 & $2.9 \,\, 10^{-28}$ (2.08) & $2.6 \,\, 10^{-28}$ (1.96) & $1.2 \,\, 10^{-26}$ (1.97) & $1.5 \,\, 10^{-26}$ (1.98) & $1.8 \,\, 10^{-23}$ (2.10) & $1.4 \,\, 10^{-23}$ (2.09) \\
6.74 & $5.2 \,\, 10^{-29}$ (0.38) & $4.6 \,\, 10^{-29}$ (0.35) & $2.1 \,\, 10^{-27}$ (0.35) & $2.7 \,\, 10^{-27}$ (0.35) & $3.3 \,\, 10^{-24}$ (0.38) & $2.5 \,\, 10^{-24}$ (0.36) \\
6.87 & $1.6 \,\, 10^{-29}$ (0.12) & $1.5 \,\, 10^{-29}$ (0.11) & $6.7 \,\, 10^{-28}$ (0.11) & $8.9 \,\, 10^{-28}$ (0.12) & $8.7 \,\, 10^{-25}$ (0.10) & $7.7 \,\, 10^{-25}$ (0.11) \\
6.96 & $1.3 \,\, 10^{-29}$ (0.09) & $1.1 \,\, 10^{-29}$ (0.09) & $5.4 \,\, 10^{-28}$ (0.09) & $7.2 \,\, 10^{-28}$ (0.09) & $8.1 \,\, 10^{-25}$ (0.09) & $6.3 \,\, 10^{-25}$ (0.09) \\
7.15 & $8.6 \,\, 10^{-29}$ (0.62) & $7.4 \,\, 10^{-29}$ (0.56) & $3.9 \,\, 10^{-27}$ (0.64) & $4.6 \,\, 10^{-27}$ (0.59) & $5.4 \,\, 10^{-24}$ (0.62) & $4.2 \,\, 10^{-24}$ (0.61) \\
7.18 & $5.8 \,\, 10^{-29}$ (0.42) & $5.2 \,\, 10^{-29}$ (0.39) & $2.6 \,\, 10^{-27}$ (0.43) & $3.3 \,\, 10^{-27}$ (0.43) & $3.5 \,\, 10^{-24}$ (0.40) & $2.9 \,\, 10^{-24}$ (0.42) \\
7.25 & $2.4 \,\, 10^{-29}$ (0.17) & $2.2 \,\, 10^{-29}$ (0.16) & $1.1 \,\, 10^{-27}$ (0.18) & $1.1 \,\, 10^{-27}$ (0.15) & $1.4 \,\, 10^{-24}$ (0.16) & $1.1 \,\, 10^{-24}$ (0.16) \\
7.29 & $2.3 \,\, 10^{-28}$ (1.65) & $2.2 \,\, 10^{-28}$ (1.65) & $1.0 \,\, 10^{-26}$ (1.67) & $1.3 \,\, 10^{-26}$ (1.65) & $1.4 \,\, 10^{-23}$ (1.67) & $1.1 \,\, 10^{-23}$ (1.64) \\
7.59 & $1.0 \,\, 10^{-28}$ (0.72) & $9.4 \,\, 10^{-29}$ (0.71) & $4.3 \,\, 10^{-27}$ (0.70) & $5.2 \,\, 10^{-27}$ (0.67) & $6.0 \,\, 10^{-24}$ (0.70) & $4.7 \,\, 10^{-24}$ (0.68) \\
7.67 & $8.5 \,\, 10^{-28}$ (6.15) & $8.0 \,\, 10^{-28}$ (6.06) & $3.8 \,\, 10^{-26}$ (6.29) & $4.8 \,\, 10^{-26}$ (6.20) & $5.1 \,\, 10^{-23}$ (5.94) & $4.1 \,\, 10^{-23}$ (5.95) \\
7.83 & $3.5 \,\, 10^{-28}$ (2.55) & $3.4 \,\, 10^{-28}$ (2.56) & $1.6 \,\, 10^{-26}$ (2.66) & $2.1 \,\, 10^{-26}$ (2.71) & $2.2 \,\, 10^{-23}$ (2.51) & $1.7 \,\, 10^{-23}$ (2.49) \\
8.07 & $1.1 \,\, 10^{-28}$ (0.82) & $1.1 \,\, 10^{-28}$ (0.83) & $5.3 \,\, 10^{-27}$ (0.88) & $6.5 \,\, 10^{-27}$ (0.84) & $6.9 \,\, 10^{-24}$ (0.79) & $5.4 \,\, 10^{-24}$ (0.79) \\
8.13 & $4.3 \,\, 10^{-29}$ (0.31) & $4.3 \,\, 10^{-29}$ (0.32) & $2.1 \,\, 10^{-27}$ (0.35) & $2.7 \,\, 10^{-27}$ (0.34) & $2.8 \,\, 10^{-24}$ (0.33) & $2.0 \,\, 10^{-24}$ (0.30) \\
8.24 & $2.9 \,\, 10^{-29}$ (0.21) & $2.9 \,\, 10^{-29}$ (0.22) & $1.3 \,\, 10^{-27}$ (0.21) & $1.7 \,\, 10^{-27}$ (0.22) & $1.7 \,\, 10^{-24}$ (0.19) & $1.3 \,\, 10^{-24}$ (0.20) \\
8.48 & $7.4 \,\, 10^{-29}$ (0.54) & $7.5 \,\, 10^{-29}$ (0.57) & $3.6 \,\, 10^{-27}$ (0.59) & $4.6 \,\, 10^{-27}$ (0.60) & $4.3 \,\, 10^{-24}$ (0.50) & $3.5 \,\, 10^{-24}$ (0.51) \\
8.56 & $8.7 \,\, 10^{-28}$ (6.26) & $8.5 \,\, 10^{-28}$ (6.46) & $4.1 \,\, 10^{-26}$ (6.74) & $5.2 \,\, 10^{-26}$ (6.79) & $5.1 \,\, 10^{-23}$ (5.92) & $4.1 \,\, 10^{-23}$ (6.00) \\
8.63 & $5.2 \,\, 10^{-30}$ (0.04) & $5.5 \,\, 10^{-30}$ (0.04) & $2.2 \,\, 10^{-28}$ (0.04) & $2.5 \,\, 10^{-28}$ (0.03) & $2.8 \,\, 10^{-25}$ (0.03) & $2.0 \,\, 10^{-25}$ (0.03) \\
10.28 & $4.2 \,\, 10^{-30}$ (0.03) & $4.9 \,\, 10^{-30}$ (0.04) & $1.9 \,\, 10^{-28}$ (0.03) & $2.7 \,\, 10^{-28}$ (0.03) & $2.4 \,\, 10^{-25}$ (0.03) & $1.8 \,\, 10^{-25}$ (0.03) \\
10.36 & $4.8 \,\, 10^{-29}$ (0.34) & $4.9 \,\, 10^{-29}$ (0.37) & $2.4 \,\, 10^{-27}$ (0.40) & $3.1 \,\, 10^{-27}$ (0.40) & $2.5 \,\, 10^{-24}$ (0.29) & $2.0 \,\, 10^{-24}$ (0.30) \\
10.98 & $1.6 \,\, 10^{-28}$ (1.13) & $1.5 \,\, 10^{-28}$ (1.16) & $7.7 \,\, 10^{-27}$ (1.27) & $1.0 \,\, 10^{-26}$ (1.34) & $8.4 \,\, 10^{-24}$ (0.97) & $6.7 \,\, 10^{-24}$ (0.98) \\
11.18 & $3.2 \,\, 10^{-27}$ (22.80) & $3.1 \,\, 10^{-27}$ (23.86) & $1.6 \,\, 10^{-25}$ (26.32) & $2.1 \,\, 10^{-25}$ (27.52) & $1.7 \,\, 10^{-22}$ (19.54) & $1.4 \,\, 10^{-22}$ (19.93) \\
11.86 & $1.9 \,\, 10^{-27}$ (13.34) & $1.9 \,\, 10^{-27}$ (14.15) & $9.5 \,\, 10^{-26}$ (15.68) & $1.3 \,\, 10^{-25}$ (16.30) & $9.7 \,\, 10^{-23}$ (11.22) & $7.8 \,\, 10^{-23}$ (11.37) \\
12.86 & $1.7 \,\, 10^{-28}$ (1.24) & $1.8 \,\, 10^{-28}$ (1.39) & $9.1 \,\, 10^{-27}$ (1.49) & $1.2 \,\, 10^{-26}$ (1.57) & $8.9 \,\, 10^{-24}$ (1.03) & $7.3 \,\, 10^{-24}$ (1.06) \\
12.89 & $1.1 \,\, 10^{-28}$ (0.81) & $1.1 \,\, 10^{-28}$ (0.86) & $5.8 \,\, 10^{-27}$ (0.95) & $7.6 \,\, 10^{-27}$ (0.99) & $5.7 \,\, 10^{-24}$ (0.66) & $4.4 \,\, 10^{-24}$ (0.65) \\
13.13 & $5.9 \,\, 10^{-29}$ (0.42) & $6 \,\, 10^{-29}$ (0.45) & $3.1 \,\, 10^{-27}$ (0.51) & $4.1 \,\, 10^{-27}$ (0.53) & $3.0 \,\, 10^{-24}$ (0.34) & $2.4 \,\, 10^{-24}$ (0.36) \\
14.20 & $5.3 \,\, 10^{-29}$ (0.38) & $5.2 \,\, 10^{-29}$ (0.39) & $2.8 \,\, 10^{-27}$ (0.46) & $3.8 \,\, 10^{-27}$ (0.49) & $2.6 \,\, 10^{-24}$ (0.30) & $2.1 \,\, 10^{-24}$ (0.31) \\
15.03 & $9.2 \,\, 10^{-30}$ (0.07) & $7.9 \,\, 10^{-30}$ (0.06) & $4.8 \,\, 10^{-28}$ (0.08) & $6.1 \,\, 10^{-28}$ (0.08) & $4.2 \,\, 10^{-25}$ (0.05) & $3.4 \,\, 10^{-25}$ (0.05) \\
15.26 & $6.1 \,\, 10^{-30}$ (0.04) & $5.3 \,\, 10^{-30}$ (0.04) & $2.8 \,\, 10^{-28}$ (0.05) & $3.8 \,\, 10^{-28}$ (0.05) & $2.2 \,\, 10^{-25}$ (0.03) & $2.0 \,\, 10^{-25}$ (0.03) \\
15.90 & $4.5 \,\, 10^{-28}$ (3.26) & $4.8 \,\, 10^{-28}$ (3.60) & $2.5 \,\, 10^{-26}$ (4.09) & $3.3 \,\, 10^{-26}$ (4.33) & $2.2 \,\, 10^{-23}$ (2.52) & $1.7 \,\, 10^{-23}$ (2.56) \\
17.45 & $7.5 \,\, 10^{-29}$ (0.54) & $7.8 \,\, 10^{-29}$ (0.59) & $4.0 \,\, 10^{-27}$ (0.67) & $5.5 \,\, 10^{-27}$ (0.71) & $3.4 \,\, 10^{-24}$ (0.40) & $2.7 \,\, 10^{-24}$ (0.40) \\
18.42 & $2.1 \,\, 10^{-28}$ (1.48) & $2.2 \,\, 10^{-28}$ (1.66) & $1.1 \,\, 10^{-26}$ (1.89) & $1.6 \,\, 10^{-26}$ (2.02) & $9.5 \,\, 10^{-24}$ (1.10) & $7.8 \,\, 10^{-24}$ (1.15) \\
20.44 & $1.1 \,\, 10^{-29}$ (0.08) & $1.1 \,\, 10^{-29}$ (0.09) & $5.9 \,\, 10^{-28}$ (0.10) & $8.4 \,\, 10^{-28}$ (0.11) & $4.8 \,\, 10^{-25}$ (0.06) & $3.9 \,\, 10^{-25}$ (0.06) \\
23.57 & $3.8 \,\, 10^{-29}$ (0.27) & $3.9 \,\, 10^{-29}$ (0.29) & $2.2 \,\, 10^{-27}$ (0.36) & $2.9 \,\, 10^{-27}$ (0.38) & $1.6 \,\, 10^{-24}$ (0.19) & $1.3 \,\, 10^{-24}$ (0.19) \\
25.61 & $9.2 \,\, 10^{-29}$ (0.66) & $1.0 \,\, 10^{-28}$ (0.76) & $5.3 \,\, 10^{-27}$ (0.87) & $7.3 \,\, 10^{-27}$ (0.95) & $3.7 \,\, 10^{-24}$ (0.43) & $3.1 \,\, 10^{-24}$ (0.46) \\
29.35 & $4.0 \,\, 10^{-30}$ (0.03) & $4.6 \,\, 10^{-30}$ (0.03) & $1.3 \,\, 10^{-28}$ (0.02) & $1.5 \,\, 10^{-28}$ (0.02) & \textemdash{} & \textemdash{} \\
29.42 & $8.0 \,\, 10^{-29}$ (0.58) & $8.9 \,\, 10^{-29}$ (0.67) & $4.7 \,\, 10^{-27}$ (0.77) & $6.6 \,\, 10^{-27}$ (0.86) & $3.4 \,\, 10^{-24}$ (0.39) & $2.7 \,\, 10^{-24}$ (0.39) \\
36.70 & $6.7 \,\, 10^{-29}$ (0.49) & $7.6 \,\, 10^{-29}$ (0.58) & $3.9 \,\, 10^{-27}$ (0.64) & $5.5 \,\, 10^{-27}$ (0.71) & $2.6 \,\, 10^{-24}$ (0.30) & $2.2 \,\, 10^{-24}$ (0.32) \\
$[37.15]$ & $6.3 \,\, 10^{-30}$ (0.05) & $7.9 \,\, 10^{-30}$ (0.06) & $1.8 \,\, 10^{-28}$ (0.03) & $2.2 \,\, 10^{-28}$ (0.03) & \textemdash{} & \textemdash{} \\
$[39.08]$ & $7.4 \,\, 10^{-30}$ (0.05) & $8.3 \,\, 10^{-30}$ (0.06) & $2.0 \,\, 10^{-28}$ (0.03) & $2.4 \,\, 10^{-28}$ (0.03) & \,\, \textemdash{} & \textemdash{} \\
47.39 & $1.0 \,\, 10^{-29}$ (0.07) & $1.2 \,\, 10^{-29}$ (0.09) & $2.1 \,\, 10^{-28}$ (0.04) & $2.4 \,\, 10^{-28}$ (0.03) & \textemdash{} & \textemdash{} \\
$[51.42]$ & $1.1 \,\, 10^{-29}$ (0.08) & $1.3 \,\, 10^{-29}$ (0.10) & $2.3 \,\, 10^{-28}$ (0.04) & $2.4 \,\, 10^{-28}$ (0.03) & \textemdash{} & \textemdash{} \\
$[65.35]$ & $1.4 \,\, 10^{-29}$ (0.10) & $1.8 \,\, 10^{-29}$ (0.13) & $2.0 \,\, 10^{-28}$ (0.03) & $2.2 \,\, 10^{-28}$ (0.03) & \textemdash{} & \textemdash{} \\
$[79.27]$ & $1.7 \,\, 10^{-29}$ (0.12) & $1.9 \,\, 10^{-29}$ (0.15) & $1.9 \,\, 10^{-28}$ (0.03) & $1.9 \,\, 10^{-28}$ (0.03) & \\
93.33 & $3.0 \,\, 10^{-29}$ (0.22) & $3.4 \,\, 10^{-29}$ (0.26) & $1.8 \,\, 10^{-27}$ (0.30) & $2.5 \,\, 10^{-27}$ (0.33) & $5.1 \,\, 10^{-25}$ (0.06) & $4.5 \,\, 10^{-25}$ (0.07) \\
160.24 & $2.6 \,\, 10^{-29}$ (0.18) & $3.1 \,\, 10^{-29}$ (0.23) & $1.5 \,\, 10^{-27}$ (0.25) & $2.2 \,\, 10^{-27}$ (0.28) & \textemdash{} & \textemdash{} \\
$[161.07]$ & $1.9 \,\, 10^{-29}$ (0.13) & $2.1 \,\, 10^{-29}$ (0.16) & $7.1 \,\, 10^{-29}$ (0.01) & $7.6 \,\, 10^{-29}$ (0.01) & \textemdash{} & \textemdash{} \\
\noalign{\smallskip} \hline 
\end{tabular}
\end{center} 
\end{table*}


\section{Comparison between detailed and approximated PAH photophysics}
\label{modelcomp}

Tables from \ref{anthracene_detailed} to \ref{pyrene_detailed}
shows the comparison between model runs for neutral anthracene, phenanthrene and pyrene, 
either taking into account the detailed relaxation channels 
available or assuming only IR 
de\textendash excitation. Neglecting fluorescence and phosphorescence leads to an 
overestimation of the absolute emission fluxes by up to a factor $\sim$2 in 
the worst case for the high energy bands. Low energy bands are almost 
unaffected. This is because fluorescence and/or phosphorescence, when 
they occur, essentially cause the molecule to ``skip'' the high excitation 
part of the vibrational cascade, in which high energy vibrational photons 
are preferentially emitted, while the low energy vibrational photons are 
emitted near the end of the emission cascade and this part is never skipped
anyway.

All closed\textendash shell (e.~g. neutral, fully hydrogenated) 
PAHs, upon excitation \mbox{$\mathrm{S}_n\gets\mathrm{S}_0$}, will almost
always (i.~e. if they don't ionise or dissociate) undergo very fast 
internal conversion (IC) to a low lying singlet electronic level
$\mathrm{S}_m$, which is usually $\mathrm{S}_1$ \citep[e.~g.]
[]{lea95a,lea95b}, with few exceptions (e.~g. fluorene, 
for which it is $\mathrm{S}_2$). From there, three relaxation channels are 
available, their branching ratios being dependent on the vibrational energy 
available: 
\begin{enumerate}
\item fluorescence \mbox{$\mathrm{S}_m\to\mathrm{S}_0$} with a permitted 
electronic transition, the remaining energy being subsequently radiated 
by vibrational transitions in $\mathrm{S}_0$; relaxation of small PAHs via 
this path is the proposed origin of the blue luminescence observed in the 
Red Rectangle \citep{vij04,vij05};
\item direct \mbox{$\mathrm{S}_m\leadsto\mathrm{T}_1$} or indirect 
\mbox{$\mathrm{S}_m\leadsto\mathrm{T}_n\leadsto\mathrm{T}_1$} intersystem crossing, 
a radiationless transition followed by the emission of a 
phosphorescence photon in a \mbox{$\mathrm{T}_1\to\mathrm{S}_0$} spin\textendash forbidden,
electronic\textendash permitted transition; the remaining energy is radiated in 
vibrational transitions either (almost always) from $\mathrm{S}_0$ after 
the phosphorescence transition or (very seldom) from the first triplet state
$\mathrm{T}_1$ before it;
\item IC \mbox{$\mathrm{S}_m\leadsto\mathrm{S}_0$}, a radiationless 
transition, after which essentially all the excitation energy is radiated 
by vibrational transitions.
\end{enumerate}
According to experimental results \citep{bre05}, the rate of fluorescence transitions 
(1) is essentially independent of the excitation energy. The rate of 
radiationless transitions 
(2) increases slightly with excitation energy for the three molecules considered. 
The relaxation path (3) 
by IC to the ground state 
is opened only when some excess vibrational energy in 
$\mathrm{S}_m$ is available, such threshold depending on the specific 
molecule, varying e.~g. from $\sim$2~$10^3$~cm$^{-1}$ for anthracene to 
$\sim$4~$10^4$~cm$^{-1}$ for pyrene \citep{bre05}. Above this threshold the rate of 
IC to $\mathrm{S}_0$ 
(3) exponentially increases, becoming by and large the 
dominant relaxation path.  

\begin{table}[t!]
\begin{center}
\caption{Comparison between model runs for neutral anthracene
assuming the detailed relaxation channels available or considering only 
IR de\textendash excitation. We list the absolute flux emitted by one molecule in the 
Red Rectangle RF and, in parentheses, the flux fraction in that band. 
Bands whose calculated flux fraction is $\leq$0.03\% are omitted. 
Electric\textendash dipole forbidden bands are enclosed in square brackets.}
\label{anthracene_detailed}
\begin{tabular}{ccc}
\hline \hline \noalign{\smallskip} 
Peak & Detailed & Approximated\\
($\mu$m) & (W$ $sr$^{-1}$) (\%) & (W$ $sr$^{-1}$) (\%) \\ 
\noalign{\smallskip} \hline \noalign{\smallskip}
3.25 & $1.4 \,\, 10^{-26}$ (22.40) & $1.6 \,\, 10^{-26}$ (22.93)\\
3.26 & $9.3 \,\, 10^{-27}$ (14.59) & $1.1 \,\, 10^{-26}$ (15.09)\\
3.28 & $3.0 \,\, 10^{-29}$ (0.05) & $3.4 \,\, 10^{-29}$ (0.05)\\
3.28 & $2.5 \,\, 10^{-27}$ (3.86) & $2.9 \,\, 10^{-27}$ (3.97)\\
3.29 & $1.3 \,\, 10^{-27}$ (2.12) & $1.6 \,\, 10^{-27}$ (2.17)\\
6.16 & $2.1 \,\, 10^{-27}$ (3.33) & $2.4 \,\, 10^{-27}$ (3.37)\\
6.51 & $6.2 \,\, 10^{-28}$ (0.98) & $6.9 \,\, 10^{-28}$ (0.96)\\
6.86 & $1.2 \,\, 10^{-27}$ (1.95) & $1.4 \,\, 10^{-27}$ (1.88)\\
6.86 & $5.7 \,\, 10^{-28}$ (0.89) & $6.6 \,\, 10^{-28}$ (0.92)\\
7.22 & $4.2 \,\, 10^{-29}$ (0.07) & $5.8 \,\, 10^{-29}$ (0.08)\\
7.43 & $7.8 \,\, 10^{-28}$ (1.22) & $9.0 \,\, 10^{-28}$ (1.25)\\
7.60 & $1.3 \,\, 10^{-27}$ (2.00) & $1.5 \,\, 10^{-27}$ (2.04)\\
7.85 & $1.6 \,\, 10^{-27}$ (2.57) & $1.8 \,\, 10^{-27}$ (2.53)\\
8.55 & $2.3 \,\, 10^{-28}$ (0.36) & $2.5 \,\, 10^{-28}$ (0.35)\\
8.62 & $5.9 \,\, 10^{-28}$ (0.93) & $6.5 \,\, 10^{-28}$ (0.91)\\
8.67 & $1.5 \,\, 10^{-27}$ (2.35) & $1.7 \,\, 10^{-27}$ (2.39)\\
9.95 & $7.3 \,\, 10^{-28}$ (1.15) & $8.1 \,\, 10^{-28}$ (1.13)\\
10.41 & $1.3 \,\, 10^{-27}$ (2.05) & $1.5 \,\, 10^{-27}$ (2.02)\\
11.00 & $2.8 \,\, 10^{-28}$ (0.44) & $3.1 \,\, 10^{-28}$ (0.43)\\
11.32 & $9.6 \,\, 10^{-27}$ (15.04) & $1.1 \,\, 10^{-26}$ (14.72)\\
13.71 & $9.6 \,\, 10^{-27}$ (15.02) & $1.1 \,\, 10^{-26}$ (14.64)\\
15.33 & $1.8 \,\, 10^{-28}$ (0.28) & $2.1 \,\, 10^{-28}$ (0.29)\\
$[15.71]$ & $4.1 \,\, 10^{-29}$ (0.06) & $4.1 \,\, 10^{-29}$ (0.06)\\
16.34 & $7.7 \,\, 10^{-28}$ (1.22) & $8.5 \,\, 10^{-28}$ (1.18)\\
$[17.11]$ & $5.3 \,\, 10^{-29}$ (0.08) & $5.0 \,\, 10^{-29}$ (0.07)\\
$[18.71]$ & $6.5 \,\, 10^{-29}$ (0.10) & $6.4 \,\, 10^{-29}$ (0.09)\\
$[20.04]$ & $7.8 \,\, 10^{-29}$ (0.12) & $7.3 \,\, 10^{-29}$ (0.10)\\
$[21.02]$ & $8.1 \,\, 10^{-29}$ (0.13) & $8.1 \,\, 10^{-29}$ (0.11)\\
21.26 & $1.4 \,\, 10^{-27}$ (2.20) & $1.5 \,\, 10^{-27}$ (2.09)\\
$[25.45]$ & $1.0 \,\, 10^{-28}$ (0.16) & $1.0 \,\, 10^{-28}$ (0.14)\\
$[25.62]$ & $1.1 \,\, 10^{-28}$ (0.17) & $1.1 \,\, 10^{-28}$ (0.15)\\
26.35 & $1.2 \,\, 10^{-28}$ (0.18) & $1.1 \,\, 10^{-28}$ (0.16)\\
$[37.41]$ & $1.5 \,\, 10^{-28}$ (0.23) & $1.5 \,\, 10^{-28}$ (0.20)\\
$[43.01]$ & $1.7 \,\, 10^{-28}$ (0.27) & $1.7 \,\, 10^{-28}$ (0.23)\\
43.68 & $2.3 \,\, 10^{-28}$ (0.36) & $2.4 \,\, 10^{-28}$ (0.33)\\
$[82.09]$ & $1.8 \,\, 10^{-28}$ (0.29) & $1.8 \,\, 10^{-28}$ (0.25)\\
110.23 & $2.8 \,\, 10^{-28}$ (0.44) & $2.8 \,\, 10^{-28}$ (0.39)\\
\noalign{\smallskip} \hline 
\end{tabular}
\end{center} 
\end{table}
\begin{table}[t!]
\begin{center}
\caption{As Table.~\ref{anthracene} for neutral phenanthrene 
(C$_{14}$H$_{10}$).}
\label{phenanthrene_detailed}
\begin{tabular}{ccc}
\hline \hline \noalign{\smallskip} 
Peak & Detailed & Approximated\\
($\mu$m) & (W$ $sr$^{-1}$) (\%) & (W$ $sr$^{-1}$) (\%) \\ 
\noalign{\smallskip} \hline \noalign{\smallskip}
3.23 & $3.6 \,\, 10^{-27}$ (6.57) & $6.2 \,\, 10^{-27}$ (7.20)\\
3.24 & $4.3 \,\, 10^{-27}$ (7.89) & $7.1 \,\, 10^{-27}$ (8.35)\\
3.25 & $3.1 \,\, 10^{-28}$ (0.56) & $5.5 \,\, 10^{-28}$ (0.64)\\
3.26 & $5.8 \,\, 10^{-27}$ (10.60) & $9.4 \,\, 10^{-27}$ (10.97)\\
3.26 & $5.1 \,\, 10^{-27}$ (9.43) & $8.6 \,\, 10^{-27}$ (10.01)\\
3.27 & $2.2 \,\, 10^{-27}$ (4.13) & $3.7 \,\, 10^{-27}$ (4.38)\\
3.28 & $3.2 \,\, 10^{-28}$ (0.58) & $4.9 \,\, 10^{-28}$ (0.57)\\
3.28 & $6.2 \,\, 10^{-28}$ (1.15) & $9.8 \,\, 10^{-28}$ (1.15)\\
3.29 & $7.0 \,\, 10^{-29}$ (0.13) & $1.0 \,\, 10^{-28}$ (0.12)\\
6.21 & $1.3 \,\, 10^{-28}$ (0.23) & $2.0 \,\, 10^{-28}$ (0.24)\\
6.23 & $8.8 \,\, 10^{-29}$ (0.16) & $1.4 \,\, 10^{-28}$ (0.16)\\
6.27 & $6.9 \,\, 10^{-28}$ (1.27) & $1.1 \,\, 10^{-27}$ (1.33)\\
6.57 & $2.6 \,\, 10^{-28}$ (0.48) & $4.8 \,\, 10^{-28}$ (0.57)\\
6.68 & $1.1 \,\, 10^{-27}$ (2.07) & $1.9 \,\, 10^{-27}$ (2.18)\\
6.84 & $2.2 \,\, 10^{-27}$ (4.05) & $3.7 \,\, 10^{-27}$ (4.30)\\
6.93 & $5.7 \,\, 10^{-28}$ (1.05) & $9.7 \,\, 10^{-28}$ (1.14)\\
7.04 & $1.0 \,\, 10^{-28}$ (0.18) & $1.8 \,\, 10^{-28}$ (0.21)\\
7.06 & $1.5 \,\, 10^{-28}$ (0.27) & $2.4 \,\, 10^{-28}$ (0.28)\\
7.45 & $3.5 \,\, 10^{-28}$ (0.64) & $5.6 \,\, 10^{-28}$ (0.66)\\
7.70 & $2.4 \,\, 10^{-28}$ (0.45) & $4.1 \,\, 10^{-28}$ (0.48)\\
8.00 & $1.5 \,\, 10^{-27}$ (2.68) & $2.3 \,\, 10^{-27}$ (2.71)\\
8.16 & $1.4 \,\, 10^{-28}$ (0.25) & $2.1 \,\, 10^{-28}$ (0.25)\\
8.31 & $3.0 \,\, 10^{-28}$ (0.55) & $4.9 \,\, 10^{-28}$ (0.57)\\
8.46 & $1.0 \,\, 10^{-28}$ (0.19) & $1.6 \,\, 10^{-28}$ (0.19)\\
8.60 & $3.7 \,\, 10^{-29}$ (0.07) & $5.8 \,\, 10^{-29}$ (0.07)\\
8.71 & $1.9 \,\, 10^{-28}$ (0.36) & $3.2 \,\, 10^{-28}$ (0.37)\\
9.15 & $1.3 \,\, 10^{-28}$ (0.23) & $2.0 \,\, 10^{-28}$ (0.23)\\
9.63 & $5.4 \,\, 10^{-28}$ (0.99) & $8.6 \,\, 10^{-28}$ (1.01)\\
9.66 & $9.5 \,\, 10^{-29}$ (0.18) & $1.4 \,\, 10^{-28}$ (0.17)\\
10.01 & $2.1 \,\, 10^{-28}$ (0.38) & $3.2 \,\, 10^{-28}$ (0.37)\\
10.53 & $6.2 \,\, 10^{-28}$ (1.15) & $9.3 \,\, 10^{-28}$ (1.09)\\
11.48 & $1.5 \,\, 10^{-27}$ (2.71) & $2.3 \,\, 10^{-27}$ (2.64)\\
11.49 & $2.2 \,\, 10^{-28}$ (0.40) & $3.1 \,\, 10^{-28}$ (0.37)\\
12.05 & $2.9 \,\, 10^{-29}$ (0.05) & $3.5 \,\, 10^{-29}$ (0.04)\\
12.24 & $7.5 \,\, 10^{-27}$ (13.83) & $1.2 \,\, 10^{-26}$ (13.46)\\
12.70 & $2.8 \,\, 10^{-29}$ (0.05) & $2.3 \,\, 10^{-29}$ (0.03)\\
13.23 & $3.0 \,\, 10^{-29}$ (0.05) & $3.2 \,\, 10^{-29}$ (0.04)\\
13.58 & $8.9 \,\, 10^{-27}$ (16.30) & $1.3 \,\, 10^{-26}$ (15.56)\\
13.95 & $3.1 \,\, 10^{-28}$ (0.58) & $4.3 \,\, 10^{-28}$ (0.50)\\
13.98 & $1.8 \,\, 10^{-28}$ (0.33) & $2.6 \,\, 10^{-28}$ (0.30)\\
14.12 & $4.6 \,\, 10^{-29}$ (0.08) & $4.8 \,\, 10^{-29}$ (0.06)\\
15.94 & $5.2 \,\, 10^{-28}$ (0.96) & $7.3 \,\, 10^{-28}$ (0.85)\\
$[16.86]$ & $6.3 \,\, 10^{-29}$ (0.12) & $6.2 \,\, 10^{-29}$ (0.07)\\
18.19 & $1.1 \,\, 10^{-28}$ (0.21) & $1.2 \,\, 10^{-28}$ (0.14)\\
$[18.73]$ & $7.6 \,\, 10^{-29}$ (0.14) & $7.5 \,\, 10^{-29}$ (0.09)\\
19.99 & $1.6 \,\, 10^{-28}$ (0.29) & $1.8 \,\, 10^{-28}$ (0.21)\\
20.07 & $4.0 \,\, 10^{-28}$ (0.74) & $5.5 \,\, 10^{-28}$ (0.64)\\
22.75 & $2.4 \,\, 10^{-28}$ (0.44) & $2.8 \,\, 10^{-28}$ (0.33)\\
23.25 & $4.8 \,\, 10^{-28}$ (0.88) & $6.3 \,\, 10^{-28}$ (0.74)\\
24.72 & $1.5 \,\, 10^{-28}$ (0.28) & $1.7 \,\, 10^{-28}$ (0.20)\\
$[25.36]$ & $1.3 \,\, 10^{-28}$ (0.23) & $1.2 \,\, 10^{-28}$ (0.14)\\
41.08 & $2.4 \,\, 10^{-28}$ (0.45) & $2.5 \,\, 10^{-28}$ (0.29)\\
$[41.68]$ & $1.8 \,\, 10^{-28}$ (0.34) & $1.9 \,\, 10^{-28}$ (0.22)\\
44.26 & $3.1 \,\, 10^{-28}$ (0.57) & $3.5 \,\, 10^{-28}$ (0.41)\\
100.13 & $3.5 \,\, 10^{-28}$ (0.65) & $3.4 \,\, 10^{-28}$ (0.40)\\
$[105.47]$ & $1.3 \,\, 10^{-28}$ (0.25) & $1.4 \,\, 10^{-28}$ (0.16)\\
\noalign{\smallskip} \hline 
\end{tabular}
\end{center} 
\end{table}

\begin{table}[ht!]
\begin{center}
\caption{As Table~\ref{anthracene} for neutral pyrene 
(C$_{16}$H$_{10}$).}
\label{pyrene_detailed}
\begin{tabular}{ccc}
\hline \hline \noalign{\smallskip} 
& \multicolumn{2}{c}{Integrated flux} \\
\noalign{\smallskip} \cline{2-3} \noalign{\smallskip}
Peak & Detailed & Approximated\\
($\mu$m) & (W$ $sr$^{-1}$) (\%) & (W$ $sr$^{-1}$) (\%) \\ 
\noalign{\smallskip} \hline \noalign{\smallskip}
3.25 & $9.6 \,\, 10^{-27}$ (17.55) & $1.4 \,\, 10^{-26}$ (19.15)\\
3.26 & $9.3 \,\, 10^{-27}$ (16.94) & $1.3 \,\, 10^{-26}$ (18.61)\\
3.27 & $2.6 \,\, 10^{-27}$ (4.70) & $3.6 \,\, 10^{-27}$ (5.05)\\
3.29 & $3.6 \,\, 10^{-28}$ (0.66) & $4.9 \,\, 10^{-28}$ (0.68)\\
6.26 & $8.4 \,\, 10^{-28}$ (1.53) & $1.1 \,\, 10^{-27}$ (1.60)\\
6.31 & $1.9 \,\, 10^{-27}$ (3.53) & $2.6 \,\, 10^{-27}$ (3.58)\\
6.78 & $6.1 \,\, 10^{-28}$ (1.12) & $7.9 \,\, 10^{-28}$ (1.10)\\
6.92 & $7.4 \,\, 10^{-29}$ (0.14) & $1.1 \,\, 10^{-28}$ (0.15)\\
7.01 & $1.8 \,\, 10^{-28}$ (0.32) & $2.5 \,\, 10^{-28}$ (0.34)\\
7.01 & $1.8 \,\, 10^{-27}$ (3.30) & $2.3 \,\, 10^{-27}$ (3.26)\\
7.61 & $9.8 \,\, 10^{-28}$ (1.78) & $1.3 \,\, 10^{-27}$ (1.77)\\
7.98 & $5.6 \,\, 10^{-28}$ (1.02) & $7.0 \,\, 10^{-28}$ (0.98)\\
8.42 & $1.5 \,\, 10^{-27}$ (2.75) & $2.0 \,\, 10^{-27}$ (2.74)\\
8.62 & $2.4 \,\, 10^{-28}$ (0.43) & $3.0 \,\, 10^{-28}$ (0.41)\\
9.16 & $7.1 \,\, 10^{-28}$ (1.29) & $8.8 \,\, 10^{-28}$ (1.23)\\
10.04 & $8.4 \,\, 10^{-29}$ (0.15) & $1.1 \,\, 10^{-28}$ (0.15)\\
10.25 & $3.7 \,\, 10^{-28}$ (0.67) & $4.9 \,\, 10^{-28}$ (0.68)\\
10.47 & $3.3 \,\, 10^{-29}$ (0.06) & $4.0 \,\, 10^{-29}$ (0.06)\\
11.79 & $1.5 \,\, 10^{-26}$ (27.79) & $1.9 \,\, 10^{-26}$ (25.96)\\
12.20 & $4.0 \,\, 10^{-28}$ (0.74) & $4.9 \,\, 10^{-28}$ (0.69)\\
13.40 & $1.2 \,\, 10^{-27}$ (2.14) & $1.4 \,\, 10^{-27}$ (1.93)\\
14.06 & $3.6 \,\, 10^{-27}$ (6.55) & $4.3 \,\, 10^{-27}$ (6.03)\\
14.43 & $4.1 \,\, 10^{-29}$ (0.07) & $4.2 \,\, 10^{-29}$ (0.06)\\
14.75 & $2.6 \,\, 10^{-29}$ (0.05) & $2.7 \,\, 10^{-29}$ (0.04)\\
$[17.26]$ & $4.5 \,\, 10^{-29}$ (0.08) & $4.5 \,\, 10^{-29}$ (0.06)\\
$[17.34]$ & $4.6 \,\, 10^{-29}$ (0.08) & $4.7 \,\, 10^{-29}$ (0.07)\\
18.20 & $2.6 \,\, 10^{-28}$ (0.48) & $3.1 \,\, 10^{-28}$ (0.43)\\
$[18.98]$ & $5.5 \,\, 10^{-29}$ (0.10) & $5.6 \,\, 10^{-29}$ (0.08)\\
$[19.82]$ & $6.1 \,\, 10^{-29}$ (0.11) & $6.2 \,\, 10^{-29}$ (0.09)\\
$[19.93]$ & $6.4 \,\, 10^{-29}$ (0.12) & $5.8 \,\, 10^{-29}$ (0.08)\\
20.00 & $2.6 \,\, 10^{-28}$ (0.48) & $2.9 \,\, 10^{-28}$ (0.40)\\
20.38 & $1.8 \,\, 10^{-28}$ (0.33) & $2.0 \,\, 10^{-28}$ (0.27)\\
$[21.96]$ & $7.2 \,\, 10^{-29}$ (0.13) & $7.5 \,\, 10^{-29}$ (0.10)\\
$[24.64]$ & $8.7 \,\, 10^{-29}$ (0.16) & $8.7 \,\, 10^{-29}$ (0.12)\\
$[25.27]$ & $9.1 \,\, 10^{-29}$ (0.17) & $8.8 \,\, 10^{-29}$ (0.12)\\
28.32 & $1.8 \,\, 10^{-28}$ (0.32) & $1.8 \,\, 10^{-28}$ (0.26)\\
$[38.63]$ & $1.4 \,\, 10^{-28}$ (0.26) & $1.4 \,\, 10^{-28}$ (0.20)\\
$[40.66]$ & $1.5 \,\, 10^{-28}$ (0.27) & $1.5 \,\, 10^{-28}$ (0.21)\\
47.80 & $3.3 \,\, 10^{-28}$ (0.61) & $3.6 \,\, 10^{-28}$ (0.50)\\
$[66.26]$ & $1.7 \,\, 10^{-28}$ (0.32) & $1.6 \,\, 10^{-28}$ (0.23)\\
101.60 & $2.6 \,\, 10^{-28}$ (0.46) & $2.5 \,\, 10^{-28}$ (0.35)\\
\noalign{\smallskip} \hline 
\end{tabular}
\end{center} 
\end{table}

\clearpage

\section{Detailed far\textendash IR spectra for all molecules in the sample}
\label{tableappendix}

In the following, we report in full detail all the calculated far\textendash IR 
spectra for our sample of molecules and RFs. In each table, we listed band 
positions and expected integrated fluxes. Only bands redwards of $\sim$15~$\mu$m 
are included. Bands whose calculated flux fraction in the ISRF is $\leq$0.05\% 
of the total IR emission were omitted. The positions of electric\textendash dipole 
forbidden bands are enclosed in square brackets. Bands corresponding to 
vibrations parallel or perpendicular to the symmetry plane of the molecule are 
marked by $\parallel$ and $\perp$, respectively. 

\begin{table}[h!]
\caption{Predicted far\textendash IR ($\lambda_\mathrm{peak}\gtrsim15~\mu\mathrm{m}$) emission 
spectrum of one molecule of neutral naphthalene (C$_{10}$H$_{8}$) in different 
exciting RFs. Under optically thin conditions total fluxes can be 
obtained simply multiplying these fluxes by the appropriate column 
density. 
}
\label{naphthalene}
\begin{center}
\begin{tabular}{cccc}
\hline \hline \noalign{\smallskip} 
& \multicolumn{3}{c}{Integrated flux} \\
\noalign{\smallskip} \cline{2-4} \noalign{\smallskip}
& & Red & IRAS \\
Peak & ISRF & Rectangle & 21282+5050 \\
($\mu$m) & (W$\,\,$sr$^{-1}$) (\%) & (W$\,\,$sr$^{-1}$) (\%)& (W$\,\,$sr$^{-1}$) (\%)\\ 
\noalign{\smallskip} \hline \noalign{\smallskip}
$\parallel$ 15.82 & $1.6\,\, 10^{-29}$ (0.51) & $2.5\,\, 10^{-28}$ (0.62) & $1.4\,\, 10^{-24}$ (0.48) \\
$[15.90]$ & $4.3\,\, 10^{-30}$ (0.14) & $7.3\,\, 10^{-29}$ (0.18) & \textemdash{} \\
$[19.50]$ & $5.4\,\, 10^{-30}$ (0.18) & $9.7\,\, 10^{-29}$ (0.24) & \textemdash{} \\
$[19.54]$ & $5.4\,\, 10^{-30}$ (0.18) & $9.4\,\, 10^{-29}$ (0.23) & \textemdash{} \\
$\perp$ 20.88 & $4.7\,\, 10^{-29}$ (1.53) & $7.5\,\, 10^{-28}$ (1.83) & $4.1\,\, 10^{-24}$ (1.45) \\
$[21.16]$ & $6.0\,\, 10^{-30}$ (0.20) & $1.0\,\, 10^{-28}$ (0.25) & \textemdash{} \\
$[25.75]$ & $7.3\,\, 10^{-30}$ (0.24) & $1.3\,\, 10^{-28}$ (0.31) & \textemdash\\
$\parallel$ 27.74 & $9.1\,\, 10^{-30}$ (0.30) & $1.6\,\, 10^{-28}$ (0.40) & $6.6\,\, 10^{-25}$ (0.23) \\
$[53.95]$ & $1.1\,\, 10^{-29}$ (0.36) & $1.8\,\, 10^{-28}$ (0.43) & \textemdash\\
$\perp$ 58.56& $1.2\,\, 10^{-29}$ (0.39) & $2.0\,\, 10^{-28}$ (0.49) & $4.9\,\, 10^{-25}$ (0.17) \\
\noalign{\smallskip} \hline 
\end{tabular}
\end{center}
\end{table}

\begin{table}
\caption{As Fig.~\ref{naphthalene} for the naphthalene cation (C$_{10}$H$_{8}^+$).}
\label{naphthalene+}
\begin{center}
\begin{tabular}{cccc}
\hline \hline \noalign{\smallskip} 
& \multicolumn{3}{c}{Integrated flux} \\
\noalign{\smallskip} \cline{2-4} \noalign{\smallskip}
 & & Red & IRAS\\
Peak & ISRF & Rectangle & 21282+5050\\
($\mu$m) & (W$\,\,$sr$^{-1}$) (\%) & (W$\,\,$sr$^{-1}$) (\%)& (W$\,\,$sr$^{-1}$) (\%)\\ 
\noalign{\smallskip} \hline \noalign{\smallskip}
$\parallel$ 16.37 & $2.4\,\, 10^{-29}$ (0.74) & $1.1\,\, 10^{-27}$ (0.98) & $1.2\,\, 10^{-24}$ (0.56) \\
$[17.83]$ & $8.1\,\, 10^{-30}$ (0.25) & $4.3\,\, 10^{-28}$ (0.38) & $1.3\,\, 10^{-26}$ (0.01) \\
$[19.96]$ & $9.6\,\, 10^{-30}$ (0.29) & $4.4\,\, 10^{-28}$ (0.40) & $1.1\,\, 10^{-26}$ (0.01) \\
$[20.72]$ & $1.0\,\, 10^{-29}$ (0.31) & $4.7\,\, 10^{-28}$ (0.42) & $1.5\,\, 10^{-26}$ (0.01) \\
$[22.77]$ & $1.1\,\, 10^{-29}$ (0.33) & $5.3\,\, 10^{-28}$ (0.48) & $1.2\,\, 10^{-26}$ (0.01) \\
$\perp$ 23.64 & $3.5\,\, 10^{-29}$ (1.06) & $1.5\,\, 10^{-27}$ (1.39) & $1.6\,\, 10^{-24}$ (0.76) \\
$[27.07]$ & $1.3\,\, 10^{-29}$ (0.38) & $6.4\,\, 10^{-28}$ (0.58) & \textemdash{} \\
$\parallel$ 28.31 & $1.5\,\, 10^{-29}$ (0.46) & $7.4\,\, 10^{-28}$ (0.67) & $5.4\,\, 10^{-25}$ (0.25) \\
$[55.16]$ & $1.9\,\, 10^{-29}$ (0.58) & $8.1\,\, 10^{-28}$ (0.73) & \textemdash{} \\
$\perp$ 63.98 & $2.1\,\, 10^{-29}$ (0.63) & $1.0\,\, 10^{-27}$ (0.95) & $5.6\,\, 10^{-25}$ (0.26) \\
\noalign{\smallskip} \hline 
\end{tabular}
\end{center}
\end{table}

\begin{table}
\caption{As Fig.~\ref{naphthalene} for fluorene (C$_{13}$H$_{10}$).}
\label{fluorene}
\begin{center}
\begin{tabular}{cccc}
\hline \hline \noalign{\smallskip} 
& \multicolumn{3}{c}{Integrated flux} \\
\noalign{\smallskip} \cline{2-4} \noalign{\smallskip}
& & Red & IRAS\\
Peak & ISRF & Rectangle & 21282+5050\\
($\mu$m) & (W$\,\,$sr$^{-1}$) (\%) & (W$\,\,$sr$^{-1}$) (\%)& (W$\,\,$sr$^{-1}$) (\%)\\ 
\noalign{\smallskip} \hline \noalign{\smallskip}
$\perp$ 15.78 & $3.7\,\, 10^{-30}$ (0.10) & $6.8\,\, 10^{-29}$ (0.11) & $2.4\,\, 10^{-25}$ (0.07) \\
$\parallel$ 16.04 & $3.7\,\, 10^{-29}$ (0.97) & $6.9\,\, 10^{-28}$ (1.17) & $3.1\,\, 10^{-24}$ (0.89) \\
$[17.69]$ & $2.5\,\, 10^{-30}$ (0.07) & $5.3\,\, 10^{-29}$ (0.09) & \textemdash{} \\
$\parallel$ 18.49 & $3.4\,\, 10^{-30}$ (0.09) & $7.3\,\, 10^{-29}$ (0.12) & $1.7\,\, 10^{-25}$ (0.05) \\
$\perp$ 20.56 & $6.6\,\, 10^{-30}$ (0.17) & $1.3\,\, 10^{-28}$ (0.22) & $4.7\,\, 10^{-25}$ (0.14) \\
$\parallel$ 20.65 & $5.3\,\, 10^{-30}$ (0.14) & $1.1\,\, 10^{-28}$ (0.18) & $3.7\,\, 10^{-25}$ (0.11) \\
$[23.72]$ & $4.4\,\, 10^{-30}$ (0.11) & $8.4\,\, 10^{-29}$ (0.14) & \textemdash{} \\
$\parallel$ 23.83 & $6.2\,\, 10^{-30}$ (0.16) & $1.3\,\, 10^{-28}$ (0.21) & $3.6\,\, 10^{-25}$ (0.11) \\
$\perp$ 23.92 & $2.6\,\, 10^{-29}$ (0.68) & $5.1\,\, 10^{-28}$ (0.85) & $2.1\,\, 10^{-24}$ (0.60) \\
$[39.45]$ & $7.3\,\, 10^{-30}$ (0.19) & $1.4\,\, 10^{-28}$ (0.23) & \textemdash{} \\
$\perp$ 40.26 & $1.7\,\, 10^{-29}$ (0.43) & $3.3\,\, 10^{-28}$ (0.55) & $1.2\,\, 10^{-24}$ (0.36) \\
$\parallel$ 45.48 & $8.2\,\, 10^{-30}$ (0.21) & $1.7\,\, 10^{-28}$ (0.29) & $9.4\,\, 10^{-26}$ (0.03) \\
$[60.73]$ & $9.3\,\, 10^{-30}$ (0.24) & $1.5\,\, 10^{-28}$ (0.25) & \textemdash{} \\
$\perp$ 135.37 & $1.2\,\, 10^{-29}$ (0.31) & $2.5\,\, 10^{-28}$ (0.43) & $5.9\,\, 10^{-26}$ (0.02) \\
\noalign{\smallskip} \hline 
\end{tabular}
\end{center}
\end{table}

\begin{table}
\caption{As Fig.~\ref{naphthalene} for the fluorene cation (C$_{13}$H$_{10}^+$).}
\label{fluorene+}
\begin{center}
\begin{tabular}{cccc}
\hline \hline \noalign{\smallskip} 
& \multicolumn{3}{c}{Integrated flux} \\
\noalign{\smallskip} \cline{2-4} \noalign{\smallskip}
 & & Red & IRAS\\
Peak & ISRF & Rectangle & 21282+5050\\
($\mu$m) & (W$\,\,$sr$^{-1}$) (\%) & (W$\,\,$sr$^{-1}$) (\%)& (W$\,\,$sr$^{-1}$) (\%)\\ 
\noalign{\smallskip} \hline \noalign{\smallskip}
$\perp$ 15.15 & $5.0\,\, 10^{-29}$ (0.90) & $2.4\,\, 10^{-27}$ (1.06) & $2.0\,\, 10^{-24}$ (0.68) \\
$\parallel$ 16.14 & $7.3\,\, 10^{-30}$ (0.13) & $3.7\,\, 10^{-28}$ (0.16) & $1.8\,\, 10^{-25}$ (0.06) \\
$\parallel$ 16.35 & $1.1\,\, 10^{-28}$ (1.91) & $5.1\,\, 10^{-27}$ (2.23) & $3.9\,\, 10^{-24}$ (1.37) \\
$\parallel$ 18.92 & $1.4\,\, 10^{-29}$ (0.26) & $7.2\,\, 10^{-28}$ (0.32) & $4.3\,\, 10^{-25}$ (0.15) \\
$[20.11]$ & $8.1\,\, 10^{-30}$ (0.15) & $3.7\,\, 10^{-28}$ (0.16) & \textemdash{} \\
$[20.38]$ & $7.8\,\, 10^{-30}$ (0.14) & $4\,\, 10^{-28}$ (0.17) & \textemdash{} \\
$[22.64]$ & $9.6\,\, 10^{-30}$ (0.17) & $4.5\,\, 10^{-28}$ (0.20) & \textemdash{} \\
$\perp$ 22.84 & $3.1\,\, 10^{-29}$ (0.56) & $1.6\,\, 10^{-27}$ (0.71) & $9.9\,\, 10^{-25}$ (0.35) \\
$\parallel$ 24.23 & $1.2\,\, 10^{-29}$ (0.22) & $6.1\,\, 10^{-28}$ (0.27) & $2.1\,\, 10^{-25}$ (0.07) \\
$\perp$ 25.54 & $1.6\,\, 10^{-29}$ (0.30) & $9.3\,\, 10^{-28}$ (0.41) & $4.3\,\, 10^{-25}$ (0.15) \\
$[40.59]$ & $1.8\,\, 10^{-29}$ (0.33) & $7.2\,\, 10^{-28}$ (0.32) & \textemdash{} \\
$\perp$ 43.62 & $2.7\,\, 10^{-29}$ (0.49) & $1.5\,\, 10^{-27}$ (0.64) & $6.7\,\, 10^{-25}$ (0.24) \\
$\parallel$ 47.31 & $2.2\,\, 10^{-29}$ (0.40) & $1.2\,\, 10^{-27}$ (0.52) & $2.5\,\, 10^{-25}$ (0.09) \\
$[78.65]$ & $2.6\,\, 10^{-29}$ (0.48) & $5.4\,\, 10^{-28}$ (0.24) & \textemdash{} \\
$\perp$ 122.23 & $3.1\,\, 10^{-29}$ (0.56) & $1.7\,\, 10^{-27}$ (0.74) & $1.3\,\, 10^{-25}$ (0.04) \\
\noalign{\smallskip} \hline 
\end{tabular}
\end{center}
\end{table}

\begin{table}
\caption{As Fig.~\ref{naphthalene} for anthracene (C$_{14}$H$_{10}$).}
\label{anthracene}
\begin{center}
\begin{tabular}{cccc}
\hline \hline \noalign{\smallskip} 
& \multicolumn{3}{c}{Integrated flux} \\
\noalign{\smallskip} \cline{2-4} \noalign{\smallskip}
 & & Red & IRAS\\
Peak & ISRF & Rectangle & 21282+5050\\
($\mu$m) & (W$\,\,$sr$^{-1}$) (\%) & (W$\,\,$sr$^{-1}$) (\%)& (W$\,\,$sr$^{-1}$) (\%)\\ 
\noalign{\smallskip} \hline \noalign{\smallskip}
$\parallel$ 15.33 & $9.6\,\, 10^{-30}$ (0.23) & $3.9\,\, 10^{-28}$ (0.31) & $7.0\,\, 10^{-25}$ (0.20) \\
$[15.71]$ & $2.2\,\, 10^{-30}$ (0.05) & $9.9\,\, 10^{-29}$ (0.08) & \textemdash{} \\
$\parallel$ 16.34 & $4.4\,\, 10^{-29}$ (1.04) & $1.7\,\, 10^{-27}$ (1.36) & $3.2\,\, 10^{-24}$ (0.93) \\
$[17.11]$ & $2.6\,\, 10^{-30}$ (0.06) & $1.1\,\, 10^{-28}$ (0.09) & \textemdash{} \\
$[18.71]$ & $3.3\,\, 10^{-30}$ (0.08) & $1.4\,\, 10^{-28}$ (0.11) & \textemdash{} \\
$[20.04]$ & $3.6\,\, 10^{-30}$ (0.09) & $1.6\,\, 10^{-28}$ (0.13) & \textemdash{} \\
$[21.02]$ & $3.9\,\, 10^{-30}$ (0.09) & $1.8\,\, 10^{-28}$ (0.14) & \textemdash{} \\
$\perp$ 21.26 & $7.7\,\, 10^{-29}$ (1.85) & $3.1\,\, 10^{-27}$ (2.49) & $5.6\,\, 10^{-24}$ (1.61) \\
$[25.45]$ & $5.5\,\, 10^{-30}$ (0.13) & $2.2\,\, 10^{-28}$ (0.18) & \textemdash{} \\
$[25.62]$ & $5.4\,\, 10^{-30}$ (0.13) & $2.3\,\, 10^{-28}$ (0.18) & \textemdash{} \\
$\perp$ 26.35 & $5.9\,\, 10^{-30}$ (0.14) & $2.6\,\, 10^{-28}$ (0.21) & $6.5\,\, 10^{-26}$ (0.02) \\
$[37.41]$ & $8.2\,\, 10^{-30}$ (0.20) & $3.2\,\, 10^{-28}$ (0.26) & \textemdash{} \\
$[43.01]$ & $8.8\,\, 10^{-30}$ (0.21) & $3.4\,\, 10^{-28}$ (0.27) & \textemdash{} \\
$\parallel$ 43.68 & $1.2\,\, 10^{-29}$ (0.29) & $5.3\,\, 10^{-28}$ (0.43) & $6.1\,\, 10^{-25}$ (0.18) \\
$[82.09]$ & $1.3\,\, 10^{-29}$ (0.30) & $3.0\,\, 10^{-28}$ (0.24) & \textemdash{} \\
$\perp$ 110.23 & $1.4\,\, 10^{-29}$ (0.33) & $6.4\,\, 10^{-28}$ (0.51) & $1.8\,\, 10^{-25}$ (0.05) \\
\noalign{\smallskip} \hline
\end{tabular}
\end{center}
\end{table}

\begin{table}
\caption{As Fig.~\ref{naphthalene} for the anthracene cation (C$_{14}$H$_{10}^+$).}
\label{anthracene+}
\begin{center}
\begin{tabular}{cccc}
\hline \hline \noalign{\smallskip} 
& \multicolumn{3}{c}{Integrated flux} \\
\noalign{\smallskip} \cline{2-4} \noalign{\smallskip}
 & & Red & IRAS\\
Peak & ISRF & Rectangle & 21282+5050\\
($\mu$m) & (W$\,\,$sr$^{-1}$) (\%) & (W$\,\,$sr$^{-1}$) (\%)& (W$\,\,$sr$^{-1}$) (\%)\\ 
\noalign{\smallskip} \hline \noalign{\smallskip}
$\parallel$ 15.08 & $9.2\,\, 10^{-30}$ (0.13) & $4.1\,\, 10^{-28}$ (0.15) & $2.5\,\, 10^{-25}$ (0.07) \\
$[15.94]$ & $5.3\,\, 10^{-30}$ (0.08) & $2.5\,\, 10^{-28}$ (0.09) & \textemdash{} \\
$\parallel$ 16.56 & $6.9\,\, 10^{-30}$ (0.10) & $3.6\,\, 10^{-28}$ (0.13) & $1.9\,\, 10^{-25}$ (0.05) \\
$[17.97]$ & $7.0\,\, 10^{-30}$ (0.10) & $3.5\,\, 10^{-28}$ (0.13) & \textemdash{} \\
$[19.41]$ & $8.3\,\, 10^{-30}$ (0.12) & $4.2\,\, 10^{-28}$ (0.16) & \textemdash{} \\
$[20.84]$ & $9.8\,\, 10^{-30}$ (0.14) & $5.0\,\, 10^{-28}$ (0.19) & \textemdash{} \\
$[22.02]$ & $1.1\,\, 10^{-29}$ (0.16) & $5.4\,\, 10^{-28}$ (0.20) & \textemdash{} \\
$\perp$ 22.81 & $7.5\,\, 10^{-29}$ (1.10) & $3.6\,\, 10^{-27}$ (1.34) & $2.7\,\, 10^{-24}$ (0.69) \\
$[26.02]$ & $1.4\,\, 10^{-29}$ (0.21) & $6.7\,\, 10^{-28}$ (0.25) & \textemdash{} \\
$\perp$ 26.51 & $1.5\,\, 10^{-29}$ (0.22) & $8.0\,\, 10^{-28}$ (0.30) & $2.0\,\, 10^{-25}$ (0.05) \\
$[26.90]$ & $1.5\,\, 10^{-29}$ (0.22) & $7.4\,\, 10^{-28}$ (0.27) & \textemdash{} \\
$[37.28]$ & $2.2\,\, 10^{-29}$ (0.33) & $9.5\,\, 10^{-28}$ (0.35) & \textemdash{} \\
$\parallel$ 44.40 & $2.9\,\, 10^{-29}$ (0.43) & $1.5\,\, 10^{-27}$ (0.55) & $6.5\,\, 10^{-25}$ (0.17) \\
$[45.89]$ & $2.6\,\, 10^{-29}$ (0.37) & $1.0\,\, 10^{-27}$ (0.38) & \textemdash{} \\
$[83.78]$ & $3.4\,\, 10^{-29}$ (0.49) & $9.5\,\, 10^{-28}$ (0.35) & \textemdash{} \\
$\perp$ 115.42 & $3.8\,\, 10^{-29}$ (0.56) & $2.0\,\, 10^{-27}$ (0.74) & $2.7\,\, 10^{-25}$ (0.07) \\
\noalign{\smallskip} \hline 
\end{tabular}
\end{center}
\end{table}

\begin{table}
\caption{As Fig.~\ref{naphthalene} for phenanthrene (C$_{14}$H$_{10}$).}
\label{phenanthrene}
\begin{center}
\begin{tabular}{cccc}
\hline \hline \noalign{\smallskip} 
& \multicolumn{3}{c}{Integrated flux} \\
\noalign{\smallskip} \cline{2-4} \noalign{\smallskip}
 & & Red & IRAS\\
Peak & ISRF & Rectangle & 21282+5050\\
($\mu$m) & (W$\,\,$sr$^{-1}$) (\%) & (W$\,\,$sr$^{-1}$) (\%)& (W$\,\,$sr$^{-1}$) (\%)\\ 
\noalign{\smallskip} \hline \noalign{\smallskip}
$\parallel$ 15.94 & $2.9\,\, 10^{-29}$ (0.70) & $7.3\,\, 10^{-28}$ (0.85) & $2.2\,\, 10^{-24}$ (0.61) \\
$[16.86]$ & $2.2\,\, 10^{-30}$ (0.05) & $6.2\,\, 10^{-29}$ (0.07) & \textemdash{} \\
$\parallel$ 18.19 & $4.5\,\, 10^{-30}$ (0.11) & $1.2\,\, 10^{-28}$ (0.14) & $2.7\,\, 10^{-25}$ (0.07) \\
$[18.73]$ & $2.8\,\, 10^{-30}$ (0.07) & $7.5\,\, 10^{-29}$ (0.09) & \textemdash{} \\
$\parallel$ 19.99 & $6.8\,\, 10^{-30}$ (0.16) & $1.8\,\, 10^{-28}$ (0.21) & $4.7\,\, 10^{-25}$ (0.13) \\
$\perp$ 20.07 & $2.0\,\, 10^{-29}$ (0.48) & $5.5\,\, 10^{-28}$ (0.64) & $1.6\,\, 10^{-24}$ (0.44) \\
$\parallel$ 22.75 & $1.0\,\, 10^{-29}$ (0.25) & $2.8\,\, 10^{-28}$ (0.33) & $7.4\,\, 10^{-25}$ (0.21) \\
$\perp$ 23.25 & $2.4\,\, 10^{-29}$ (0.57) & $6.3\,\, 10^{-28}$ (0.74) & $1.8\,\, 10^{-24}$ (0.50) \\
$\parallel$ 24.72 & $6.1\,\, 10^{-30}$ (0.15) & $1.7\,\, 10^{-28}$ (0.20) & $3.0\,\, 10^{-25}$ (0.08) \\
$[25.36]$ & $4.9\,\, 10^{-30}$ (0.12) & $1.2\,\, 10^{-28}$ (0.14) & \textemdash{} \\
$\parallel$ 41.08 & $8.7\,\, 10^{-30}$ (0.21) & $2.5\,\, 10^{-28}$ (0.29) & $2.4\,\, 10^{-25}$ (0.07) \\
$[41.68]$ & $7.8\,\, 10^{-30}$ (0.19) & $1.9\,\, 10^{-28}$ (0.22) & \textemdash{} \\
$\perp$ 44.26 & $1.2\,\, 10^{-29}$ (0.30) & $3.5\,\, 10^{-28}$ (0.41) & $7.5\,\, 10^{-25}$ (0.21) \\
$\perp$ 100.13 & $1.2\,\, 10^{-29}$ (0.30) & $3.4\,\, 10^{-28}$ (0.40) & $1.3\,\, 10^{-25}$ (0.04) \\
$[105.47]$ & $1.1\,\, 10^{-29}$ (0.27) & $1.4\,\, 10^{-28}$ (0.16) & \textemdash{} \\
\noalign{\smallskip} \hline 
\end{tabular}
\end{center}
\end{table}

\begin{table}
\caption{As Fig.~\ref{naphthalene} for the phenanthrene cation 
(C$_{14}$H$_{10}^+$).}
\label{phenanthrene+}
\begin{center}
\begin{tabular}{cccc}
\hline \hline \noalign{\smallskip} 
& \multicolumn{3}{c}{Integrated flux} \\
\noalign{\smallskip} \cline{2-4} \noalign{\smallskip}
 & & Red & IRAS\\
Peak & ISRF & Rectangle & 21282+5050\\
($\mu$m) & (W$\,\,$sr$^{-1}$) (\%) & (W$\,\,$sr$^{-1}$) (\%)& (W$\,\,$sr$^{-1}$) (\%)\\ 
\noalign{\smallskip} \hline \noalign{\smallskip}
$\parallel$ 16.49 & $7.1\,\, 10^{-29}$ (1.12) & $3.6\,\, 10^{-28}$ (1.35) & $2.4\,\, 10^{-24}$ (0.70) \\
$\parallel$ 18.23 & $1.1\,\, 10^{-29}$ (0.18) & $5.9\,\, 10^{-29}$ (0.22) & $2.4\,\, 10^{-25}$ (0.07) \\
$[18.34]$ & $7.0\,\, 10^{-30}$ (0.11) & $3.8\,\, 10^{-29}$ (0.14) & \textemdash{} \\
$\parallel$ 20.26 & $1.3\,\, 10^{-29}$ (0.20) & $7.8\,\, 10^{-29}$ (0.29) & $3.3\,\, 10^{-25}$ (0.10) \\
$[20.82]$ & $9.0\,\, 10^{-30}$ (0.14) & $4.6\,\, 10^{-29}$ (0.17) & \textemdash{} \\
$\perp$ 21.17 & $1.1\,\, 10^{-29}$ (0.17) & $5.7\,\, 10^{-29}$ (0.21) & $1.4\,\, 10^{-25}$ (0.04) \\
$\parallel$ 22.98 & $1.2\,\, 10^{-29}$ (0.19) & $6.6\,\, 10^{-29}$ (0.25) & $1.8\,\, 10^{-25}$ (0.05) \\
$\perp$ 24.51 & $3.0\,\, 10^{-29}$ (0.48) & $1.6\,\, 10^{-28}$ (0.60) & $7.9\,\, 10^{-25}$ (0.23) \\
$\parallel$ 25.38 & $1.4\,\, 10^{-29}$ (0.22) & $7.2\,\, 10^{-29}$ (0.27) & $1.4\,\, 10^{-25}$ (0.04) \\
$[26.47]$ & $1.4\,\, 10^{-29}$ (0.22) & $7.8\,\, 10^{-29}$ (0.29) & \textemdash{} \\
$\parallel$ 41.63 & $2.3\,\, 10^{-29}$ (0.37) & $1.3\,\, 10^{-28}$ (0.48) & $3.7\,\, 10^{-26}$ (0.01) \\
$[42.88]$ & $2.4\,\, 10^{-29}$ (0.38) & $1.2\,\, 10^{-28}$ (0.45) & \textemdash{} \\
$\perp$ 47.27 & $3.1\,\, 10^{-29}$ (0.48) & $1.7\,\, 10^{-28}$ (0.63) & $6.3\,\, 10^{-25}$ (0.18) \\
$\perp$ 98.06 & $3.6\,\, 10^{-29}$ (0.57) & $2.0\,\, 10^{-28}$ (0.75) & $2.3\,\, 10^{-25}$ (0.07) \\
$[121.17]$ & $3.6\,\, 10^{-29}$ (0.57) & $1.6\,\, 10^{-28}$ (0.59) & \textemdash{} \\
\noalign{\smallskip} \hline 
\end{tabular}
\end{center}
\end{table}

\begin{table}
\caption{As Fig.~\ref{naphthalene} for pyrene (C$_{16}$H$_{10}$).}
\label{pyrene}
\begin{center}
\begin{tabular}{cccc}
\hline \hline \noalign{\smallskip} 
& \multicolumn{3}{c}{Integrated flux} \\
\noalign{\smallskip} \cline{2-4} \noalign{\smallskip}
 & & Red & IRAS\\
Peak & ISRF & Rectangle & 21282+5050\\
($\mu$m) & (W$\,\,$sr$^{-1}$) (\%) & (W$\,\,$sr$^{-1}$) (\%)& (W$\,\,$sr$^{-1}$) (\%)\\ 
\noalign{\smallskip} \hline \noalign{\smallskip}
$[17.26]$ & $3.2\,\, 10^{-30}$ (0.06) & $1.6\,\, 10^{-28}$ (0.09) & \textemdash{} \\
$[17.34]$ & $3.3\,\, 10^{-30}$ (0.06) & $1.6\,\, 10^{-28}$ (0.09) & \textemdash{} \\
$\parallel$ 18.20 & $2.0\,\, 10^{-29}$ (0.40) & $9.4\,\, 10^{-28}$ (0.53) & $1.2\,\, 10^{-24}$ (0.32) \\
$[18.98]$ & $3.5\,\, 10^{-30}$ (0.07) & $1.8\,\, 10^{-28}$ (0.10) & \textemdash{} \\
$[19.82]$ & $4.3\,\, 10^{-30}$ (0.08) & $1.9\,\, 10^{-28}$ (0.11) & \textemdash{} \\
$[19.93]$ & $4.0\,\, 10^{-30}$ (0.08) & $1.9\,\, 10^{-28}$ (0.11) & \textemdash{} \\
$\parallel$ 20.00 & $1.9\,\, 10^{-29}$ (0.38) & $9.0\,\, 10^{-28}$ (0.51) & $1.1\,\, 10^{-24}$ (0.30) \\
$\perp$ 20.38 & $1.3\,\, 10^{-29}$ (0.26) & $6.3\,\, 10^{-28}$ (0.36) & $7.8\,\, 10^{-25}$ (0.21) \\
$[21.96]$ & $4.7\,\, 10^{-30}$ (0.09) & $2.3\,\, 10^{-28}$ (0.13) & \textemdash{} \\
$[24.64]$ & $6.2\,\, 10^{-30}$ (0.12) & $2.7\,\, 10^{-28}$ (0.16) & \textemdash{} \\
$[25.27]$ & $6.2\,\, 10^{-30}$ (0.12) & $2.9\,\, 10^{-28}$ (0.17) & \textemdash{} \\
$\parallel$ 28.32 & $1.2\,\, 10^{-29}$ (0.24) & $6.0\,\, 10^{-28}$ (0.34) & $6.1\,\, 10^{-25}$ (0.16) \\
$[38.63]$ & $1.0\,\, 10^{-29}$ (0.19) & $4.1\,\, 10^{-28}$ (0.23) & \textemdash{} \\
$[40.66]$ & $1.0\,\, 10^{-29}$ (0.20) & $4.2\,\, 10^{-28}$ (0.24) & \textemdash{} \\
$\perp$ 47.80 & $2.4\,\, 10^{-29}$ (0.46) & $1.2\,\, 10^{-27}$ (0.66) & $1.3\,\, 10^{-24}$ (0.33) \\
$[66.26]$ & $1.4\,\, 10^{-29}$ (0.27) & $3.9\,\, 10^{-28}$ (0.22) & \textemdash{} \\
$\perp$ 101.60 & $1.7\,\, 10^{-29}$ (0.32) & $8.6\,\, 10^{-28}$ (0.49) & $9.0\,\, 10^{-26}$ (0.02) \\
\noalign{\smallskip} \hline
\end{tabular}
\end{center}
\end{table}

\begin{table}
\caption{As Fig.~\ref{naphthalene} for the pyrene cation (C$_{16}$H$_{10}^+$).}
\label{pyrene+}
\begin{center}
\begin{tabular}{cccc}
\hline \hline \noalign{\smallskip} 
& \multicolumn{3}{c}{Integrated flux} \\
\noalign{\smallskip} \cline{2-4} \noalign{\smallskip}
 & & Red & IRAS\\
Peak & ISRF & Rectangle & 21282+5050\\
($\mu$m) & (W$\,\,$sr$^{-1}$) (\%) & (W$\,\,$sr$^{-1}$) (\%)& (W$\,\,$sr$^{-1}$) (\%)\\ 
\noalign{\smallskip} \hline \noalign{\smallskip}
$[15.02]$ & $3.2\,\, 10^{-30}$ (0.05) & $2.1\,\, 10^{-28}$ (0.07) & $1.7\,\, 10^{-27}$ (0.00) \\
$[17.57]$ & $5.3\,\, 10^{-30}$ (0.08) & $3.4\,\, 10^{-28}$ (0.11) & $4.6\,\, 10^{-27}$ (0.00) \\
$[18.30]$ & $6.4\,\, 10^{-30}$ (0.10) & $3.7\,\, 10^{-28}$ (0.12) & $3.1\,\, 10^{-27}$ (0.00) \\
$\parallel$ 18.61 & $1.7\,\, 10^{-29}$ (0.25) & $1\,\, 10^{-27}$ (0.32) & $5.9\,\, 10^{-25}$ (0.15) \\
\hline 
\end{tabular}
\end{center}
\end{table}

\begin{table}
\caption{As Fig.~\ref{naphthalene} for fluoranthene (C$_{16}$H$_{10}$).}
\label{fluoranthene}
\begin{center}
\begin{tabular}{cccc}
\hline \hline \noalign{\smallskip} 
& \multicolumn{3}{c}{Integrated flux} \\
\noalign{\smallskip} \cline{2-4} \noalign{\smallskip}
 & & Red & IRAS\\
Peak & ISRF & Rectangle & 21282+5050\\
($\mu$m) & (W$\,\,$sr$^{-1}$) (\%) & (W$\,\,$sr$^{-1}$) (\%)& (W$\,\,$sr$^{-1}$) (\%)\\ 
\noalign{\smallskip} \hline \noalign{\smallskip}
$\parallel$ 14.99 & $1.5\,\, 10^{-29}$ (0.30) & $5.5\,\, 10^{-28}$ (0.38) & $7.5\,\, 10^{-25}$ (0.27) \\
$\perp$ 15.51 & $3.6\,\, 10^{-29}$ (0.71) & $1.2\,\, 10^{-27}$ (0.81) & $1.8\,\, 10^{-24}$ (0.66) \\
$\parallel$ 16.04 & $3.2\,\, 10^{-30}$ (0.06) & $1.5\,\, 10^{-28}$ (0.11) & $1.2\,\, 10^{-25}$ (0.04) \\
$[16.99]$ & $3.0\,\, 10^{-30}$ (0.06) & $1.3\,\, 10^{-28}$ (0.09) & \textemdash{} \\
\hline 
\end{tabular}
\end{center}
\end{table}

\begin{table}
\caption{As Fig.~\ref{naphthalene} for the fluoranthene cation 
(C$_{16}$H$_{10}^+$).}
\label{fluoranthene+}
\begin{center}
\begin{tabular}{cccc}
\hline \hline \noalign{\smallskip} 
& \multicolumn{3}{c}{Integrated flux} \\
\noalign{\smallskip} \cline{2-4} \noalign{\smallskip}
 & & Red & IRAS\\
Peak & ISRF & Rectangle & 21282+5050\\
($\mu$m) & (W$\,\,$sr$^{-1}$) (\%) & (W$\,\,$sr$^{-1}$) (\%)& (W$\,\,$sr$^{-1}$) (\%)\\ 
\noalign{\smallskip} \hline \noalign{\smallskip}
$[15.58]$ & $5.5\,\, 10^{-30}$ (0.07) & $2.5\,\, 10^{-28}$ (0.07) & $5.9\,\, 10^{-28}$ (0.00) \\
$\parallel$ 16.05 & $7.5\,\, 10^{-30}$ (0.09) & $4.1\,\, 10^{-28}$ (0.12) & $6.7\,\, 10^{-26}$ (0.02) \\
$\perp$ 16.18 & $1.9\,\, 10^{-28}$ (2.31) & $9.8\,\, 10^{-27}$ (2.82) & $5.1\,\, 10^{-24}$ (1.33) \\
$[17.63]$ & $8.1\,\, 10^{-30}$ (0.10) & $3.3\,\, 10^{-28}$ (0.09) & $1.2\,\, 10^{-28}$ (0.00) \\
\hline 
\end{tabular}
\end{center}
\end{table}

\begin{table}
\caption{As Fig.~\ref{naphthalene} for tetracene (C$_{18}$H$_{12}$).}
\label{tetracene}
\begin{center}
\begin{tabular}{cccc}
\hline \hline \noalign{\smallskip} 
& \multicolumn{3}{c}{Integrated flux} \\
\noalign{\smallskip} \cline{2-4} \noalign{\smallskip}
 & & Red & IRAS\\
Peak & ISRF & Rectangle & 21282+5050\\
($\mu$m) & (W$\,\,$sr$^{-1}$) (\%) & (W$\,\,$sr$^{-1}$) (\%)& (W$\,\,$sr$^{-1}$) (\%)\\ 
\noalign{\smallskip} \hline \noalign{\smallskip}
$\parallel$ 15.63 & $1.6\,\, 10^{-29}$ (0.23) & $7.1\,\, 10^{-29}$ (0.29) & $4.4\,\, 10^{-25}$ (0.18) \\
$\parallel$ 16.31 & $5.3\,\, 10^{-29}$ (0.78) & $2.3\,\, 10^{-28}$ (0.93) & $1.5\,\, 10^{-24}$ (0.62) \\
$\parallel$ 17.93 & $8.5\,\, 10^{-29}$ (1.25) & $3.8\,\, 10^{-28}$ (1.54) & $2.5\,\, 10^{-24}$ (0.99) \\
$[19.98]$ & $2.6\,\, 10^{-30}$ (0.04) & $1.2\,\, 10^{-29}$ (0.05) & \textemdash{} \\
\hline
\end{tabular}
\end{center}
\end{table}

\begin{table}
\caption{As Fig.~\ref{naphthalene} for the tetracene cation (C$_{18}$H$_{12}^+$).}
\label{tetracene+}
\begin{center}
\begin{tabular}{cccc}
\hline \hline \noalign{\smallskip} 
& \multicolumn{3}{c}{Integrated flux} \\
\noalign{\smallskip} \cline{2-4} \noalign{\smallskip}
 & & Red & IRAS\\
Peak & ISRF & Rectangle & 21282+5050\\
($\mu$m) & (W$\,\,$sr$^{-1}$) (\%) & (W$\,\,$sr$^{-1}$) (\%)& (W$\,\,$sr$^{-1}$) (\%)\\ 
\noalign{\smallskip} \hline \noalign{\smallskip}
$\parallel$ 16.23 & $9.9\,\, 10^{-30}$ (0.08) & $5.1\,\, 10^{-28}$ (0.09) & $2.1\,\, 10^{-25}$ (0.04) \\
$\parallel$ 18.16 & $8.7\,\, 10^{-30}$ (0.07) & $4.9\,\, 10^{-28}$ (0.09) & $1.4\,\, 10^{-25}$ (0.03) \\
$[19.72]$ & $6.7\,\, 10^{-30}$ (0.05) & $3.2\,\, 10^{-28}$ (0.06) & \textemdash{} \\
$[20.37]$ & $7.8\,\, 10^{-30}$ (0.06) & $3.6\,\, 10^{-28}$ (0.07) & \textemdash{} \\
\hline 
\end{tabular}
\end{center}
\end{table}

\begin{table}
\caption{As Fig.~\ref{naphthalene} for chrysene (C$_{18}$H$_{12}$).}
\label{chrysene}
\begin{center}
\begin{tabular}{cccc}
\hline \hline \noalign{\smallskip} 
& \multicolumn{3}{c}{Integrated flux} \\
\noalign{\smallskip} \cline{2-4} \noalign{\smallskip}
 & & Red & IRAS\\
Peak & ISRF & Rectangle & 21282+5050\\
($\mu$m) & (W$\,\,$sr$^{-1}$) (\%) & (W$\,\,$sr$^{-1}$) (\%)& (W$\,\,$sr$^{-1}$) (\%) \\ 
\noalign{\smallskip} \hline \noalign{\smallskip}
$\perp$ 17.26 & $4.1\,\, 10^{-29}$ (0.71) & $1.4\,\, 10^{-27}$ (0.88) & $2.7\,\, 10^{-24}$ (0.62) \\
$\parallel$ 17.53 & $7.2\,\, 10^{-30}$ (0.13) & $2.3\,\, 10^{-28}$ (0.15) & $4.3\,\, 10^{-25}$ (0.10) \\
$\perp$ 18.05 & $1.9\,\, 10^{-29}$ (0.33) & $6.3\,\, 10^{-28}$ (0.41) & $1.2\,\, 10^{-24}$ (0.27) \\
$\parallel$ 18.57 & $1.6\,\, 10^{-29}$ (0.29) & $5.9\,\, 10^{-28}$ (0.38) & $1.1\,\, 10^{-24}$ (0.25) \\
$\parallel$ 20.83 & $5.6\,\, 10^{-29}$ (0.99) & $2.0\,\, 10^{-27}$ (1.26) & $3.6\,\, 10^{-24}$ (0.83) \\
$\perp$ 23.22 & $4.7\,\, 10^{-29}$ (0.83) & $1.6\,\, 10^{-27}$ (1.05) & $3.0\,\, 10^{-24}$ (0.70) \\
$[26.04]$ & $2.8\,\, 10^{-30}$ (0.05) & $8.6\,\, 10^{-29}$ (0.06) & \textemdash{} \\
$[26.45]$ & $3.0\,\, 10^{-30}$ (0.05) & $9.4\,\, 10^{-29}$ (0.06) & \textemdash{} \\
$\perp$ 34.63 & $5.1\,\, 10^{-30}$ (0.09) & $2.0\,\, 10^{-28}$ (0.13) & $7.7\,\, 10^{-26}$ (0.02) \\
$[34.66]$ & $4.6\,\, 10^{-30}$ (0.08) & $1.4\,\, 10^{-28}$ (0.09) & \textemdash{} \\
$\perp$ 43.01 & $1.6\,\, 10^{-29}$ (0.29) & $6.0\,\, 10^{-28}$ (0.39) & $9.4\,\, 10^{-25}$ (0.22) \\
$\parallel$ 54.14 & $8.5\,\, 10^{-30}$ (0.15) & $3.3\,\, 10^{-28}$ (0.21) & $1.6\,\, 10^{-25}$ (0.04) \\
$[57.60]$ & $8.3\,\, 10^{-30}$ (0.15) & $1.8\,\, 10^{-28}$ (0.12) & \textemdash{} \\
$[75.03]$ & $9.6\,\, 10^{-30}$ (0.17) & $1.7\,\, 10^{-28}$ (0.11) & \textemdash{} \\
$\perp$ 131.58 & $1.3\,\, 10^{-29}$ (0.22) & $4.7\,\, 10^{-28}$ (0.30) & \textemdash{} \\
$\perp$ 206.15 & $1.4\,\, 10^{-29}$ (0.24) & $4.8\,\, 10^{-28}$ (0.31) & \textemdash{} \\
\noalign{\smallskip} \hline 
\end{tabular}
\end{center}
\end{table}

\begin{table}
\caption{As Fig.~\ref{naphthalene} for the chrysene cation (C$_{18}$H$_{12}^+$).}
\label{chrysene+}
\begin{center}
\begin{tabular}{cccc}
\hline \hline \noalign{\smallskip} 
& \multicolumn{3}{c}{Integrated flux} \\
\noalign{\smallskip} \cline{2-4} \noalign{\smallskip}
 &  & Red & IRAS\\
Peak & ISRF & Rectangle & 21282+5050\\
($\mu$m) & (W$\,\,$sr$^{-1}$) (\%) & (W$\,\,$sr$^{-1}$) (\%)& (W$\,\,$sr$^{-1}$) (\%)\\ 
\noalign{\smallskip} \hline \noalign{\smallskip}
$\parallel$ 17.87 & $5.5\,\, 10^{-29}$ (0.45) & $3.8\,\, 10^{-27}$ (0.60) & $1.1\,\, 10^{-24}$ (0.20) \\
$\perp$ 18.38 & $8.2\,\, 10^{-30}$ (0.07) & $5.8\,\, 10^{-28}$ (0.09) & $1.2\,\, 10^{-25}$ (0.02) \\
$\parallel$ 18.54 & $2.3\,\, 10^{-29}$ (0.19) & $1.4\,\, 10^{-27}$ (0.22) & $3.6\,\, 10^{-25}$ (0.07) \\
$\perp$ 19.44 & $5.1\,\, 10^{-29}$ (0.42) & $3.2\,\, 10^{-27}$ (0.51) & $8.8\,\, 10^{-25}$ (0.17) \\
\hline 
\end{tabular}
\end{center}
\end{table}

\clearpage

\begin{table}
\caption{As Fig.~\ref{naphthalene} for perylene (C$_{20}$H$_{12}$).}
\label{perylene}
\begin{center}
\begin{tabular}{cccc}
\hline \hline \noalign{\smallskip} 
& \multicolumn{3}{c}{Integrated flux} \\
\noalign{\smallskip} \cline{2-4} \noalign{\smallskip}
 &  & Red & IRAS\\
Peak & ISRF & Rectangle & 21282+5050\\
($\mu$m) & (W$\,\,$sr$^{-1}$) (\%) & (W$\,\,$sr$^{-1}$) (\%)& (W$\,\,$sr$^{-1}$) (\%) \\ 
\noalign{\smallskip} \hline \noalign{\smallskip}
$\parallel$ 17.21 & $4.6\,\, 10^{-29}$ (0.59) & $3.0\,\, 10^{-27}$ (0.79) & $1.9\,\, 10^{-24}$ (0.42) \\
$\perp$ 18.36 & $9.4\,\, 10^{-29}$ (1.20) & $6.0\,\, 10^{-27}$ (1.60) & $3.8\,\, 10^{-24}$ (0.83) \\
$\parallel$ 18.63 & $1.6\,\, 10^{-29}$ (0.20) & $8.9\,\, 10^{-28}$ (0.24) & $6.0\,\, 10^{-25}$ (0.13) \\
$\perp$ 21.19 & $5.1\,\, 10^{-30}$ (0.06) & $3.4\,\, 10^{-28}$ (0.09) & $1.5\,\, 10^{-25}$ (0.03) \\
\hline 
\end{tabular}
\end{center}
\end{table}

\begin{table}
\caption{As Fig.~\ref{naphthalene} for the perylene cation (C$_{20}$H$_{12}^+$).}
\label{perylene+}
\begin{center}
\begin{tabular}{cccc}
\hline \hline \noalign{\smallskip} 
& \multicolumn{3}{c}{Integrated flux} \\
\noalign{\smallskip} \cline{2-4} \noalign{\smallskip}
 &  & Red & IRAS\\
Peak & ISRF & Rectangle & 21282+5050\\
($\mu$m) & (W$\,\,$sr$^{-1}$) (\%) & (W$\,\,$sr$^{-1}$) (\%)& (W$\,\,$sr$^{-1}$) (\%) \\ 
\noalign{\smallskip} \hline \noalign{\smallskip}
$\parallel$ 17.41 & $7.4\,\, 10^{-29}$ (0.71) & $4.1\,\, 10^{-27}$ (0.86) & $3.0\,\, 10^{-24}$ (0.48) \\
$\perp$ 18.75 & $1.0\,\, 10^{-28}$ (1.00) & $6.0\,\, 10^{-27}$ (1.24) & $4.1\,\, 10^{-24}$ (0.65) \\
$\parallel$ 18.99 & $2.7\,\, 10^{-29}$ (0.26) & $1.6\,\, 10^{-27}$ (0.32) & $1.1\,\, 10^{-24}$ (0.17) \\
$\perp$ 21.54 & $9.8\,\, 10^{-30}$ (0.09) & $5.8\,\, 10^{-28}$ (0.12) & $3.0\,\, 10^{-25}$ (0.05) \\
\hline
\end{tabular}
\end{center}
\end{table}

\begin{table}
\caption{As Fig.~\ref{naphthalene} for pentacene (C$_{22}$H$_{14}$).}
\label{pentacene}
\begin{center}
\begin{tabular}{cccc}
\hline \hline \noalign{\smallskip} 
& \multicolumn{3}{c}{Integrated flux} \\
\noalign{\smallskip} \cline{2-4} \noalign{\smallskip}
 &  & Red & IRAS\\
Peak & ISRF & Rectangle & 21282+5050\\
($\mu$m) & (W$\,\,$sr$^{-1}$) (\%) & (W$\,\,$sr$^{-1}$) (\%)& (W$\,\,$sr$^{-1}$) (\%) \\ 
\noalign{\smallskip} \hline \noalign{\smallskip}
$\parallel$ 15.78 & $5.8\,\, 10^{-29}$ (0.65) & $2.6\,\, 10^{-27}$ (0.75) & $2.6\,\, 10^{-24}$ (0.55) \\
$\parallel$ 17.39 & $1.3\,\, 10^{-28}$ (1.45) & $6.1\,\, 10^{-27}$ (1.74) & $5.5\,\, 10^{-24}$ (1.16) \\
$\parallel$ 20.75 & $8.6\,\, 10^{-29}$ (0.95) & $3.9\,\, 10^{-27}$ (1.12) & $3.4\,\, 10^{-24}$ (0.72) \\
$\perp$ 21.38 & $6.4\,\, 10^{-29}$ (0.71) & $3.0\,\, 10^{-27}$ (0.85) & $2.6\,\, 10^{-24}$ (0.55) \\
\hline
\end{tabular}
\end{center}
\end{table}

\begin{table}
\caption{As Fig.~\ref{naphthalene} for the pentacene cation (C$_{22}$H$_{14}^+$).}
\label{pentacene+}
\begin{center}
\begin{tabular}{cccc}
\hline \hline \noalign{\smallskip} 
& \multicolumn{3}{c}{Integrated flux} \\
\noalign{\smallskip} \cline{2-4} \noalign{\smallskip}
 &  & Red & IRAS\\
Peak & ISRF & Rectangle & 21282+5050\\
($\mu$m) & (W$\,\,$sr$^{-1}$) (\%) & (W$\,\,$sr$^{-1}$) (\%)& (W$\,\,$sr$^{-1}$) (\%) \\ 
\noalign{\smallskip} \hline \noalign{\smallskip}
$\parallel$ 17.37 & $3.3\,\, 10^{-29}$ (0.15) & $1.8\,\, 10^{-27}$ (0.16) & $4.7\,\, 10^{-25}$ (0.06) \\
$\perp$ 20.63 & $1.6\,\, 10^{-28}$ (0.76) & $9.7\,\, 10^{-27}$ (0.89) & $2.3\,\, 10^{-24}$ (0.30) \\
$\perp$ 21.94 & $6.9\,\, 10^{-28}$ (3.20) & $4.0\,\, 10^{-26}$ (3.71) & $9.4\,\, 10^{-24}$ (1.22) \\
$\perp$ 25.93 & $1.1\,\, 10^{-29}$ (0.05) & $6.3\,\, 10^{-28}$ (0.06) & $5.1\,\, 10^{-26}$ (0.01) \\
$\parallel$ 27.66 & $1.5\,\, 10^{-29}$ (0.07) & $9.1\,\, 10^{-28}$ (0.08) & $1\,\, 10^{-25}$ (0.01) \\
$[27.87]$ & $1.2\,\, 10^{-29}$ (0.05) & $2.3\,\, 10^{-28}$ (0.02) & \textemdash{} \\
$[35.10]$ & $2.1\,\, 10^{-29}$ (0.10) & $3.8\,\, 10^{-28}$ (0.03) & \textemdash{} \\
$[40.11]$ & $2.6\,\, 10^{-29}$ (0.12) & $4.5\,\, 10^{-28}$ (0.04) & \textemdash{} \\
$[40.71]$ & $2.8\,\, 10^{-29}$ (0.13) & $4.5\,\, 10^{-28}$ (0.04) & \textemdash{} \\
$[41.70]$ & $2.8\,\, 10^{-29}$ (0.13) & $4.2\,\, 10^{-28}$ (0.04) & \textemdash{} \\
$\perp$ 52.73 & $6.1\,\, 10^{-29}$ (0.28) & $3.5\,\, 10^{-27}$ (0.32) & $4.6\,\, 10^{-25}$ (0.06) \\
$[62.41]$ & $4.7\,\, 10^{-29}$ (0.22) & $4.9\,\, 10^{-28}$ (0.05) & \textemdash{} \\
$\parallel$ 86.88 & $7.9\,\, 10^{-29}$ (0.37) & $4.7\,\, 10^{-27}$ (0.43) & $4.5\,\, 10^{-25}$ (0.06) \\
$[100.30]$ & $6.5\,\, 10^{-29}$ (0.30) & $3.1\,\, 10^{-28}$ (0.03) & \textemdash{} \\
$[110.73]$ & $6.6\,\, 10^{-29}$ (0.31) & $2.9\,\, 10^{-28}$ (0.03) & \textemdash{} \\
$\perp$ 474.50 & $1.1\,\, 10^{-28}$ (0.49) & $3.7\,\, 10^{-27}$ (0.34) & \textemdash{} \\
\noalign{\smallskip} \hline 
\end{tabular}
\end{center}
\end{table}

\begin{table}
\caption{As Fig.~\ref{naphthalene} for benzo[g,h,i]perylene (C$_{22}$H$_{12}$)}
\label{benzoperylene}
\begin{center}
\begin{tabular}{cccc}
\hline \hline \noalign{\smallskip} 
& \multicolumn{3}{c}{Integrated flux} \\
\noalign{\smallskip} \cline{2-4} \noalign{\smallskip}
 &  & Red & IRAS\\
Peak & ISRF & Rectangle & 21282+5050\\
($\mu$m) & (W$\,\,$sr$^{-1}$) (\%) & (W$\,\,$sr$^{-1}$) (\%)& (W$\,\,$sr$^{-1}$) (\%)\\ 
\noalign{\smallskip} \hline \noalign{\smallskip}
$\perp$ 15.41 & $1.1\,\, 10^{-28}$ (1.49) & $5.8\,\, 10^{-27}$ (1.94) & $6.0\,\, 10^{-24}$ (1.19) \\
$\parallel$ 17.68 & $3.0\,\, 10^{-29}$ (0.41) & $1.6\,\, 10^{-27}$ (0.55) & $1.6\,\, 10^{-24}$ (0.31) \\
$\perp$ 18.40 & $7.2\,\, 10^{-29}$ (0.98) & $3.9\,\, 10^{-27}$ (1.31) & $3.9\,\, 10^{-24}$ (0.77) \\
$\parallel$ 18.62 & $5.2\,\, 10^{-30}$ (0.07) & $2.4\,\, 10^{-28}$ (0.08) & $2.3\,\, 10^{-25}$ (0.05) \\
\hline
\end{tabular}
\end{center}
\end{table}

\begin{table}
\caption{As Fig.~\ref{naphthalene} for the benzo[g,h,i]perylene cation
(C$_{22}$H$_{12}^+$)}
\label{benzoperylene+}
\begin{center}
\begin{tabular}{cccc}
\hline \hline \noalign{\smallskip} 
& \multicolumn{3}{c}{Integrated flux} \\
\noalign{\smallskip} \cline{2-4} \noalign{\smallskip}
 &  & Red & IRAS\\
Peak & ISRF & Rectangle & 21282+5050\\
($\mu$m) & (W$\,\,$sr$^{-1}$) (\%) & (W$\,\,$sr$^{-1}$) (\%)& (W$\,\,$sr$^{-1}$) (\%) \\ 
\noalign{\smallskip} \hline \noalign{\smallskip}
$\perp$ 15.56 & $1.9\,\, 10^{-28}$ (1.63) & $1.0\,\, 10^{-26}$ (1.99) & $7.2\,\, 10^{-24}$ (1.07) \\
$\parallel$ 15.96 & $5.3\,\, 10^{-30}$ (0.05) & $2.6\,\, 10^{-28}$ (0.05) & $1.3\,\, 10^{-25}$ (0.02) \\
$\parallel$ 17.82 & $2.6\,\, 10^{-29}$ (0.22) & $1.3\,\, 10^{-27}$ (0.25) & $8.6\,\, 10^{-25}$ (0.13) \\
$\perp$ 18.62 & $1.2\,\, 10^{-28}$ (1.02) & $6.4\,\, 10^{-27}$ (1.28) & $4.3\,\, 10^{-24}$ (0.63) \\
\hline
\end{tabular}
\end{center}
\end{table}

\begin{table}
\caption{As Fig.~\ref{naphthalene} for anthanthrene (C$_{22}$H$_{12}$)}
\label{anthanthrene}
\begin{center}
\begin{tabular}{cccc}
\hline \hline \noalign{\smallskip} 
& \multicolumn{3}{c}{Integrated flux} \\
\noalign{\smallskip} \cline{2-4} \noalign{\smallskip}
 &  & Red & IRAS\\
Peak & ISRF & Rectangle & 21282+5050\\
($\mu$m) & (W$\,\,$sr$^{-1}$) (\%) & (W$\,\,$sr$^{-1}$) (\%)& (W$\,\,$sr$^{-1}$) (\%)\\ 
\noalign{\smallskip} \hline \noalign{\smallskip}
$\parallel$ 16.59 & $7.4\,\, 10^{-29}$ (0.87) & $4.5\,\, 10^{-27}$ (1.07) & $3.1\,\, 10^{-24}$ (0.62) \\
$\perp$ 17.73 & $2.5\,\, 10^{-29}$ (0.29) & $1.5\,\, 10^{-27}$ (0.35) & $9.7\,\, 10^{-25}$ (0.20) \\
$\parallel$ 17.90 & $4.0\,\, 10^{-29}$ (0.47) & $2.5\,\, 10^{-27}$ (0.59) & $1.6\,\, 10^{-24}$ (0.33) \\
$\perp$ 18.41 & $5.5\,\, 10^{-29}$ (0.64) & $3.5\,\, 10^{-27}$ (0.82) & $2.3\,\, 10^{-24}$ (0.46) \\
\hline
\end{tabular}
\end{center}
\end{table}

\begin{table}
\caption{As Fig.~\ref{naphthalene} for the anthanthrene cation 
(C$_{22}$H$_{12}^+$)}
\label{anthanthrene+}
\begin{center}
\begin{tabular}{cccc}
\hline \hline \noalign{\smallskip} 
& \multicolumn{3}{c}{Integrated flux} \\
\noalign{\smallskip} \cline{2-4} \noalign{\smallskip}
 &  & Red & IRAS\\
Peak & ISRF & Rectangle & 21282+5050\\
($\mu$m) & (W$\,\,$sr$^{-1}$) (\%) & (W$\,\,$sr$^{-1}$) (\%)& (W$\,\,$sr$^{-1}$) (\%) \\ 
\noalign{\smallskip} \hline \noalign{\smallskip}
$\parallel$ 16.80 & $4.0\,\, 10^{-29}$ (0.33) & $2.3\,\, 10^{-27}$ (0.41) & $1.2\,\, 10^{-24}$ (0.18) \\
$\parallel$ 17.85 & $1.0\,\, 10^{-29}$ (0.09) & $5.6\,\, 10^{-28}$ (0.10) & $3\,\, 10^{-25}$ (0.04) \\
$\perp$ 17.98 & $4.3\,\, 10^{-29}$ (0.36) & $2.6\,\, 10^{-27}$ (0.46) & $1.3\,\, 10^{-24}$ (0.19) \\
$\parallel$ 20.08 & $3.3\,\, 10^{-29}$ (0.27) & $1.9\,\, 10^{-27}$ (0.34) & $9.3\,\, 10^{-25}$ (0.14) \\
\hline
\end{tabular}
\end{center}
\end{table}

\begin{table}
\caption{As Fig.~\ref{naphthalene} for coronene (C$_{24}$H$_{12}$)}
\label{coronene}
\begin{center}
\begin{tabular}{cccc}
\hline \hline \noalign{\smallskip} 
& \multicolumn{3}{c}{Integrated flux} \\
\noalign{\smallskip} \cline{2-4} \noalign{\smallskip}
 &  & Red & IRAS\\
Peak & ISRF & Rectangle & 21282+5050\\
($\mu$m) & (W$\,\,$sr$^{-1}$) (\%) & (W$\,\,$sr$^{-1}$) (\%)& (W$\,\,$sr$^{-1}$) (\%) \\ 
\noalign{\smallskip} \hline \noalign{\smallskip}
$\perp$ 18.21 & $3.4\,\, 10^{-28}$ (4.24) & $1.2\,\, 10^{-26}$ (5.29) & $1.9\,\, 10^{-23}$ (3.41) \\
$[20.07]$ & $3.2\,\, 10^{-30}$ (0.04) & $1.3\,\, 10^{-28}$ (0.06) & \textemdash{} \\
$[22.28]$ & $5.3\,\, 10^{-30}$ (0.07) & $1.8\,\, 10^{-28}$ (0.08) & \textemdash{} \\
$\parallel$ 26.20 & $2.5\,\, 10^{-29}$ (0.31) & $9.7\,\, 10^{-28}$ (0.42) & $1.2\,\, 10^{-24}$ (0.22) \\
$[27.34]$ & $8.3\,\, 10^{-30}$ (0.10) & $3.1\,\, 10^{-28}$ (0.13) & \textemdash{} \\
$[33.59]$ & $1.2\,\, 10^{-29}$ (0.15) & $4.1\,\, 10^{-28}$ (0.18) & \textemdash{} \\
$[34.34]$ & $1.3\,\, 10^{-29}$ (0.17) & $4.4\,\, 10^{-28}$ (0.19) & \textemdash{} \\
$[44.57]$ & $8.8\,\, 10^{-30}$ (0.11) & $2.9\,\, 10^{-28}$ (0.12) & \textemdash{} \\
$[61.10]$ & $1.2\,\, 10^{-29}$ (0.15) & $3.1\,\, 10^{-28}$ (0.13) & \textemdash{} \\
$\perp$ 80.60 & $2.2\,\, 10^{-29}$ (0.27) & $8.4\,\, 10^{-28}$ (0.36) & $7.2\,\, 10^{-25}$ (0.13) \\
$[113.37]$ & $2.8\,\, 10^{-29}$ (0.36) & $4.2\,\, 10^{-28}$ (0.18) & \textemdash\\
\noalign{\smallskip} \hline
\end{tabular}
\end{center}
\end{table}

\begin{table}
\caption{As Fig.~\ref{naphthalene} for the coronene cation (C$_{24}$H$_{12}^+$)}
\label{coronene+}
\begin{center}
\begin{tabular}{cccc}
\hline \hline \noalign{\smallskip} 
& \multicolumn{3}{c}{Integrated flux} \\
\noalign{\smallskip} \cline{2-4} \noalign{\smallskip}
 &  & Red & IRAS\\
Peak & ISRF & Rectangle & 21282+5050\\
($\mu$m) & (W$\,\,$sr$^{-1}$) (\%) & (W$\,\,$sr$^{-1}$) (\%)& (W$\,\,$sr$^{-1}$) (\%) \\ 
\noalign{\smallskip} \hline \noalign{\smallskip}
$\parallel$ 15.13 & $8.5\,\, 10^{-29}$ (0.68) & $4.4\,\, 10^{-27}$ (0.88) & $2.8\,\, 10^{-24}$ (0.40) \\
$\parallel$ 18.08 & $1.4\,\, 10^{-28}$ (1.14) & $7.4\,\, 10^{-27}$ (1.50) & $4.3\,\, 10^{-24}$ (0.62) \\
$\perp$ 18.14 & $1.4\,\, 10^{-28}$ (1.12) & $7.1\,\, 10^{-27}$ (1.44) & $4.1\,\, 10^{-24}$ (0.59) \\
$\perp$ 19.39 & $1.1\,\, 10^{-29}$ (0.09) & $5.9\,\, 10^{-28}$ (0.12) & $2.8\,\, 10^{-25}$ (0.04) \\
\hline
\end{tabular}
\end{center}
\end{table}

\begin{table}
\caption{As Fig.~\ref{naphthalene} for dibenzo[cd,lm]perylene (C$_{26}$H$_{14}$)}
\label{dibenzoperylene}
\begin{center}
\begin{tabular}{cccc}
\hline \hline \noalign{\smallskip} 
& \multicolumn{3}{c}{Integrated flux} \\
\noalign{\smallskip} \cline{2-4} \noalign{\smallskip}
 &  & Red & IRAS\\
Peak & ISRF & Rectangle & 21282+5050\\
($\mu$m) & (W$\,\,$sr$^{-1}$) (\%) & (W$\,\,$sr$^{-1}$) (\%)& (W$\,\,$sr$^{-1}$) (\%)\\ 
\noalign{\smallskip} \hline \noalign{\smallskip}
$\parallel$ 15.05 & $6.1\,\, 10^{-29}$ (0.42) & $5.2\,\, 10^{-27}$ (0.52) & $1.5\,\, 10^{-24}$ (0.27) \\
$\perp$ 15.10 & $5.9\,\, 10^{-28}$ (4.12) & $4.8\,\, 10^{-26}$ (4.85) & $1.4\,\, 10^{-23}$ (2.56) \\
$\parallel$ 18.30 & $1.4\,\, 10^{-28}$ (0.99) & $1.2\,\, 10^{-26}$ (1.20) & $2.9\,\, 10^{-24}$ (0.52) \\
$\parallel$ 19.45 & $8.0\,\, 10^{-30}$ (0.06) & $5.8\,\, 10^{-28}$ (0.06) & $1.3\,\, 10^{-25}$ (0.02) \\
\hline
\end{tabular}
\end{center}
\end{table}

\begin{table}
\caption{As Fig.~\ref{naphthalene} for the dibenzo[cd,lm]perylene cation
(C$_{26}$H$_{14}^+$)}
\label{dibenzoperylene+}
\begin{center}
\begin{tabular}{cccc}
\hline \hline \noalign{\smallskip} 
& \multicolumn{3}{c}{Integrated flux} \\
\noalign{\smallskip} \cline{2-4} \noalign{\smallskip}
 &  & Red & IRAS\\
Peak & ISRF & Rectangle & 21282+5050\\
($\mu$m) & (W$\,\,$sr$^{-1}$) (\%) & (W$\,\,$sr$^{-1}$) (\%)& (W$\,\,$sr$^{-1}$) (\%)\\ 
\noalign{\smallskip} \hline \noalign{\smallskip}
$\perp$ 15.12 & $2.6\,\, 10^{-28}$ (1.48) & $1.5\,\, 10^{-26}$ (1.74) & $6.2\,\, 10^{-24}$ (0.78) \\
$\parallel$ 18.01 & $1.7\,\, 10^{-29}$ (0.10) & $1.1\,\, 10^{-27}$ (0.13) & $3.6\,\, 10^{-25}$ (0.04) \\
$\parallel$ 19.53 & $2.6\,\, 10^{-29}$ (0.15) & $1.7\,\, 10^{-27}$ (0.20) & $5.5\,\, 10^{-25}$ (0.07) \\
$\perp$ 20.78 & $5.2\,\, 10^{-29}$ (0.30) & $3.3\,\, 10^{-27}$ (0.39) & $1.1\,\, 10^{-24}$ (0.13) \\
\hline
\end{tabular}
\end{center}
\end{table}

\begin{table}
\caption{As Fig.~\ref{naphthalene} for bisanthene (C$_{28}$H$_{14}$)}
\label{bisanthene}
\begin{center}
\begin{tabular}{cccc}
\hline \hline \noalign{\smallskip} 
& \multicolumn{3}{c}{Integrated flux} \\
\noalign{\smallskip} \cline{2-4} \noalign{\smallskip}
 &  & Red & IRAS\\
Peak & ISRF & Rectangle & 21282+5050\\
($\mu$m) & (W$\,\,$sr$^{-1}$) (\%) & (W$\,\,$sr$^{-1}$) (\%)& (W$\,\,$sr$^{-1}$) (\%)\\ 
\noalign{\smallskip} \hline \noalign{\smallskip}
$\parallel$ 15.10 & $1.7\,\, 10^{-28}$ (1.12) & $8.5\,\, 10^{-27}$ (1.33) & $3.7\,\, 10^{-24}$ (0.63) \\
$\parallel$ 15.78 & $7.1\,\, 10^{-29}$ (0.47) & $3.4\,\, 10^{-27}$ (0.53) & $1.5\,\, 10^{-24}$ (0.26) \\
$\perp$ 16.97 & $1.7\,\, 10^{-29}$ (0.11) & $8.5\,\, 10^{-28}$ (0.13) & $3.0\,\, 10^{-25}$ (0.05) \\
$\parallel$ 17.93 & $5.3\,\, 10^{-29}$ (0.35) & $2.7\,\, 10^{-27}$ (0.42) & $9.9\,\, 10^{-25}$ (0.17) \\
\hline
\end{tabular}
\end{center}
\vspace{-0.5cm}
\end{table}

\begin{table}
\caption{As Fig.~\ref{naphthalene} for the bisanthene cation
(C$_{28}$H$_{14}^+$)}
\label{bisanthene+}
\begin{center}
\begin{tabular}{cccc}
\hline \hline \noalign{\smallskip} 
& \multicolumn{3}{c}{Integrated flux} \\
\noalign{\smallskip} \cline{2-4} \noalign{\smallskip}
 &  & Red & IRAS\\
Peak & ISRF & Rectangle & 21282+5050\\
($\mu$m) & (W$\,\,$sr$^{-1}$) (\%) & (W$\,\,$sr$^{-1}$) (\%)& (W$\,\,$sr$^{-1}$) (\%)\\ 
\noalign{\smallskip} \hline \noalign{\smallskip}
$\parallel$ 15.04 & $1.2\,\, 10^{-28}$ (0.68) & $7.1\,\, 10^{-27}$ (0.88) & $2.8\,\, 10^{-24}$ (0.29) \\
$\perp$ 16.42 & $4.8\,\, 10^{-29}$ (0.27) & $2.8\,\, 10^{-27}$ (0.35) & $1.1\,\, 10^{-24}$ (0.12) \\
$\parallel$ 17.07 & $3.6\,\, 10^{-28}$ (2.03) & $2.2\,\, 10^{-26}$ (2.71) & $7.9\,\, 10^{-24}$ (0.84) \\
$\perp$ 18.02 & $1.7\,\, 10^{-28}$ (0.94) & $1.0\,\, 10^{-26}$ (1.28) & $3.4\,\, 10^{-24}$ (0.36) \\
\hline
\end{tabular}
\end{center}
\end{table}

\begin{table}
\caption{As Fig.~\ref{naphthalene} for terrylene (C$_{30}$H$_{16}$)}
\label{terrylene}
\begin{center}
\begin{tabular}{cccc}
\hline \hline \noalign{\smallskip} 
& \multicolumn{3}{c}{Integrated flux} \\
\noalign{\smallskip} \cline{2-4} \noalign{\smallskip}
 &  & Red & IRAS\\
Peak & ISRF & Rectangle & 21282+5050\\
($\mu$m) & (W$\,\,$sr$^{-1}$) (\%) & (W$\,\,$sr$^{-1}$) (\%)& (W$\,\,$sr$^{-1}$) (\%)\\ 
\noalign{\smallskip} \hline \noalign{\smallskip}
$\perp$ 17.82 & $2.8\,\, 10^{-28}$ (1.22) & $1.4\,\, 10^{-26}$ (1.32) & $3.6\,\, 10^{-24}$ (0.55)\\ 
$[18.82]$ & $5.9\,\, 10^{-29}$ (0.26) & $3.0\,\, 10^{-27}$ (0.29) & $7.3\,\, 10^{-25}$ (0.11) \\
$\parallel$ 19.47 & $2.3\,\, 10^{-28}$ (0.98) & $1.2\,\, 10^{-26}$ (1.14) & $2.7\,\, 10^{-24}$ (0.42) \\
$\parallel$ 21.27 & $2.5\,\, 10^{-28}$ (1.10) & $1.3\,\, 10^{-26}$ (1.24) & $2.7\,\, 10^{-24}$ (0.42) \\
$\perp$ 37.56 & $1.1\,\, 10^{-28}$ (0.48) & $5.8\,\, 10^{-27}$ (0.55) & $7.3\,\, 10^{-25}$ (0.11) \\
$[41.19]$ & $8.2\,\, 10^{-30}$ (0.04) & $7.7\,\, 10^{-29}$ (0.01) & \textemdash{} \\
$[45.95]$ & $1.1\,\, 10^{-29}$ (0.05) & $9.2\,\, 10^{-29}$ (0.01) & \\
$[47.77]$ & $1.2\,\, 10^{-29}$ (0.05) & $9.6\,\, 10^{-29}$ (0.01) & \textemdash{} \\
$\perp$ 57.55 & $2.6\,\, 10^{-28}$ (1.12) & $1.3\,\, 10^{-26}$ (1.26) & $1.4\,\, 10^{-24}$ (0.21) \\
$[61.21]$ & $2.0\,\, 10^{-29}$ (0.09) & $8.9\,\, 10^{-29}$ (0.01) & \textemdash{} \\
$\parallel$ 67.52 & $4.8\,\, 10^{-29}$ (0.21) & $2.5\,\, 10^{-27}$ (0.24) & $1.3\,\, 10^{-25}$ (0.02) \\
$\perp$ 88.43 & $3.8\,\, 10^{-29}$ (0.17) & $1.7\,\, 10^{-27}$ (0.16) & \textemdash{} \\
$[91.34]$ & $2.6\,\, 10^{-29}$ (0.11) & $5.0\,\, 10^{-29}$ (0.00) & \textemdash{} \\
$\perp$ 271.11 & $6.5\,\, 10^{-29}$ (0.28) & $1.5\,\, 10^{-27}$ (0.14) & \textemdash{} \\
$[306.54]$ & $1.5\,\, 10^{-29}$ (0.07) & \textemdash{} & \textemdash{} \\
\noalign{\smallskip} \hline
\end{tabular}
\end{center}
\end{table}

\begin{table}
\caption{As Fig.~\ref{naphthalene} for the terrylene cation (C$_{30}$H$_{16}^+$)}
\label{terrylene+}
\begin{center}
\begin{tabular}{cccc}
\hline \hline \noalign{\smallskip} 
& \multicolumn{3}{c}{Integrated flux} \\
\noalign{\smallskip} \cline{2-4} \noalign{\smallskip}
 &  & Red & IRAS\\
Peak & ISRF & Rectangle & 21282+5050\\
($\mu$m) & (W$\,\,$sr$^{-1}$) (\%) & (W$\,\,$sr$^{-1}$) (\%)& (W$\,\,$sr$^{-1}$) (\%)\\ 
\noalign{\smallskip} \hline \noalign{\smallskip}
$\parallel$ 15.00 & $8.0\,\, 10^{-29}$ (0.28) & $3.9\,\, 10^{-27}$ (0.31) & $1.4\,\, 10^{-24}$ (0.13) \\
$\perp$ 18.03 & $1.6\,\, 10^{-28}$ (0.57) & $7.9\,\, 10^{-27}$ (0.64) & $2.5\,\, 10^{-24}$ (0.22) \\
$\parallel$ 19.02 & $1.8\,\, 10^{-29}$ (0.06) & $7.9\,\, 10^{-28}$ (0.06) & $2.9\,\, 10^{-25}$ (0.03) \\
$\parallel$ 19.36 & $6.8\,\, 10^{-29}$ (0.24) & $3.2\,\, 10^{-27}$ (0.25) & $9.1\,\, 10^{-25}$ (0.08) \\
\hline
\end{tabular}
\end{center}
\end{table}

\clearpage

\begin{table}
\caption{As Fig.~\ref{naphthalene} for ovalene (C$_{32}$H$_{14}$)}
\label{ovalene}
\begin{center}
\begin{tabular}{cccc}
\hline \hline \noalign{\smallskip} 
& \multicolumn{3}{c}{Integrated flux} \\
\noalign{\smallskip} \cline{2-4} \noalign{\smallskip}
 &  & Red & IRAS\\
Peak & ISRF & Rectangle & 21282+5050\\
($\mu$m) & (W$\,\,$sr$^{-1}$) (\%) & (W$\,\,$sr$^{-1}$) (\%)& (W$\,\,$sr$^{-1}$) (\%)\\ 
\noalign{\smallskip} \hline \noalign{\smallskip}
$\parallel$ 15.03 & $7.9\,\, 10^{-30}$ (0.06) & $6.1\,\, 10^{-28}$ (0.08) & $3.4\,\, 10^{-25}$ (0.05) \\
$\parallel$ 15.26 & $5.3\,\, 10^{-30}$ (0.04) & $3.8\,\, 10^{-28}$ (0.05) & $2.0\,\, 10^{-25}$ (0.03) \\
$\perp$ 15.90 & $4.8\,\, 10^{-28}$ (3.60) & $3.3\,\, 10^{-26}$ (4.33) & $1.7\,\, 10^{-23}$ (2.56) \\
$\parallel$ 17.45 & $7.8\,\, 10^{-29}$ (0.59) & $5.5\,\, 10^{-27}$ (0.71) & $2.7\,\, 10^{-24}$ (0.40) \\
\hline
\end{tabular}
\end{center}
\end{table}

\begin{table}
\caption{As Fig.~\ref{naphthalene} for the ovalene cation (C$_{32}$H$_{14}^+$)}
\label{ovalene+}
\begin{center}
\begin{tabular}{cccc}
\hline \hline \noalign{\smallskip} 
& \multicolumn{3}{c}{Integrated flux} \\
\noalign{\smallskip} \cline{2-4} \noalign{\smallskip}
 &  & Red & IRAS\\
Peak & ISRF & Rectangle & 21282+5050\\
($\mu$m) & (W$\,\,$sr$^{-1}$) (\%) & (W$\,\,$sr$^{-1}$) (\%)& (W$\,\,$sr$^{-1}$) (\%)\\ 
\noalign{\smallskip} \hline \noalign{\smallskip}
$\parallel$ 15.01 & $3.2\,\, 10^{-28}$ (1.59) & $2.1\,\, 10^{-26}$ (1.91) & $1.0\,\, 10^{-23}$ (0.90) \\
$\perp$ 15.80 & $4.5\,\, 10^{-28}$ (2.20) & $2.9\,\, 10^{-26}$ (2.71) & $1.4\,\, 10^{-23}$ (1.25) \\
$\parallel$ 18.29 & $2.7\,\, 10^{-29}$ (0.13) & $1.8\,\, 10^{-27}$ (0.17) & $7.2\,\, 10^{-25}$ (0.06) \\
$\perp$ 18.47 & $3.7\,\, 10^{-28}$ (1.80) & $2.4\,\, 10^{-26}$ (2.26) & $1.0\,\, 10^{-23}$ (0.92) \\
\hline
\end{tabular}
\end{center}
\end{table}

\begin{table}
\caption{As Fig.~\ref{naphthalene} for circumbiphenyl (C$_{38}$H$_{16}$)}
\label{circumbiphenyl}
\begin{center}
\begin{tabular}{cccc}
\hline \hline \noalign{\smallskip} 
& \multicolumn{3}{c}{Integrated flux} \\
\noalign{\smallskip} \cline{2-4} \noalign{\smallskip}
 &  & Red & IRAS\\
Peak & ISRF & Rectangle & 21282+5050\\
($\mu$m) & (W$\,\,$sr$^{-1}$) (\%) & (W$\,\,$sr$^{-1}$) (\%)& (W$\,\,$sr$^{-1}$) (\%)\\ 
\noalign{\smallskip} \hline \noalign{\smallskip}
$\parallel$ 15.29 & $1.7\,\, 10^{-28}$ (1.12) & $1.4\,\, 10^{-26}$ (1.31) & $6.4\,\, 10^{-24}$ (0.80) \\
$\perp$ 16.00 & $2.8\,\, 10^{-28}$ (1.82) & $2.3\,\, 10^{-26}$ (2.15) & $1.0\,\, 10^{-23}$ (1.30) \\
$\perp$ 17.63 & $2.6\,\, 10^{-28}$ (1.71) & $2.2\,\, 10^{-26}$ (2.07) & $9.2\,\, 10^{-24}$ (1.16) \\
$\parallel$ 18.78 & $5.8\,\, 10^{-29}$ (0.38) & $5.2\,\, 10^{-27}$ (0.48) & $2.1\,\, 10^{-24}$ (0.26) \\
\hline
\end{tabular}
\end{center}
\end{table}

\begin{table}
\caption{As Fig.~\ref{naphthalene} for the circumbiphenyl cation 
(C$_{38}$H$_{16}^+$)}
\label{circumbiphenyl+}
\begin{center}
\begin{tabular}{cccc}
\hline \hline \noalign{\smallskip} 
& \multicolumn{3}{c}{Integrated flux} \\
\noalign{\smallskip} \cline{2-4} \noalign{\smallskip}
 &  & Red & IRAS\\
Peak & ISRF & Rectangle & 21282+5050\\
($\mu$m) & (W$\,\,$sr$^{-1}$) (\%) & (W$\,\,$sr$^{-1}$) (\%)& (W$\,\,$sr$^{-1}$) (\%)\\ 
\noalign{\smallskip} \hline \noalign{\smallskip}
$\parallel$ 15.03 & $1.6\,\, 10^{-29}$ (0.06) & $1.2\,\, 10^{-27}$ (0.07) & $3.8\,\, 10^{-25}$ (0.03) \\
$\parallel$ 15.25 & $3.3\,\, 10^{-29}$ (0.12) & $2.2\,\, 10^{-27}$ (0.13) & $7.2\,\, 10^{-25}$ (0.06) \\
$\perp$ 15.94 & $3.1\,\, 10^{-28}$ (1.14) & $2.2\,\, 10^{-26}$ (1.34) & $6.8\,\, 10^{-24}$ (0.52) \\
$\parallel$ 16.75 & $6.6\,\, 10^{-28}$ (2.46) & $5.0\,\, 10^{-26}$ (3.01) & $1.4\,\, 10^{-23}$ (1.06) \\
\hline
\end{tabular}
\end{center}
\end{table}

\begin{table}
\caption{As Fig.~\ref{naphthalene} for quaterrylene (C$_{40}$H$_{20}$)}
\label{quaterrylene}
\begin{center}
\begin{tabular}{cccc}
\hline \hline \noalign{\smallskip} 
& \multicolumn{3}{c}{Integrated flux} \\
\noalign{\smallskip} \cline{2-4} \noalign{\smallskip}
 &  & Red & IRAS\\
Peak & ISRF & Rectangle & 21282+5050\\
($\mu$m) & (W$\,\,$sr$^{-1}$) (\%) & (W$\,\,$sr$^{-1}$) (\%)& (W$\,\,$sr$^{-1}$) (\%)\\ 
\noalign{\smallskip} \hline \noalign{\smallskip}
$\perp$ 15.06 & $8.4\,\, 10^{-28}$ (1.61) & $4.1\,\, 10^{-26}$ (1.66) & $6.5\,\, 10^{-24}$ (0.77) \\
$\parallel$ 17.49 & $4.1\,\, 10^{-28}$ (0.78) & $2.0\,\, 10^{-26}$ (0.82) & $2.7\,\, 10^{-24}$ (0.32) \\
$\perp$ 17.58 & $4.2\,\, 10^{-28}$ (0.80) & $2.0\,\, 10^{-26}$ (0.81) & $2.6\,\, 10^{-24}$ (0.31) \\
$\parallel$ 18.52 & $3.7\,\, 10^{-28}$ (0.71) & $1.9\,\, 10^{-26}$ (0.79) & $2.3\,\, 10^{-24}$ (0.27) \\
$\parallel$ 19.38 & $9.4\,\, 10^{-28}$ (1.80) & $4.9\,\, 10^{-26}$ (1.98) & $5.4\,\, 10^{-24}$ (0.64) \\
$\parallel$ 22.64 & $1.1\,\, 10^{-27}$ (2.15) & $5.8\,\, 10^{-26}$ (2.33) & $5.3\,\, 10^{-24}$ (0.63) \\
$\parallel$ 24.63 & $3.5\,\, 10^{-29}$ (0.07) & $2.0\,\, 10^{-27}$ (0.08) & $1.3\,\, 10^{-25}$ (0.02) \\
$\parallel$ 28.32 & $6.3\,\, 10^{-28}$ (1.22) & $3.2\,\, 10^{-26}$ (1.28) & $2.1\,\, 10^{-24}$ (0.24) \\
$\perp$ 32.03 & $3.1\,\, 10^{-28}$ (0.60) & $1.5\,\, 10^{-26}$ (0.60) & $8.6\,\, 10^{-25}$ (0.10) \\
$\parallel$ 34.14 & $7.2\,\, 10^{-29}$ (0.14) & $3.6\,\, 10^{-27}$ (0.15) & $1.8\,\, 10^{-25}$ (0.02) \\
$\perp$ 57.30 & $1.5\,\, 10^{-27}$ (2.89) & $7.5\,\, 10^{-26}$ (3.03) & $2.0\,\, 10^{-24}$ (0.24) \\
$\perp$ 87.48 & $7.0\,\, 10^{-29}$ (0.13) & $3.2\,\, 10^{-27}$ (0.13) & $4.5\,\, 10^{-26}$ (0.01) \\
$\parallel$ 102.85 & $2.2\,\, 10^{-28}$ (0.42) & $1.0\,\, 10^{-26}$ (0.41) & $1.6\,\, 10^{-25}$ (0.02) \\
$\perp$ 471.64 & $7.7\,\, 10^{-29}$ (0.15) & $2.7\,\, 10^{-28}$ (0.01) & \textemdash{} \\
\noalign{\smallskip} \hline 
\end{tabular}
\end{center}
\end{table}

\begin{table}
\caption{As Fig.~\ref{naphthalene} for the quaterrylene cation
(C$_{40}$H$_{20}^+$)}
\label{quaterrylene+}
\begin{center}
\begin{tabular}{cccc}
\hline \hline \noalign{\smallskip} 
& \multicolumn{3}{c}{Integrated flux} \\
\noalign{\smallskip} \cline{2-4} \noalign{\smallskip}
 &  & Red & IRAS\\
Peak & ISRF & Rectangle & 21282+5050\\
($\mu$m) & (W$\,\,$sr$^{-1}$) (\%) & (W$\,\,$sr$^{-1}$) (\%)& (W$\,\,$sr$^{-1}$) (\%)\\ 
\noalign{\smallskip} \hline \noalign{\smallskip}
$\perp$ 15.20 & $5.1\,\, 10^{-28}$ (0.82) & $2.6\,\, 10^{-26}$ (0.86) & $4.3\,\, 10^{-24}$ (0.28) \\
$\parallel$ 17.54 & $1.7\,\, 10^{-28}$ (0.27) & $8.4\,\, 10^{-27}$ (0.29) & $1.1\,\, 10^{-24}$ (0.07) \\
$\perp$ 17.87 & $1.5\,\, 10^{-28}$ (0.24) & $7.7\,\, 10^{-27}$ (0.26) & $9.8\,\, 10^{-25}$ (0.06) \\
$\parallel$ 18.53 & $1.8\,\, 10^{-28}$ (0.29) & $9.8\,\, 10^{-27}$ (0.33) & $1.2\,\, 10^{-24}$ (0.08) \\
\hline
\end{tabular}
\end{center}
\end{table}

\begin{table}
\caption{As Fig.~\ref{naphthalene} for dicoronylene (C$_{48}$H$_{20}$).}
\label{dicoronylene}
\begin{center}
\begin{tabular}{cccc}
\hline \hline \noalign{\smallskip} 
& \multicolumn{3}{c}{Integrated flux} \\
\noalign{\smallskip} \cline{2-4} \noalign{\smallskip}
 &  & Red & IRAS\\
Peak & ISRF & Rectangle & 21282+5050\\
($\mu$m) & (W$\,\,$sr$^{-1}$) (\%) & (W$\,\,$sr$^{-1}$) (\%)& (W$\,\,$sr$^{-1}$) (\%)\\ 
\noalign{\smallskip} \hline \noalign{\smallskip}
$\perp$ 17.44 & $1.2\,\, 10^{-27}$ (3.61) & $7.4\,\, 10^{-26}$ (3.94) & $1.9\,\, 10^{-23}$ (1.89) \\
$\perp$ 18.44 & $1.3\,\, 10^{-27}$ (4.00) & $8.4\,\, 10^{-26}$ (4.45) & $2.1\,\, 10^{-23}$ (2.03) \\
$\parallel$ 18.75 & $2.6\,\, 10^{-28}$ (0.80) & $1.7\,\, 10^{-26}$ (0.89) & $4.1\,\, 10^{-24}$ (0.40) \\
$\parallel$ 19.53 & $8.3\,\, 10^{-29}$ (0.26) & $5.6\,\, 10^{-27}$ (0.30) & $1.2\,\, 10^{-24}$ (0.12) \\
\hline
\end{tabular}
\end{center} 
\end{table}

\begin{table}
\caption{As Fig.~\ref{naphthalene} for the dicoronylene cation
(C$_{48}$H$_{20}^+$)}
\label{dicoronylene+}
\begin{center}
\begin{tabular}{cccc}
\hline \hline \noalign{\smallskip} 
& \multicolumn{3}{c}{Integrated flux} \\
\noalign{\smallskip} \cline{2-4} \noalign{\smallskip}
&  & Red & IRAS\\
Peak & ISRF & Rectangle & 21282+5050\\
($\mu$m) & (W$\,\,$sr$^{-1}$) (\%) & (W$\,\,$sr$^{-1}$) (\%)& (W$\,\,$sr$^{-1}$) (\%)\\ 
\noalign{\smallskip} \hline \noalign{\smallskip}
$\parallel$ 15.01 & $4.4\,\, 10^{-28}$ (0.82) & $3.0\,\, 10^{-26}$ (0.95) & $5.3\,\, 10^{-24}$ (0.32) \\
$\parallel$ 16.60 & $1.4\,\, 10^{-27}$ (2.58) & $9.2\,\, 10^{-26}$ (2.97) & $1.5\,\, 10^{-23}$ (0.89) \\
$\perp$ 17.59 & $1.1\,\, 10^{-27}$ (2.12) & $7.6\,\, 10^{-26}$ (2.43) & $1.1\,\, 10^{-23}$ (0.66) \\
$\perp$ 18.61 & $7.3\,\, 10^{-28}$ (1.34) & $4.9\,\, 10^{-26}$ (1.57) & $6.7\,\, 10^{-24}$ (0.40) \\
\hline 
\end{tabular}
\end{center} 
\end{table}

\end{document}